\documentclass{article}

\usepackage[a4paper, total={6.2in, 9in}]{geometry}
\usepackage{authblk}
\usepackage{graphicx}
\usepackage{color}
\usepackage{amsmath}
\usepackage[hidelinks, colorlinks=true, citecolor=blue, linkcolor=blue]{hyperref}

\begin{document}

\title{\textbf{Scientific and technological knowledge grows linearly over time}}

% Use letters for affiliations, numbers to show equal authorship (if applicable) and to indicate the corresponding author
\author[a]{Huquan Kang}
\author[a]{Luoyi Fu}
\author[b]{Russell J. Funk}
\author[a,1]{Xinbing Wang}
\author[a]{Jiaxin Ding}
\author[a]{Shiyu Liang}
\author[c]{Jianghao Wang}
\author[a]{Lei Zhou}
\author[c]{Chenghu Zhou}

\affil[a]{School of Electronic Information and Electrical Engineering, Shanghai Jiao Tong University, Shanghai 200240, China}
\affil[b]{Carlson School of Management, University of Minnesota, Minneapolis, MN 55455}
\affil[c]{State Key Laboratory of Resources and Environmental Information System, Institute of Geographic Sciences and Natural Resources Research, Chinese Academy of Sciences, Beijing 100101, China}
\affil[1]{Corresponding author. E-mail: xwang8@sjtu.edu.cn}

\date{}

\maketitle

\begin{abstract}
The past few centuries have witnessed a dramatic growth in scientific and technological knowledge \cite{RN697,RN698}. However, the nature of that growth—whether exponential or otherwise—remains controversial, perhaps partly due to the lack of quantitative characterizations \cite{RN645,RN640,RN688,RN641,RN687,RN636,RN637,RN658,suh2009singularity,RN572,RN696,RN638,horgan2015end,RN658,RN648}. We evaluated knowledge as a collective thinking structure, using citation networks as a representation, by examining extensive datasets that include 213 million publications (1800–2020) and 7.6 million patents (1976–2020). We found that knowledge—which we conceptualize as the reduction of uncertainty in a knowledge network \cite{RN660,RN692,RN693,dretske1981knowledge,RN695}—grew linearly over time in naturally formed citation networks that themselves expanded exponentially. Moreover, our results revealed inflection points in the growth of knowledge that often corresponded to important developments within fields, such as major breakthroughs, new paradigms, or the emergence of entirely new areas of study. Around these inflection points, knowledge may grow rapidly or exponentially on a local scale, although the overall growth rate remains linear when viewed globally. Previous studies concluding an exponential growth of knowledge may have focused primarily on these local bursts of rapid growth around key developments, leading to the misconception of a global exponential trend. Our findings help to reconcile the discrepancy between the perceived exponential growth and the actual linear growth of knowledge by highlighting the distinction between local and global growth patterns. Overall, our findings reveal major science development trends for policymaking, showing that producing knowledge is far more challenging than producing papers.
\vspace{1cm}
\end{abstract}

% Use \firstpage to indicate which paragraph and line will start the second page and subsequent formatting. In this example, there are a total of 11 paragraphs on the first page, counting the first level heading as a paragraph. The value {12} represents the number of the paragraph starting the second page. If a paragraph runs over onto the second page, include a bracket with the paragraph line number starting the second page, followed by the paragraph number in curly brackets, e.g. "\firstpage[4]{11}".

% If your first paragraph (i.e. with the \dropcap) contains a list environment (quote, quotation, theorem, definition, enumerate, itemize...), the line after the list may have some extra indentation. If this is the case, add \parshape=0 to the end of the list environment.
Science and technology have grown significantly over the past decades~\cite{RN698,RN697}, with various metrics such as publications, scientists, inventors, and journals all showing exponential growth~\cite{RN699,RN700,RN701}. However, the question of whether the true object of science and technology—the expansion of knowledge—is also growing at the same explosive rates remains controversial. Many presume that there has been an explosion in knowledge, suggesting exponential growth, and this stance is supported by rapid technological advancement and the proliferation of online information~\cite{RN645,RN640,RN688,RN641,RN687}. The notion of an exponential growth of knowledge was first introduced into academic literature as early as 1961~\cite{RN645}, and since then, has been frequently discussed in diverse fields such as education~\cite{RN640}, politics~\cite{RN688}, and scientific research~\cite{RN641,RN687}. However, others argue that the view of exponential knowledge growth may be overly simplistic~\cite{RN636, RN637,RN658,suh2009singularity,RN572,RN696,RN638,horgan2015end,RN658}. For instance, the expansion of Wikipedia, which encompasses vast knowledge, has decelerated~\cite{suh2009singularity}. As early as 1962, Nobel laureate Fritz Albert Lippmann suggested that knowledge could not grow exponentially due to cognitive limitations and the slow pace of human evolution~\cite{RN638}. Subsequently, numerous studies have indicated that the rate of critical scientific discoveries is declining~\cite{horgan2015end,RN658} and that the extensive existing literature may impede the creation of new knowledge~\cite{RN572}.

The controversy surrounding the rate of knowledge expansion arises from the use of various indicators not specifically designed for quantifying knowledge. In particular, distinguishing between knowledge and publication is essential. While these two concepts are correlated, they are not the same, and although publication has grown exponentially, the growth of knowledge remains unclear. The question of whether knowledge is exponentially growing has been raised, but due to the difficulty of measuring knowledge, no definitive answer exists~\cite{RN648}. Understanding the actual rate of knowledge expansion is crucial, as this understanding informs future policy decisions within the scientific community and guides efforts to foster innovation and drive progress in various fields. Accurately assessing the growth of knowledge can help prioritize research efforts, allocate resources effectively, and develop strategies to facilitate the creation and dissemination of new knowledge.

To better understand and quantitatively characterize knowledge growth, we analyzed 213 million publications (1800–2020) from the Microsoft Academic Graph (MAG) and 7.6 million patents (1976–2020) from the United States Patent and Trademark Office's (USPTO) Patents View database (see \textcolor{blue}{\textit{Methods}}). The MAG data include a field classification of the publications into 19 major disciplines and 292 secondary subjects. We constructed yearly citation network snapshots using MAG data, Patents View data, and subject data obtained by partitioning MAG. Then, we applied a previously established metric—the knowledge quantification index (KQI)~\cite{RN660}—in analyses of each snapshot to observe changes in knowledge over time (see \textcolor{blue}{\textit{Methods}}). As an additional measure to corroborate our findings, we analyzed the evolution of the time required to prove mathematical conjectures over the past six decades. This investigation offers a window into the pace of knowledge creation and confirmation within the field of mathematics, which can be used to support our broader conclusions about the growth rate of knowledge across disciplines. Our analysis of these data revealed universal patterns in the growth of knowledge and shed light on possible reasons behind the illusion of exponential growth.

\section*{Results}

\subsection*{Measurement of knowledge}

\begin{figure*}%[tbhp]
	\centering
	\includegraphics[width=6.2in]{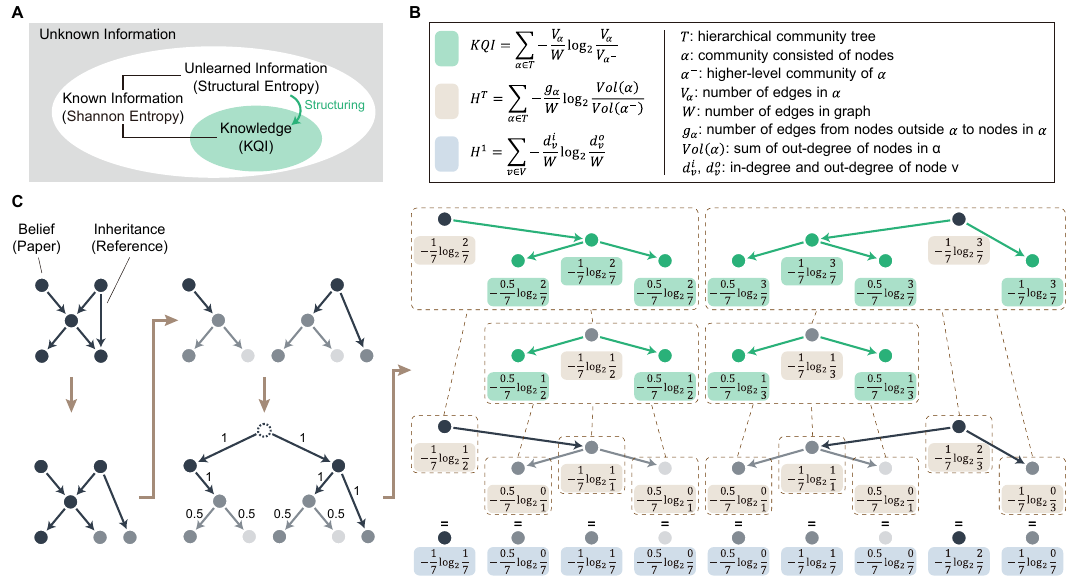}
	\caption{Overview of the measurement approach. (\textit{A}), The relationship between knowledge, structure, and entropy. Humans discover unknown information to stretch the boundaries of known information, which is measured by Shannon entropy~\cite{shannon1948a} in information bits. By structuring known information, humans learn knowledge, which emerges through transforming isolated, disordered data into interconnected, ordered data. Structural entropy~\cite{angsheng2016structural} has been proven capable of quantifying information organized with a network structure, which is a hierarchical community structure. By quantifying information with and without structure, the knowledge represented by structure is quantified by the difference between the two. (\textit{B}), The formula of Knowledge Quantification Index (KQI), structural entropy, and Shannon entropy. In the algorithm implementation of KQI, we use the equivalent form of the formula to calculate (see \textit{Methods}). (\textit{C}), The principle of the KQI applied to a hypothetical citation network. The citation network is first decomposed into multiple trees, and a virtual node acting as the source of all knowledge connects all the roots. The tree structure represents the knowledge formed in ideological inheritance. The KQI quantifies the potential inheritance structure of beliefs within the network. The mathematical expressions of KQI (green) are equal to the difference between Shannon entropy (blue) and structural entropy (brown).}
	\label{fig:kqi}
\end{figure*}

Knowledge is a network structure of interconnected beliefs, intertwined with various pieces of information~\cite{RN692,RN693,dretske1981knowledge,RN695}. Humans acquire knowledge by organizing and structuring the information they have learned. Building on this idea, we quantify knowledge using a measure—the knowledge quantification index (KQI)~\cite{RN660}—that conceptualizes knowledge as a network structure of beliefs and embodies the certainty of information the structure brings (Fig. \ref{fig:kqi}). KQI is defined as the difference between Shannon entropy~\cite{shannon1948a} of the node (i.e., paper) degree distribution and the structural entropy~\cite{angsheng2016structural} of the citation graph, where Shannon entropy measures the information of discrete data and structural entropy measures the information of the network. Fig. \ref{fig:kqi}\textit{B} shows the formula of KQI and how it is calculated. KQI follows the traditional definition of knowledge, justified true belief (JTB) theory~\cite{sep-knowledge-analysis}, and is closer to the essence of knowledge than counting publications, scientists, and concepts. The effectiveness of KQI in evaluating scientific knowledge has been thoroughly validated through multiple approaches, including by examining its correlation with the impact of classical literature and the work of Nobel laureates~\cite{RN660}. By treating each publication as a belief and each citation as an inheritance, KQI quantifies the uncertainty reduced by the collective thinking structure constructed from the scientific publications. This approach allows us to systematically examine the growth rate of knowledge over time.

\subsection*{Linear growth of knowledge}

\begin{figure*}%[tbhp]
	\centering
	\includegraphics[width=6.2in]{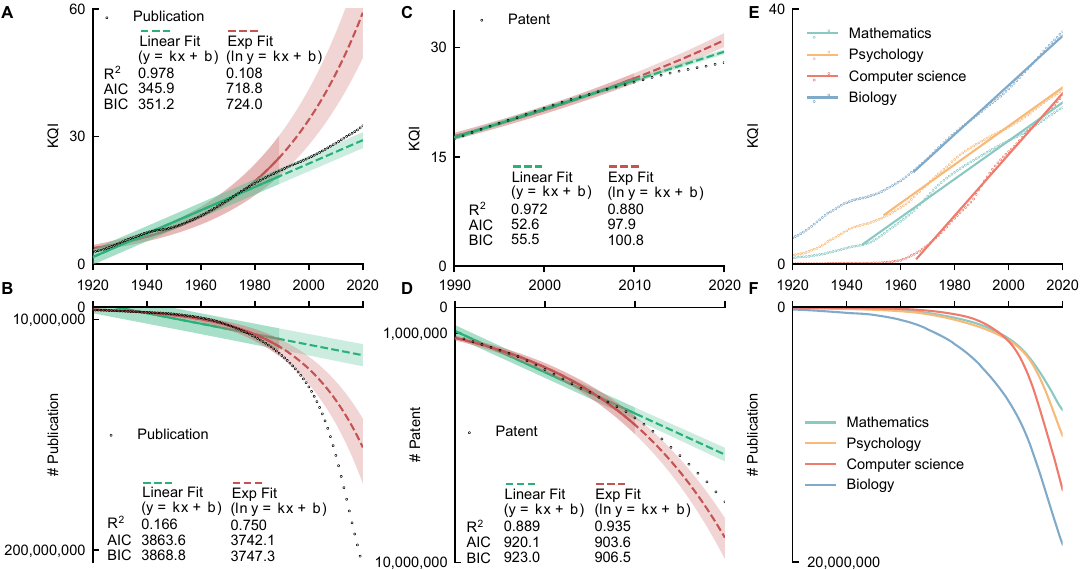}
	\caption{Linear growth of knowledge. (\textit{A-D}), Growth of KQI and the number of publications or patents in MAG and Patents View. The green lines and red curves provide regression fittings for linear and exponential models. The first 70\% of the data is for regression fitting (solid line), and the last 30\% is for forecasting (dashed line). The shaded bands represent the 95\% confidence interval. Coefficient of determination and information criterion validate trends more suitable for a given data series. (\textit{E-F}), Growth of KQI and number of publications in the disciplines of mathematics (green), psychology (orange), computer science (red), and biology (blue). Straight lines exhibit trends approaching linearity starting from certain years.}
	\label{fig:linear}
\end{figure*}

The number of publications has been increasing exponentially (Fig. \ref{fig:linear}\textit{B}), in line with the idea of an “information explosion.” When analyzing the growth of knowledge, however, we found that it was not increasing at the same rate. Instead, the growth of KQI was almost linear (Fig. \ref{fig:linear}\textit{A}), suggesting that scientific productivity—conceptualized as the number of papers required to grow knowledge—may be declining~\cite{horgan2015end,RN650,RN651}. Unlike the explosive growth in the number of publications, the number of patents was relatively slower (Fig. \ref{fig:linear}\textit{D}). Remarkably, despite the significant difference between the growth rates of patents and publications, the KQI of patents exhibited the same linear growth pattern (Fig. \ref{fig:linear}\textit{C}). We used linear regression and exponential regression to fit the above growth trends, and both the coefficient of determination and information criterion show that the linear trend reflects knowledge growth better than the exponential trend (see \textcolor{blue}{\textit{Methods}}). To verify this finding in more realistic scenarios, we split publications by disciplines and subjects. Although the rate of knowledge growth varied across different fields, they all exhibited patterns of linear growth (Fig. \ref{fig:linear}\textit{E,F}). The increase in knowledge was notably different from the growth pattern of scientific productivity.

\begin{figure*}%[tbhp]
	\centering
	\includegraphics[width=6.2in]{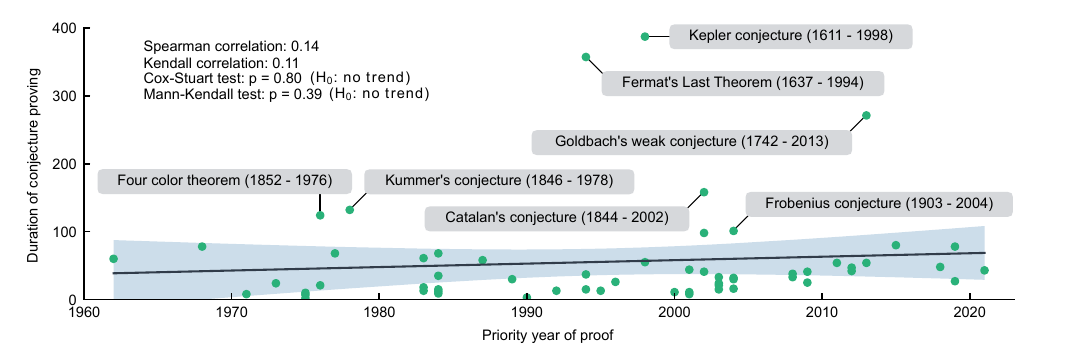}
	\caption{Duration of mathematical conjecture proving. The green scatter shows the duration (from the formulation to the proof completion, in years) of mathematical conjectures proved since 1960, with several notable examples highlighted. The solid black line is the least square linear regression, and the blue shaded band represents the 95\% confidence interval. Spearman and Kendall rank correlation coefficients indicate a weak relationship between the duration of conjecture proving and the priority year of proof. The two-sided Cox-Stuart and Mann-Kendall hypothesis tests show that the duration of the mathematical conjectures' proofs has not changed significantly.}
	\label{fig:math}
\end{figure*}

We also find evidence of a linear growth pattern in knowledge in alternative analyses that do not rely on KQI. In particular, we also observe this trend in the development of mathematical conjectures~\cite{RN684,wigner1990unreasonable}. The formulation and proof of mathematical conjectures reflects the level of understanding and mastery of knowledge, with the duration from formulation to proof corresponding to the time required for knowledge to move from understanding to mastery. We therefore collected data on 61 proven conjectures since 1960 and calculated the average duration per year. The results of correlation analysis and hypothesis testing indicate that there has been no significant change in the rate of human knowledge acquisition (Fig. \ref{fig:math}, see \textcolor{blue}{\textit{Methods}}). This finding suggests that we have been consistently building and expanding our understanding of the world and our place within it at a steady pace.

\subsection*{Inflection points in knowledge growth}
\begin{figure*}%[tbhp]
	\centering
	\includegraphics[width=6.2in]{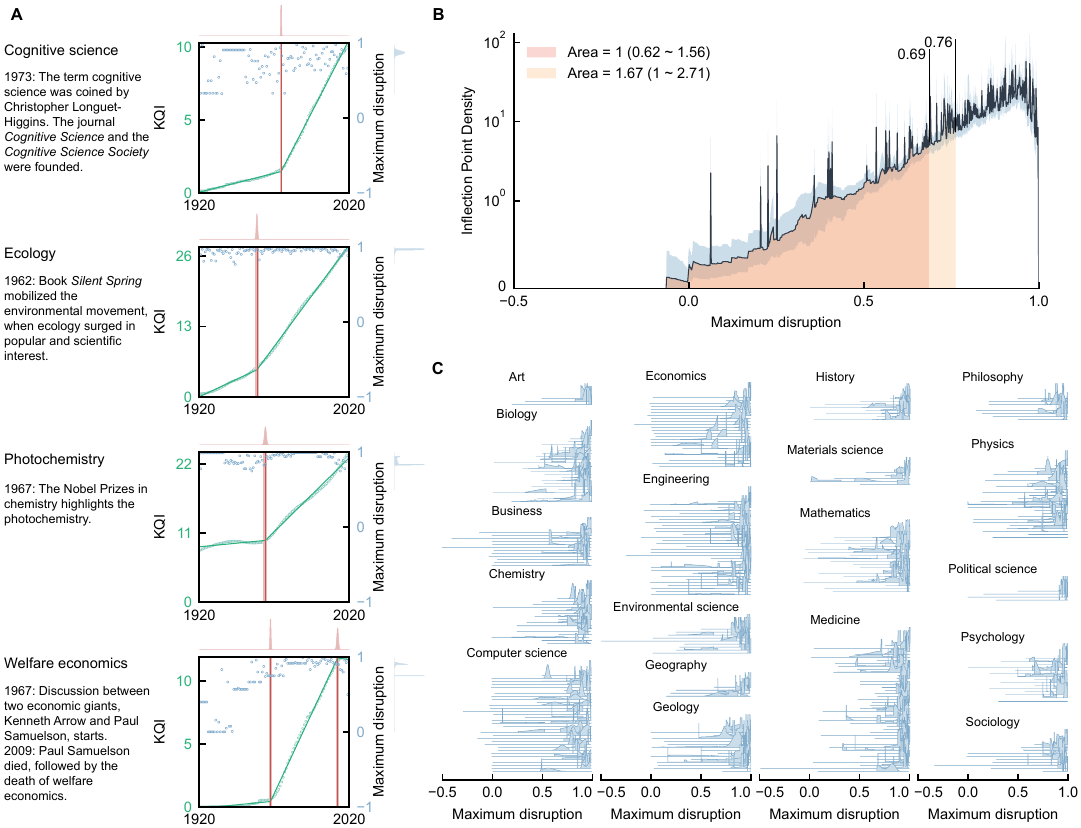}
	\caption{Inflection points in KQI evolution. (\textit{A}), Inflection points in distinct disciplines. The green circles represent the KQIs computed for the entire network at different years. The blue circles represent the maximum disruption of papers published at each year. The green lines are the segmented linear regression results. The red lines denote the estimated inflection points, and the red shaded bands represent the standard deviations. The upper insets show the norm distribution of inflection points that we assumed. The right insets show the transformed distribution of inflection points with maximum disruption. The potentially relevant key events are also marked. (\textit{B}), Distribution of inflection points with maximum disruption. The inflection density (see Methods) is plotted as a black line, while a blue-shaded band indicates the 95\% confidence interval. The y-axis is scaled using symlog, producing a linear plot within the specified range of values near zero (<1). Two regions of interest in the density curve are highlighted using different colored shaded regions. During the evolution of the network, on average, one inflection point occurs as the maximum disruption increases from 0 to 0.69. There is a 95\% probability of experiencing at least one inflection as the maximum disruption increases from 0 to 0.76. (\textit{C}), The distribution of inflection points for 311 disciplines taken in b. Each line represents the evolving network of a discipline, and its extent on the x-axis corresponds to the range of maximum disruption for the evolving network.}
	\label{fig:inflection}
\end{figure*}

Our examination of 19 major disciplines and 292 secondary subjects revealed that knowledge generally follows a linear growth pattern. However, we also observed significant inflection points between different stages of linear growth (Fig. \ref{fig:inflection}\textit{B-E} and \textcolor{blue}{\textit{SI Appendix}, Fig. S1} and \textcolor{blue}{Table S2}). Before an inflection point, the growth of knowledge increases linearly at a specific rate, which changes to another rate after the inflection point. Interestingly, the number of publications around these inflection points, which correspond to important developments in the field, does not show a noticeable or distinctive change.

Intuitively, inflection points seem likely to be associated with major discoveries or breakthroughs in a field that catalyze the growth of future knowledge. Consequently, we hypothesized that inflection points will tend to be accompanied by disruptive research within a field, as breakthrough scientific products lead to a paradigm shift in scientific research. To measure disruptiveness, we compute the CD index of each paper (see \textcolor{blue}{\textit{Methods}}), which captures how subsequent papers cite a focal paper while disregarding its references, and has been used to identify scientific breakthroughs~\cite{RN509,RN658}. We used the maximum CD-index of papers to characterize the disruptiveness of a field at a particular time. We found that when the field had a high maximum disruption, meaning that the field encountered a scientific breakthrough, the probability of inflection points occurring was high (Fig. \ref{fig:inflection}\textit{A}). Conversely, when the maximum disruption was low, inflection points rarely occur. This finding validates the relationship between the inflection points and scientific breakthroughs, and also confirms the reliability of KQI as a measure of knowledge.

\subsection*{Diminishing returns of knowledge}
\begin{figure*}%[tbhp]
	\centering
	\includegraphics[width=6.2in]{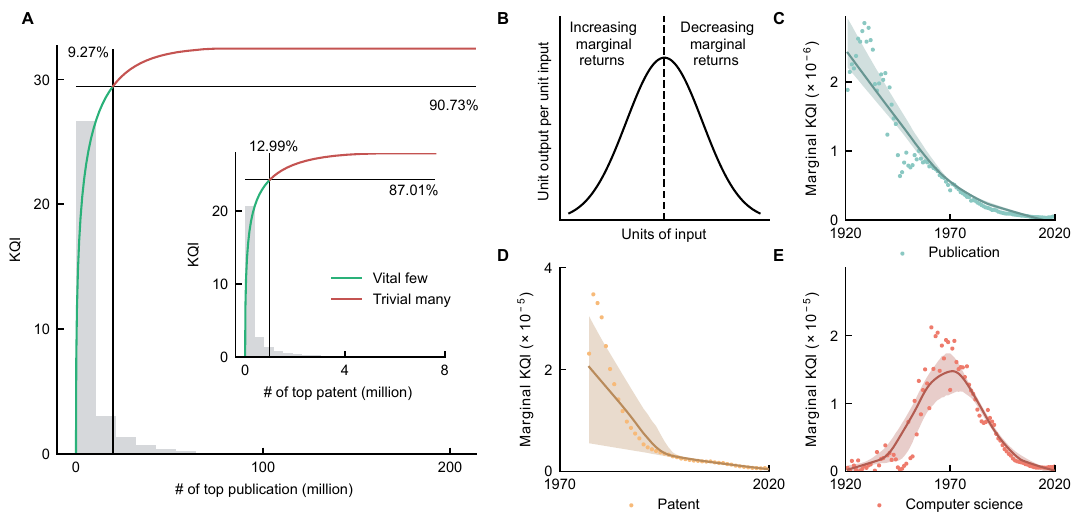}
	\caption{Pareto principle and diminishing returns in KQI. (\textit{A}), Cumulative KQI distribution after sorting publications by descending KQI order. The grey bar represents a histogram of the publications sorted in descending order according to KQI with KQI as its weight, expressing the contribution of the respective range of publications on KQI. The green and red lines are the cumulative curves of the KQI distribution; at their critical point, the ratio of publications in the vital few equals the ratio of trivial many to the total KQI contribution. The main plot displays statistics for publications in MAG, while the inset plot displays statistics for patents in Patents View. (\textit{B}), Demo illustrating the law of diminishing returns. (\textit{C-E}), Marginal KQI (see Methods) increment per publication (or patent) over time. The circle markers represent the average incremental KQI from each publication (or patent) in that year. The solid curves show the local regression, and the shaded bands indicate the 95\% confidence interval.}
	\label{fig:marginal}
\end{figure*}

The large discrepancy between the growth rate of knowledge and publication implies that the knowledge contained in each publication is declining, as implied in previous work~\cite{RN658,RN572}. The Pareto principle, also known as the 80/20 rule, describes the general imbalance of wealth distribution among individuals. We found a similar but even more imbalanced knowledge distribution in academia, where 90.73\% of knowledge comes from 9.27\% of scientific productivity; in comparison, 9.27\% of knowledge comes from 90.73\% of scientific productivity (Fig. \ref{fig:marginal}\textit{A}). With abundant literature as a backdrop, the increased knowledge resulting from a newly published article is minimal, identifying with the previously reported stagnation caused by massive literature~\cite{RN572}. This imbalance also suggests that the increased knowledge begins to diminish after establishing a certain level of background knowledge (Fig. \ref{fig:marginal}\textit{B-E}), consistent with the disruptiveness decline~\cite{RN658}. As new areas of knowledge expansion become increasingly complex, each incremental advancement appears almost trivial compared to the vast body of existing knowledge. At this stage, only marginal gains can be achieved without substantial investments of time and resources. Despite the disproportionate gap between effort and outcomes, maintaining and enhancing scientific productivity remains essential for the incremental progress of knowledge, perhaps unless a paradigm shift can be achieved.

\section*{Discussion}

This study presents novel findings that elucidate the patterns of knowledge growth and, for the first time, explain the origins of the perceived exponential growth. Our results suggest that linear rather than exponential growth better describes knowledge growth. This linear trend in knowledge growth implies that the emergence of new scientific breakthroughs may occur at more steady intervals than previously assumed. Our findings contribute to a deeper understanding of the dynamics of knowledge growth and its potential impact on the scientific community. By clarifying the true nature of knowledge growth and the factors that contribute to the illusion of exponential growth, this study provides valuable insights for policymakers, researchers, and other stakeholders involved in the production and dissemination of scientific knowledge.

Whether knowledge is experiencing an explosion has been questioned in previous studies~\cite{RN638,RN649,RN648,horgan2015end,milojevic2015quantifying}. Quantifying knowledge, however, was challenging before introducing KQI as a metric~\cite{RN648}. Our results immaculately validate the previous skepticism that knowledge did not grow exponentially. Recent studies that suggest decreasing disruptiveness~\cite{RN658} and slowed canonical progress~\cite{RN572} in scientific fields imply a slowdown in knowledge growth, yet leave unanswered how much this slowdown occurs. Our findings, on the one hand, discard earlier optimistic claims of ``knowledge explosion'' and, on the other hand, clarify recent pessimistic claims of a scientific slowdown. Despite the slowdown of knowledge growth relative to the proliferation of publications, overall knowledge remains steadily increasing. Our concern in slowing scientific activities should be more on whether the expansion of science is sustainable and how to improve scientific efficiency.

Our findings regarding the prevalence of inflection points explain the perceived exponential growth of knowledge, i.e., the sudden disparity between pre- and post-transition knowledge growth rates creates an illusion of exponential growth. These transition points suggest that fundamental changes occur, like knowledge acquisition, as scientific research reaches a certain level of expertise. Understanding this shift could provide valuable insights into the process of scientific discovery. Based on our preliminary observation of the distribution pattern of inflection points, it is easier to observe an inflection when the maximum CD index is high when it is possible that a major scientific breakthrough has occurred. This relation also confirms the reliability of KQI as a measure of knowledge.

Some studies have shown inequality in academia~\cite{RN676,RN677,RN678,RN679}. Our reported inequality in knowledge contributions further reveals that the continuously expanding citation network causes such inequality. The accompanying diminishing returns of knowledge postulates that the more we acquire knowledge, the more cognizant we become of the vastness of the unknown and the more necessary it is to reassess our earlier perceptions. This implies that the growth of knowledge is subject to similar constraints as other complex systems. 

The present study is not without limitations. The KQI is a novel metric for quantifying knowledge and has yet to gain widespread acceptance within the scientific community. Further investigation into the characteristics and value of KQI will contribute to its gradual recognition. Additionally, our research focuses primarily on academic publications and patents, limiting our analysis's scope. Future work involving various media types may deepen our understanding of knowledge evolution. As knowledge network construction is complicated, particularly in defining concepts and relationships, the present study utilized a citation network constructed using publications or patents as nodes without consideration for semantics. In the future, incorporating more diverse knowledge network types, such as knowledge graphs, may further enrich the results of analyzing knowledge evolution patterns.

In conclusion, the findings of this study have significant implications for understanding the growth of knowledge. The use of KQI as a metric for measuring knowledge growth provides a valuable tool for quantifying this complex phenomenon. Moreover, our results support emergence theories arising from inflection points in knowledge growth and the concept of a complexity brake. Further investigation into the causes of these inflection points and the constraints on knowledge growth is necessary to gain a deeper understanding of the dynamics at play. Our findings also suggest that we should not be overwhelmed by the perceived ``explosion'' of knowledge but rather focus on managing the rapid growth of information. The actual growth of knowledge appears to be more measured and steadier than the exponential growth in publications might suggest. 

% \begin{figure}%[tbhp]
% \centering
% \includegraphics[width=.8\linewidth]{frog.pdf}
% \caption{Placeholder image of a frog with a long example legend to show justification setting.}
% \label{fig:frog}
% \end{figure}

% \begin{table}[t!]
% \centering
% \caption{Comparison of the fitted potential energy surfaces and ab initio benchmark electronic energy calculations}
% \begin{tabular}{lrrr}
% Species & CBS & CV & G3 \\
% \midrule
% 1. Acetaldehyde & 0.0 & 0.0 & 0.0 \\
% 2. Vinyl alcohol & 9.1 & 9.6 & 13.5 \\
% 3. Hydroxyethylidene & 50.8 & 51.2 & 54.0\\
% \bottomrule
% \end{tabular}

% \addtabletext{nomenclature for the TSs refers to the numbered species in the table.}
% \end{table}

% \begin{SCfigure*}[\sidecaptionrelwidth][t!]
% \centering
% \includegraphics[width=11.4cm,height=11.4cm]{frog.pdf}
% \caption{This legend would be placed at the side of the figure, rather than below it.}\label{fig:side}
% \end{SCfigure*}

% \begin{figure*}[bt!]
% \begin{align*}
% (x+y)^3&=(x+y)(x+y)^2\\
%        &=(x+y)(x^2+2xy+y^2) \numberthis \label{eqn:example} \\
%        &=x^3+3x^2y+3xy^3+x^3.
% \end{align*}
% \end{figure*}

\section*{Materials and Methods}
\label{sec:method}

\subsection*{Publications data}
Our data is derived from the Microsoft Academic Graph data, which archives publications from 1800 to 2021. The publications cover 292 secondary subjects in 19 major disciplines, including but not limited to Economics, Biology, Computer science, and Physics. We excluded patents, datasets, and repositories, utilizing the doctype field in the MAG data. We limited our focus to publications up to 2020 because recent literature was not sufficiently collected. Although we used literature from as early as 1800, the KQI was only calculated from 1920 because the citations were too sparse to be interconnected in the early years. We removed possible errors in the data, including self-citations, duplicate citations, and citations violating time order. After eliminating potentially incorrect publications and closing the data up to 2020, the analytical sample consisted of 213,715,816 publications and 1,762,008,545 citations. The MAG data provides subject classifications for papers, and we used this information to filter papers in a specific subject to create a smaller subject citation network. While processing data from a particular subject, we only preserved citation relationships that both article and reference are on the same subject, thus guaranteeing that all nodes within the network are from the same subject.

\subsection*{Patents data}
The Patents View data collect 8.1 million patents granted between 1976 and 2022 and their corresponding 126 million citations. We limited our focus to citations made to U.S. granted patents by U.S. patents up to 2020 because recent patents were not yet sufficiently collected. Although we used patents from 1976, the KQI was only calculated from 1990 because the citations were too sparse to be interconnected in the early years. We removed possible errors in the data, including self-citations, duplicate citations, and citations violating time order. After eliminating potentially incorrect patents and closing the data up to 2020, the analytical sample consisted of 7,627,229 patents and 101,148,606 citations.

\subsection*{Derivation of KQI}
KQI~\cite{RN660} is the difference between Shannon entropy~\cite{shannon1948a} and structural entropy~\cite{angsheng2016structural}. Shannon entropy was initially defined on a discrete probability distribution, i.e., $H(X) = \sum_x{-p_x log_2 {p_x}}$, where $X$ is a discrete random variable and $p_x$ is the probability. The essence of Shannon entropy is the expected amount of information, where $-log_2 {p_x}$ is the amount of information and $p_x$ is the weight. The Shannon entropy of an undirected graph usually uses a degree probability distribution. In a directed graph, we similarly define Shannon entropy as the expected information of out-degree probability distribution weighted by the in-degree. Let $V$ be nodes set of graph, $W$ be number of edges, $d_v^i$ and $d_v^o$ be the in-degree and out-degree of node $v$, the formula is shown as follow:
$$
H^1 = \sum_{v \in V}{-\frac{d_v^i}{W} log_2 {\frac{d_v^o}{W}}}.
$$
Structural entropy is defined as follow:
$$
H^T = \sum_{\alpha \in T}{-\frac{g_\alpha}{W} log_2 {\frac{Vol(\alpha)}{Vol(\alpha^-)}}},
$$
where $T$ is the hierarchical community tree, $\alpha$ is the community consisted of nodes, $\alpha^-$ is the farther community of $\alpha$, $g_\alpha$ is the number of edges from nodes outside $\alpha$ to nodes in $\alpha$, $Vol(\alpha)$ is the sum of out-degree of nodes in $\alpha$.
The unsimplified KQI formula is expressed as:
$$
K = \sum_{\alpha \in T}{-\frac{V_\alpha}{W} log_2 {\frac{V_\alpha}{V_{\alpha^-}}}},
$$
where $V_\alpha$ is the number of edges in $\alpha$. Therefore,
\begin{align*}
K &= \sum_{\alpha \in T}{-\frac{V_\alpha}{W} log_2 {\frac{Vol(\alpha)}{Vol(\alpha^-)}} }\\
  &= \sum_{\alpha \in T}{-\frac{\sum_{v \in \alpha}d_v^i -g_\alpha}{W} log_2 {\frac{Vol(\alpha)}{Vol(\alpha^-)}} } \\
  &= \sum_{\alpha \in T}{-\frac{\sum_{v \in \alpha}d_v^i}{W} log_2 {\frac{Vol(\alpha)}{Vol(\alpha^-)}} }-\sum_{\alpha \in T}{-\frac{g_\alpha}{W} log_2 {\frac{Vol(\alpha)}{Vol(\alpha^-)}} }\\
  &= \sum_{v \in V}{-\frac{d_v^i}{W} \sum_{\alpha \in \{\alpha|v \in \alpha\}}log_2 {\frac{Vol(\alpha)}{Vol(\alpha^-)}} }-H^T\\
  &= \sum_{v \in V}{-\frac{d_v^i}{W} log_2 {\frac{d_v^o}{W}} }-H^T\\
  &= H^1-H^T.
\end{align*}

\subsection*{KQI calculation}
KQI~\cite{RN660} is a metric that quantifies knowledge from the perspective of information structurization. KQI quantifies knowledge by utilizing the hierarchical structure of citation networks. Unlike traditional metrics focusing on counting, the KQI considers the entire citation network following the justified true belief (JTB) theory~\cite{sep-knowledge-analysis}, a classical knowledge definition. The calculation of KQI requires input from a directed acyclic graph, which contains knowledge inheritance relationships. The algorithm to calculate KQI involves the following steps:

\begin{enumerate}
    \item \textit{Calculate the volume of each node.} The volume of each node is obtained by the sum of its out-degree and the volumes of its children. Before adding the volumes of its children, they should be divided by their in-degrees.
	\item \textit{Calculate the KQI of each node following the formula:}
	$$
	K_\alpha=\sum_{\beta\rightarrow\alpha}{-\frac{V_\alpha}{dW} log_2{\frac{V_\alpha}{dV_\beta}}},
	$$
	where node $\beta$ is the father of node $\alpha$, $V$ is the volume, $d$ is the in-degree, and $W$ is the total volume of the graph.
	\item \textit{Sum over KQIs of all nodes:} $K= \sum {K_\alpha}$.
\end{enumerate}

The code for calculating KQI is open sourced and provided with the paper. The computational complexity of the algorithm is linear and can be applied to large-scale data. In order to observe the growth trend of KQI, we constructed year-by-year publication citation networks or patent citation networks, which are transformed into directed acyclic graphs. We calculated the KQI of each node in a citation network and added them up to obtain the KQI of the citation network for each year, exploiting the additivity of the KQI as pointed out by the proposers.

\subsection*{Test of growth trend}
We performed linear and exponential regression fittings to test which growth trend is more suitable to describe a data series. We used ordinary least squares to determine the best parameters. For exponential regression, we calculate the logarithm of the data and then perform linear regression. R-squared, Akaike information criterion (AIC), and Bayesian information criterion (BIC) are adjusted to the original data for exponential regression.

\subsection*{Analysis of mathematical conjectures}
We collected 61 mathematical conjectures proven to be correct since 1960 (\textcolor{blue}{\textit{SI Appendix}, Table 1}) with the help of a Wikipedia list and expert advice. Conjectures proved wrong and those not yet proven are excluded because their durations from understanding to mastery are unavailable. Some conjectures do not have a specific formulation or proof year, so we chose a median of possible time ranges for such cases instead. Due to certain mathematical conjectures proved in the same year but with different durations and difficulty in determining their temporal order, we took the average duration of proofs within the same year. We used a time series of 32 average durations from different proof years for correlation analyses and hypothesis testing. Spearman and Kendall~\cite{RN668} rank correlation coefficients are non-parametric measures of the strength of monotonic association between two variables and are calculated by measuring the rank correlation between two variables. The range of these two coefficients is from -1 to 1, with values closer to 0 indicating a weaker relationship between the two variables. Cox-Stuart~\cite{cox1955some} and Mann-Kendall~\cite{mann1945nonparametric} hypothesis tests assess whether there is a monotonic increasing or decreasing trend over time in a time series data. The null hypothesis for both hypothesis tests is the absence of a monotonic trend, so a p-value greater than 0.05 indicates the lack of a significant trend.

% \subsection*{Limitation of knowledge growth}
% We prove that the growth rate of KQI has an upper bound regarding the graph size through a theoretical derivation from the KQI formula. We rewrite the formula of KQI:
% $$
% K_{\alpha} = \sum_{\beta \rightarrow \alpha}{\frac{d_\alpha^i V_{\beta}-V_{\alpha}}{d_\alpha^i W} \log{{\left(1+\frac{1}{\frac{V_{\alpha}}{d_\alpha^i V_{\beta}-V_{\alpha}}}\right)}^{\frac{V_{\alpha}}{d_\alpha^i V_{\beta}-V_{\alpha}}}}}.
% $$
% Applying the Euler limit formula, it is simplified as follows:
% $$
% K_{\alpha} = \frac{a}{W} \left(\sum_{\beta \rightarrow \alpha}{V_{\beta} - V_{\alpha}}\right), (0<a<\log{e}).
% $$
% We sum KQI over all nodes and note that W is the sum of the out-degrees of all nodes. The relation between K and W is thus derived:
% \begin{align*}
% K \triangleq \sum_{\alpha}{K_{\alpha}} &= \frac{a}{W} \sum_{\alpha}{\left(\sum_{\beta \rightarrow \alpha}{V_{\beta} - V_{\alpha}}\right)} \\
% 									   &= a \sum_{\alpha}{\frac{\left(d_\alpha^o - 1\right)}{W} V_{\alpha}} < a \mathbb{E}{\left(V_{\alpha}\right)} < W \log{e}.
% \end{align*}

\subsection*{Calculation of the CD index}
The CD index~\cite{RN658,RN509} quantifies the degree to which a paper disrupts or advances existing literature. Calculating the CD index involves dividing subsequent papers citing a focal paper into three types:
\begin{enumerate}
	\item Type $i$: Papers that exclusively cite the focal paper but disregard its references.
	\item Type $j$: Papers that cite both the focal paper and its references.
	\item Type $k$: Papers that solely cite the references of the focal paper but not the focal paper itself.
\end{enumerate}

The formula is as follows, with $n$ denoting the number of each type of paper:
$$
CD = \frac{n_i-n_j}{n_i+n_j+n_k}.
$$
Following the convention, we calculate the CD index for all papers with at least one forward and backward citation. The maximum CD index is the largest value selected from papers published in a given year in a field.

\subsection*{Discovery of inflection points}
We discover the inflection points using segmented regression models developed by Vito M. R. Muggeo~\cite{RN667}. The segmented regression was performed on the curve of KQI over time, with the regression line breakpoint considered the inflection point. We started with the null hypothesis of no breakpoint and performed a score test to determine if there was an additional breakpoint~\cite{RN666}. This process repeated until there was no additional breakpoint. The significance level was 0.01. To counteract the multiple comparisons problem, we employed Bonferroni correction, requiring that the p-values for each of the first k tests be smaller than 0.01/k. Once the number of breakpoints was determined, we used the segmented method to estimate their positions. 

\subsection*{Estimation of inflection density}
The inflection density is a quantity that characterizes the distribution of the network state at its transition between two different regimes. The area under an inflection density curve represents the average times finding the system in the inflection state. The main text investigates the inflection density per unit maximum CD index. Due to the estimation error of inflection points, we map the probability densities of normal distributions centered around the inflection points to the maximum CD index using linear interpolation and summing in cases with multiple mapping values. The estimated standard deviation of the inflection point determines the standard deviation of the normal distribution. We estimated the inflection density of a discipline by summing the probability densities with respect to maximum CD index during network evolution. The total inflection density is the mean of densities for 311 disciplines. 1000 bootstrap resamplings are employed to estimate confidence intervals.

\subsection*{Calculation of marginal KQI}
The marginal KQI is calculated by subtracting the KQI for a given graph of the previous year from the current year and dividing it by the number of nodes added in the current year. We applied the LOWESS (locally weighted scatterplot smoothing) nonparametric regression method to perform local regression of marginal KQI. To estimate the 95\% confidence interval of the LOWESS fit, we performed 1000 bootstrap resamplings. The fraction of data used when estimating was 2/3.

\subsection*{Data availability}
Data from MAG and Patents View are publicly available. Data from MAG are requested from Acemap (https://www.acemap.info/) at Shanghai Jiao Tong University and are available at \\ https://zenodo.org/record/7878551. Data from Patents View are available at https://patentsview.org/. Other source data are provided with this paper. 

\subsection*{Code availability}
Open-source code for calculating KQI is available at https://github.com/Girafboy/KQI.

%Use \bibsplit to split the references from the body of the text. Value "[2]" represents the number of reference in the left column (Note: Please avoid single column figures & tables on this page.)

% Bibliography
\bibliographystyle{plain}
\bibliography{pnas-sample}

\end{document}

\title{{\Huge\sffamily\bfseries Supporting Information for\par}
\textbf{Scientific and technological knowledge grows linearly over time}}

% Use letters for affiliations, numbers to show equal authorship (if applicable) and to indicate the corresponding author
\author[a]{Huquan Kang}
\author[a]{Luoyi Fu}
\author[b]{Russell J. Funk}
\author[a,1]{Xinbing Wang}
\author[a]{Jiaxin Ding}
\author[a]{Shiyu Liang}
\author[c]{Jianghao Wang}
\author[a]{Lei Zhou}
\author[c]{Chenghu Zhou}

\affil[a]{School of Electronic Information and Electrical Engineering, Shanghai Jiao Tong University, Shanghai 200240, China}
\affil[b]{Carlson School of Management, University of Minnesota, Minneapolis, MN 55455}
\affil[c]{State Key Laboratory of Resources and Environmental Information System, Institute of Geographic Sciences and Natural Resources Research, Chinese Academy of Sciences, Beijing 100101, China}
\affil[1]{Corresponding author. E-mail: xwang8@sjtu.edu.cn}

\date{}

\maketitle

\subsection*{This PDF file includes:}
\begin{itemize}
    \item[] Fig. S1
    \item[] Tables S1 to S2
    \item[] Legends for Dataset S1 to S4
\end{itemize}

\subsection*{Other supporting materials for this manuscript include the following:}
\begin{itemize}
    \item[] Datasets S1 to S4
\end{itemize}
%% Comment out or remove this line before generating final copy for submission; this will also remove the warning re: "Consecutive odd pages found".
% \instructionspage  

\maketitle

%% Adds the main heading for the SI text. Comment out this line if you do not have any supporting information text.
% \SItext

% \subsection*{Subhead}
% Type or paste text here. This should be additional explanatory text such as an extended technical description of results, full details of mathematical models, etc.   

% \section*{Heading}
% \subsection*{Subhead}
% Type or paste text here. You may break this section up into subheads as needed (e.g., one section on ``Materials'' and one on ``Methods'').

% \subsection*{Materials}
% Add a materials subsection if you need to.

% \subsection*{Methods}
% Add a methods subsection if you need to.

%%% Each figure should be on its own page
% \begin{figure}
% \centering
% \includegraphics[width=\textwidth]{example-image}
% \caption{First figure}
% \end{figure}

% \begin{figure}
% \centering
% \includegraphics[width=\textwidth]{frog}
% \caption{Second figure}
% \end{figure}

% \begin{table}\centering
% \caption{This is a table}

% \begin{tabular}{lrrr}
% Species & CBS & CV & G3 \\
% \midrule
% 1. Acetaldehyde & 0.0 & 0.0 & 0.0 \\
% 2. Vinyl alcohol & 9.1 & 9.6 & 13.5 \\
% 3. Hydroxyethylidene & 50.8 & 51.2 & 54.0\\
% \bottomrule
% \end{tabular}
% \end{table}

\begin{table}[H]
    \centering
    \begin{threeparttable}
        \caption{Mathematical conjectures since 1960, sorted by priority date.}
        \scriptsize
        \begin{tabular}{p{0.5in}p{0.5in}p{2.8in}p{1.9in}}
        \toprule
        Priority date &  Conjecture date &                                                            Proved by &                                        Former name \\
        \midrule
                1962 &             1902 &                                    Walter Feit, and John G. Thompson &                                Burnside conjecture \\
                1968 &             1890 &                                Gerhard Ringel, and John W. T. Youngs &                                 Heawood conjecture \\
                1971 &             1963 &                                                       Daniel Quillen &                                   Adams conjecture \\
                1973 &             1949 &                                                       Pierre Deligne &                                   Weil conjectures \\
                1975 &             1968 &                                    Henryk Hecht, and Wilfried Schmid &                              Blattner's conjecture \\
                1975 &             1965 &                                                      William Haboush &                                 Mumford conjecture \\
                1975 &             1973 &                                                       Václav Chvátal &                                Art Gallery Problem \\
                1975 &             1936 &                                           Paul Erdős, and Paul Turán &                                Szemerédi's theorem \\
                1976 &             1955 &                   Daniel Quillen, and independently by Andrei Suslin &                                 Serre's conjecture \\
                1976 &             1852 &                                    Kenneth Appel, and Wolfgang Haken &                                 Four color theorem \\
                1977 &             1909 &                                                     Alberto Calderón &                                Denjoy's conjecture \\
                1978 &             1846 &                           Roger Heath-Brown, and Samuel J. Patterson &                                Kummer's conjecture \\
                1983 &             1922 &                                                        Gerd Faltings &                                 Mordell conjecture \\
                1983 &             1970 &                                  Neil Robertson, and Paul D. Seymour &                                Wagner's conjecture \\
                1983 &             1965 &                                                       Michel Raynaud &                           Manin-Mumford conjecture \\
                1984 &             1969 &                                        Barry Mazur, and Andrew Wiles &                   Main conjecture (Iwasawa theory) \\
                1984 &             1972 &                                                        Haynes Miller &                                Sullivan conjecture \\
                1984 &             1949 &                                          John Morgan, and Hyman Bass &                                   Smith conjecture \\
                1984 &             1916 &                                          Louis de Branges de Bourcia &                              Bieberbach conjecture \\
                1984 &             1975 &                                                      Gunnar Carlsson &                                 Segal's conjecture \\
                1987 &             1929 &                                                     Grigory Margulis &                               Oppenheim conjecture \\
                1989 &             1959 &                                               Vladimir I. Chernousov &                                  Weil's conjecture \\
                1990 &             1987 &                                                            Ken Ribet &                                 Epsilon conjecture \\
                1992 &             1979 &                                                    Richard Borcherds &                           Conway-Norton conjecture \\
                1992 &             1978 &  Charles P. Boyer, Jacques C. Hurtubise, and Benjamin M. Mann et al. &                            Atiyah-Jones conjecture \\
                1994 &             1957 &                                   David Harbater, and Michel Raynaud &                             Abhyankar's conjecture \\
                1994 &             1637 &                                                         Andrew Wiles &                              Fermat's Last Theorem \\
                1994 &             1979 &                                                          Fred Galvin &                                  Dinitz conjecture \\
                1995 &             1982 &                                                     Doron Zeilberger &                 Alternating sign matrix conjecture \\
                1996 &             1970 &                                                   Vladimir Voevodsky &                                  Milnor conjecture \\
                1998 &             1943 &                                 Thomas C. Hales, and Sean McLaughlin &                            Dodecahedral conjecture \\
                1998 &             1611 &                                                      Thomas C. Hales &                                  Kepler conjecture \\
                2000 &             1989 &            Krzysztof Kurdyka, Tadeusz Mostowski, and Adam Parusiński &                                Gradient conjecture \\
                2001 &             1957 &    Christophe Breuil, Brian Conrad, Fred Diamond, and Richard Taylor &                        Taniyama-Shimura conjecture \\
                2001 &             1993 &                                                          Mark Haiman &                                      n! conjecture \\
                2001 &             1990 &                                     Daniel Frohardt, and Kay Magaard &                      Guralnick-Thompson conjecture \\
                2002 &             1844 &                                                     Preda Mihăilescu &                               Catalan's conjecture \\
                2002 &             1961 &  Maria Chudnovsky, Neil Robertson, Paul D. Seymour, and Robin Thomas &                    Strong perfect graph conjecture \\
                2002 &             1904 &                                                     Grigori Perelman &                                Poincaré conjecture \\
                2003 &             1982 &                                                     Grigori Perelman &                          Geometrization conjecture \\
                2003 &             1988 &                 Ben Green, and independently by Alexander Sapozhenko &                           Cameron-Erdős conjecture \\
                2003 &             1970 &                                                         Nils Dencker &                        Nirenberg-Treves conjecture \\
                2003 &             1979 &          Luca Barbieri-Viale, Andreas Rosenschon, and Morihiko Saito &                               Deligne's conjecture \\
                2004 &             1972 &                            Ualbai U. Umirbaev, and Ivan P. Shestakov &                                Nagata's conjecture \\
                2004 &             1974 &            Ian Agol, and independently by Danny Calegari-David Gabai &                                tameness conjecture \\
                2004 &             1903 &                                     Nobuo Iiyori, and Hiroshi Yamaki &                               Frobenius conjecture \\
                2004 &             1988 &                                        Adam Marcus, and Gábor Tardos &                            Stanley-Wilf conjecture \\
                2008 &             1970 &                                                     Avraham Trahtman &                           Road coloring conjecture \\
                2008 &             1975 &                   Chandrashekhar Khare, and Jean-Pierre Wintenberger &                      Serre's modularity conjecture \\
                2009 &             1968 &                                   Jeremy Kahn, and Vladimir Markovic &                        surface subgroup conjecture \\
                2009 &             1984 &                               Jeremie Chalopin, and Daniel Gonçalves &                           Scheinerman's conjecture \\
                2011 &             1957 &                     Joel Friedman, and independently by Igor Mineyev &                           Hanna Neumann conjecture \\
                2012 &             1970 &                                                        Simon Brendle &                         Hsiang-Lawson's conjecture \\
                2012 &             1965 &                                 Fernando C. Marques, and André Neves &                                Willmore conjecture \\
                2013 &             1959 &                  Adam Marcus, Daniel Spielman, and Nikhil Srivastava &                             Kadison-Singer problem \\
                2013 &             1742 &                                                      Harald Helfgott &                         Goldbach's weak conjecture \\
                2015 &             1935 &                       Jean Bourgain, Ciprian Demeter, and Larry Guth &  Main conjecture (Vinogradov's mean-value theorem) \\
                2018 &             1970 &                                                     Karim Adiprasito &                                       g-conjecture \\
                2019 &             1992 &                                                            Hao Huang &                             Sensitivity conjecture \\
                2019 &             1941 &                           Dimitris Koukoulopoulos, and James Maynard &                        Duffin-Schaeffer conjecture \\
                2021 &             1978 &                                 Alexander Dunn, and Maksym Radziwill &                             Patterson's conjecture \\
        \bottomrule
        \end{tabular}
        \begin{tablenotes}
            \item[a] Wagner's conjecture is proved in a series of twenty papers spanning over 500 pages from 1983 to 2004. For ease of handling, we use the earliest year 1983.
            \item[b] The Segal's conjecture was made in the mid 1970s by Graeme Segal. For ease of handling, we use 1975 to approximate it.
            \item[c] The Stanley-Wilf conjecture is formulated independently by Richard P. Stanley and Herbert Wilf in the late 1980s. For ease of handling, we use 1988 to approximate it. 
        \end{tablenotes}
    \end{threeparttable}
\end{table}

\begin{table}[H]
    \centering
    \begin{threeparttable}
        \caption{Breakpoints in the evolution of KQI across disciplines, sorted in alphabetical order.}
        \scriptsize
        \begin{tabular}{p{1.7in}p{4in}}
        \toprule
                                   Discipline &                                                                                      Breakpoints Year \\
        \midrule
                           Accounting &                                                                      1976.36 ($\pm$0.42), 1994.74 ($\pm$0.43) \\
                            Acoustics &                                                                                       1961.51 ($\pm$0.43) \\
                    Actuarial science &                                                                      1973.57 ($\pm$0.52), 1990.74 ($\pm$0.63) \\
                          Advertising &                                                     1972.21 ($\pm$0.51), 1984.96 ($\pm$0.58), 2000.25 ($\pm$0.36) \\
                          Aeronautics &                                                                                       1948.54 ($\pm$0.89) \\
                Aerospace engineering &                                                                      1951.42 ($\pm$0.57), 1998.69 ($\pm$1.92) \\
                           Aesthetics &                                                     1966.49 ($\pm$0.55), 1983.72 ($\pm$0.41), 2003.46 ($\pm$0.24) \\
               Agricultural economics &                   1936.60 ($\pm$3.71), 1963.30 ($\pm$0.58), 1972.23 ($\pm$0.67), 1985.16 ($\pm$2.28), 1999.92 ($\pm$0.37) \\
             Agricultural engineering &                                                     1970.52 ($\pm$1.14), 2000.63 ($\pm$0.74), 2013.94 ($\pm$0.30) \\
                 Agricultural science &                                                                      1975.73 ($\pm$0.67), 1997.52 ($\pm$0.29) \\
                         Agroforestry &                   1948.63 ($\pm$0.79), 1960.63 ($\pm$0.97), 1980.17 ($\pm$0.36), 1992.43 ($\pm$0.79), 2001.91 ($\pm$0.31) \\
                             Agronomy &                                                                      1956.73 ($\pm$0.36), 2006.60 ($\pm$0.52) \\
                              Algebra &                                    1948.75 ($\pm$0.31), 1965.39 ($\pm$0.18), 1980.42 ($\pm$0.25), 1999.74 ($\pm$0.73) \\
                            Algorithm &                                    1959.33 ($\pm$0.33), 1969.43 ($\pm$0.29), 1985.54 ($\pm$0.28), 1995.66 ($\pm$0.19) \\
                 Analytical chemistry &                                                                                       1949.43 ($\pm$0.48) \\
                              Anatomy &                                    1939.45 ($\pm$0.33), 1962.48 ($\pm$0.18), 1989.33 ($\pm$0.42), 2005.51 ($\pm$0.79) \\
                      Ancient history &                                                                      1958.67 ($\pm$0.31), 1993.64 ($\pm$1.42) \\
                            Andrology &                                    1940.47 ($\pm$2.36), 1965.64 ($\pm$0.49), 1977.52 ($\pm$0.36), 2004.65 ($\pm$1.12) \\
                           Anesthesia &                                                                      1963.56 ($\pm$0.28), 1990.48 ($\pm$0.77) \\
                       Animal science &                                    1948.85 ($\pm$0.33), 1974.78 ($\pm$0.37), 1988.43 ($\pm$0.70), 2000.46 ($\pm$0.57) \\
                         Anthropology &                                                     1951.56 ($\pm$0.76), 1972.38 ($\pm$0.70), 1992.84 ($\pm$0.48) \\
                  Applied mathematics &                                                     1947.49 ($\pm$1.16), 1964.44 ($\pm$0.74), 2012.61 ($\pm$0.28) \\
                   Applied psychology &                                                                      1937.93 ($\pm$1.37), 1969.85 ($\pm$0.46) \\
                          Archaeology &  1930.72 ($\pm$1.64), 1951.64 ($\pm$0.77), 1963.50 ($\pm$0.33), 1980.52 ($\pm$0.60), 1992.26 ($\pm$0.42), 2002.68 ($\pm$0.22) \\
            Architectural engineering &                                                     1966.25 ($\pm$1.27), 1985.47 ($\pm$0.89), 2009.61 ($\pm$0.11) \\
                           Arithmetic &                                                     1962.39 ($\pm$0.49), 1985.80 ($\pm$0.63), 1998.80 ($\pm$0.41) \\
                                  Art &                                                                      1966.54 ($\pm$0.71), 1996.75 ($\pm$0.37) \\
                          Art history &                                                     1959.53 ($\pm$1.12), 1981.68 ($\pm$0.76), 1997.35 ($\pm$0.32) \\
              Artificial intelligence &                                                                      1960.78 ($\pm$0.87), 1976.51 ($\pm$0.55) \\
                         Astrobiology &                                                                                       1955.58 ($\pm$1.29) \\
                            Astronomy &                                    1954.60 ($\pm$0.67), 1965.37 ($\pm$0.24), 1977.32 ($\pm$0.27), 1997.03 ($\pm$0.18) \\
                         Astrophysics &                                    1952.53 ($\pm$1.17), 1965.70 ($\pm$0.32), 1976.65 ($\pm$0.43), 1995.64 ($\pm$0.24) \\
                 Atmospheric sciences &                                    1947.81 ($\pm$0.25), 1961.57 ($\pm$0.33), 1975.40 ($\pm$0.20), 2006.19 ($\pm$0.50) \\
                       Atomic physics &                                                     1950.28 ($\pm$0.25), 1973.48 ($\pm$0.51), 1986.63 ($\pm$0.52) \\
                            Audiology &                                                                      1951.43 ($\pm$1.18), 1969.60 ($\pm$0.65) \\
               Automotive engineering &                                                     1964.73 ($\pm$0.22), 2000.58 ($\pm$0.36), 2009.40 ($\pm$0.14) \\
              Biochemical engineering &                                                     1977.47 ($\pm$0.92), 2000.86 ($\pm$0.58), 2010.58 ($\pm$0.18) \\
                         Biochemistry &                                                                                                       \\
                       Bioinformatics &                                                                                       1994.82 ($\pm$0.26) \\
                    Biological system &                                                                      1972.88 ($\pm$0.54), 2000.42 ($\pm$0.24) \\
                              Biology &                                                                                       1963.81 ($\pm$0.89) \\
               Biomedical engineering &                                    1956.43 ($\pm$0.85), 1974.33 ($\pm$0.41), 1993.56 ($\pm$0.38), 2004.87 ($\pm$0.11) \\
                           Biophysics &                                                     1961.52 ($\pm$0.31), 1987.59 ($\pm$0.70), 2012.72 ($\pm$0.39) \\
                        Biotechnology &                                                     1971.49 ($\pm$1.77), 1985.99 ($\pm$0.57), 2004.63 ($\pm$0.21) \\
                               Botany &                                                                                       1960.54 ($\pm$1.73) \\
                             Business &                                                                      1972.55 ($\pm$0.78), 1991.55 ($\pm$0.44) \\
              Business administration &                                                                                       1989.62 ($\pm$0.59) \\
                             Calculus &                                                                      1960.50 ($\pm$0.85), 1983.11 ($\pm$0.75) \\
                      Cancer research &                                                                      1971.89 ($\pm$0.27), 1993.29 ($\pm$0.17) \\
                           Cardiology &                                                                                       1994.58 ($\pm$1.35) \\
                          Cartography &                                                     1952.48 ($\pm$1.11), 1976.62 ($\pm$0.83), 1998.57 ($\pm$0.21) \\
                         Cell biology &  1936.47 ($\pm$0.92), 1952.30 ($\pm$0.73), 1957.70 ($\pm$0.36), 1977.69 ($\pm$0.70), 1994.60 ($\pm$0.12), 2007.64 ($\pm$0.26) \\
                    Ceramic materials &                                                                      1956.76 ($\pm$0.89), 2009.67 ($\pm$0.32) \\
                 Chemical engineering &                                    1963.73 ($\pm$1.72), 1993.97 ($\pm$0.77), 2004.67 ($\pm$0.36), 2015.00 ($\pm$0.22) \\
                     Chemical physics &                   1950.58 ($\pm$0.82), 1969.60 ($\pm$0.33), 1984.61 ($\pm$0.54), 2001.73 ($\pm$0.22), 2015.00 ($\pm$0.33) \\
                            Chemistry &                                    1936.80 ($\pm$1.42), 1957.42 ($\pm$0.22), 1982.32 ($\pm$0.41), 2007.29 ($\pm$0.35) \\
                       Chromatography &                                    1945.73 ($\pm$0.14), 1960.64 ($\pm$0.38), 1983.28 ($\pm$0.24), 2007.16 ($\pm$0.51) \\
                    Civil engineering &                                                     1963.20 ($\pm$0.45), 1997.81 ($\pm$0.39), 2010.56 ($\pm$0.16) \\
                  Classical economics &                                                                                       1972.58 ($\pm$0.47) \\
                  Classical mechanics &                                    1933.76 ($\pm$0.76), 1957.59 ($\pm$0.22), 1974.53 ($\pm$0.33), 2007.00 ($\pm$1.69) \\
                             Classics &                                                     1959.33 ($\pm$0.88), 1980.55 ($\pm$0.80), 1995.46 ($\pm$0.55) \\
                          Climatology &                   1932.41 ($\pm$2.77), 1948.74 ($\pm$0.29), 1965.45 ($\pm$0.29), 1977.74 ($\pm$0.46), 1996.64 ($\pm$0.32) \\
        \bottomrule
        \end{tabular}        
        \begin{tablenotes}
            \item[] Continued on next page...
        \end{tablenotes}
    \end{threeparttable}
\end{table}

\begin{table*}
    \centering
    \begin{threeparttable}
        \scriptsize
        \begin{tabular}{p{1.7in}p{4in}}
        \toprule
                                   Discipline &                                                                                      Breakpoints Year \\
        \midrule
                  Clinical psychology &                   1936.67 ($\pm$0.34), 1961.55 ($\pm$0.30), 1975.27 ($\pm$0.25), 1987.49 ($\pm$0.66), 2003.26 ($\pm$0.50) \\
                 Cognitive psychology &                                                     1954.30 ($\pm$0.85), 1985.54 ($\pm$0.94), 2002.65 ($\pm$0.47) \\
                    Cognitive science &                                                                                       1974.61 ($\pm$0.20) \\
              Combinatorial chemistry &                                                                      1965.45 ($\pm$1.38), 1992.70 ($\pm$0.17) \\
                        Combinatorics &                                                     1952.12 ($\pm$0.51), 1964.58 ($\pm$0.26), 1988.56 ($\pm$0.39) \\
                             Commerce &                                                                      1975.84 ($\pm$1.81), 1992.17 ($\pm$0.49) \\
                        Communication &                                                                      1962.33 ($\pm$0.20), 1981.74 ($\pm$0.44) \\
                   Composite material &                                                     1943.64 ($\pm$0.99), 1966.51 ($\pm$0.28), 2009.31 ($\pm$0.25) \\
                Computational biology &                                                    1940.00 ($\pm$10.12), 1985.53 ($\pm$1.28), 2000.34 ($\pm$0.43) \\
              Computational chemistry &                                                                                                       \\
                Computational physics &                                                     1936.29 ($\pm$0.95), 1985.73 ($\pm$0.46), 2004.90 ($\pm$0.30) \\
                Computational science &                                                     1965.64 ($\pm$0.90), 1983.61 ($\pm$0.34), 2008.39 ($\pm$0.26) \\
                Computer architecture &                                                                      1968.61 ($\pm$0.22), 2009.32 ($\pm$0.42) \\
                 Computer engineering &                                                                                       1956.58 ($\pm$0.44) \\
           Computer graphics (images) &                                                                                       1977.79 ($\pm$0.70) \\
                    Computer hardware &                                                                                       1959.82 ($\pm$0.95) \\
                     Computer network &                                                                      1971.20 ($\pm$0.37), 1984.67 ($\pm$0.93) \\
                     Computer science &                                                                                       1965.65 ($\pm$0.31) \\
                    Computer security &                                                                      1977.34 ($\pm$0.77), 1995.56 ($\pm$0.29) \\
                      Computer vision &                                                                      1969.29 ($\pm$0.50), 1980.65 ($\pm$0.35) \\
             Condensed matter physics &                                                                                                       \\
             Construction engineering &                                                     1962.34 ($\pm$1.63), 1988.54 ($\pm$0.72), 2010.75 ($\pm$0.18) \\
                  Control engineering &                                                                      1966.71 ($\pm$1.12), 1981.71 ($\pm$0.73) \\
                       Control theory &                                                                                       1959.75 ($\pm$0.28) \\
                          Criminology &                                                                                       1972.48 ($\pm$0.21) \\
                      Crystallography &                                                                      1953.40 ($\pm$0.22), 1975.59 ($\pm$0.34) \\
                          Data mining &                                                                      1972.39 ($\pm$0.44), 1988.50 ($\pm$0.54) \\
                         Data science &                                                     1981.69 ($\pm$0.91), 1999.64 ($\pm$0.28), 2008.68 ($\pm$0.15) \\
                             Database &                                                                      1971.58 ($\pm$0.14), 1993.63 ($\pm$0.26) \\
                Demographic economics &                                                                      1968.39 ($\pm$0.52), 1995.72 ($\pm$0.34) \\
                           Demography &                                                     1964.23 ($\pm$0.35), 1976.63 ($\pm$0.54), 1994.60 ($\pm$0.40) \\
                            Dentistry &                                                     1948.70 ($\pm$1.35), 1978.29 ($\pm$0.81), 2001.55 ($\pm$1.42) \\
                          Dermatology &                                                                                       1980.30 ($\pm$0.41) \\
                Development economics &                                                     1956.80 ($\pm$1.07), 1971.75 ($\pm$0.64), 1991.41 ($\pm$0.68) \\
             Developmental psychology &                                                                                                       \\
                 Discrete mathematics &                                                     1950.72 ($\pm$0.56), 1962.46 ($\pm$0.41), 1989.41 ($\pm$0.57) \\
                Distributed computing &                                                                                       1967.71 ($\pm$0.21) \\
                        Earth science &                                                                      1956.55 ($\pm$0.47), 2003.82 ($\pm$0.37) \\
                              Ecology &                                                                                       1958.44 ($\pm$0.47) \\
                         Econometrics &                                    1935.30 ($\pm$0.48), 1948.71 ($\pm$0.49), 1962.75 ($\pm$0.74), 1985.49 ($\pm$0.72) \\
                   Economic geography &                                                                      1962.15 ($\pm$0.39), 1999.42 ($\pm$0.21) \\
                      Economic growth &                                                                      1968.41 ($\pm$0.49), 1989.51 ($\pm$0.55) \\
                     Economic history &                                                     1956.54 ($\pm$1.09), 1975.55 ($\pm$1.57), 1999.00 ($\pm$1.13) \\
                      Economic policy &                                                                      1970.51 ($\pm$1.40), 1985.41 ($\pm$0.73) \\
                      Economic system &                                                                      1962.32 ($\pm$0.53), 1977.33 ($\pm$0.43) \\
                            Economics &                                                                      1965.60 ($\pm$0.19), 2007.82 ($\pm$1.09) \\
                              Economy &                                                                      1971.62 ($\pm$0.39), 1994.37 ($\pm$0.68) \\
               Electrical engineering &                                                     1950.60 ($\pm$0.33), 1986.54 ($\pm$0.46), 2001.75 ($\pm$0.54) \\
               Electronic engineering &                                                     1943.51 ($\pm$0.44), 1965.00 ($\pm$0.67), 1985.89 ($\pm$0.33) \\
                      Embedded system &                                                                      1972.59 ($\pm$0.54), 1993.42 ($\pm$0.19) \\
                   Emergency medicine &                   1937.61 ($\pm$3.42), 1965.75 ($\pm$0.58), 1982.64 ($\pm$0.33), 1996.65 ($\pm$0.16), 2008.51 ($\pm$0.24) \\
                        Endocrinology &                                                     1937.68 ($\pm$0.31), 1965.23 ($\pm$0.26), 1986.68 ($\pm$1.25) \\
                          Engineering &                                                                      1985.33 ($\pm$0.31), 2005.65 ($\pm$0.26) \\
                  Engineering drawing &                                                     1963.61 ($\pm$0.82), 1985.36 ($\pm$0.42), 2010.15 ($\pm$0.72) \\
                   Engineering ethics &                                                     1968.42 ($\pm$0.84), 1990.51 ($\pm$0.39), 2004.39 ($\pm$0.34) \\
               Engineering management &                                                     1965.65 ($\pm$0.68), 1987.39 ($\pm$0.66), 1995.81 ($\pm$0.42) \\
                  Engineering physics &                                                                      1963.36 ($\pm$1.93), 2010.32 ($\pm$0.23) \\
              Environmental chemistry &                                                                      1966.68 ($\pm$0.32), 2012.25 ($\pm$0.84) \\
              Environmental economics &                                                     1969.45 ($\pm$0.66), 1994.77 ($\pm$0.46), 2011.61 ($\pm$0.24) \\
            Environmental engineering &                                                     1961.55 ($\pm$1.13), 1980.54 ($\pm$0.42), 1999.32 ($\pm$0.24) \\
                 Environmental ethics &                                                                      1979.30 ($\pm$0.97), 2004.50 ($\pm$0.35) \\
                 Environmental health &                                                     1966.49 ($\pm$0.64), 1983.51 ($\pm$0.29), 1998.26 ($\pm$0.18) \\
        \bottomrule
        \end{tabular}        
        \begin{tablenotes}
            \item[] Continued on next page...
        \end{tablenotes}
    \end{threeparttable}
\end{table*}

\begin{table*}
    \centering
    \begin{threeparttable}
        \scriptsize
        \begin{tabular}{p{1.7in}p{4in}}
        \toprule
                                   Discipline &                                                                                      Breakpoints Year \\
        \midrule
               Environmental planning &                                                                      1971.34 ($\pm$0.52), 1997.37 ($\pm$0.17) \\
             Environmental protection &                                    1955.37 ($\pm$2.40), 1974.52 ($\pm$0.59), 1989.76 ($\pm$0.43), 2004.52 ($\pm$0.21) \\
    Environmental resource management &                                                                      1972.54 ($\pm$0.59), 1997.50 ($\pm$0.29) \\
                Environmental science &                                                     1959.78 ($\pm$0.45), 1976.68 ($\pm$0.69), 1998.68 ($\pm$0.72) \\
                         Epistemology &                                    1949.62 ($\pm$0.69), 1966.45 ($\pm$0.23), 1976.49 ($\pm$0.75), 1999.00 ($\pm$0.92) \\
                            Ethnology &                                                                      1967.48 ($\pm$0.64), 1997.81 ($\pm$0.28) \\
                 Evolutionary biology &                                                                                       1990.47 ($\pm$0.75) \\
                      Family medicine &                   1932.32 ($\pm$1.85), 1962.59 ($\pm$0.63), 1975.43 ($\pm$0.30), 1986.41 ($\pm$0.34), 1997.40 ($\pm$0.12) \\
                              Finance &                                                                      1967.76 ($\pm$0.44), 1979.61 ($\pm$1.04) \\
                  Financial economics &                                                                      1966.69 ($\pm$0.13), 2004.68 ($\pm$0.48) \\
                     Financial system &                                                                      1969.63 ($\pm$0.87), 1990.36 ($\pm$0.38) \\
                              Fishery &                                                                                       1963.41 ($\pm$0.28) \\
                         Food science &                                    1931.36 ($\pm$0.40), 1950.31 ($\pm$0.37), 1971.90 ($\pm$0.64), 2004.32 ($\pm$0.14) \\
                 Forensic engineering &                                    1945.16 ($\pm$3.16), 1967.35 ($\pm$0.52), 1983.35 ($\pm$0.99), 2006.48 ($\pm$0.26) \\
                             Forestry &                                                                      1961.50 ($\pm$1.97), 1992.20 ($\pm$0.62) \\
                     Gastroenterology &                                                                                       1978.29 ($\pm$0.58) \\
                       Gender studies &                                                                      1969.41 ($\pm$0.83), 1986.39 ($\pm$0.37) \\
                            Genealogy &                                                                      1958.57 ($\pm$0.63), 1997.55 ($\pm$0.36) \\
                      General surgery &                                                     1950.86 ($\pm$2.15), 1991.45 ($\pm$0.69), 2001.58 ($\pm$0.48) \\
                             Genetics &                                                                                       1972.68 ($\pm$0.75) \\
                         Geochemistry &                                    1949.83 ($\pm$0.41), 1960.94 ($\pm$0.35), 1981.36 ($\pm$0.27), 2010.51 ($\pm$0.81) \\
                              Geodesy &                                                     1956.42 ($\pm$0.27), 1978.26 ($\pm$0.71), 2011.00 ($\pm$1.37) \\
                            Geography &                                                                      1960.45 ($\pm$0.55), 1997.84 ($\pm$0.30) \\
                              Geology &                                    1950.57 ($\pm$0.37), 1961.53 ($\pm$0.25), 1982.71 ($\pm$0.36), 2009.43 ($\pm$1.02) \\
                             Geometry &                                                                                       1967.70 ($\pm$0.38) \\
                        Geomorphology &                                                     1950.52 ($\pm$0.38), 1968.40 ($\pm$0.65), 2006.00 ($\pm$1.15) \\
                           Geophysics &                                                                      1955.18 ($\pm$0.20), 1976.50 ($\pm$0.30) \\
             Geotechnical engineering &                                                     1949.68 ($\pm$0.89), 1965.68 ($\pm$0.25), 2009.60 ($\pm$0.18) \\
                          Gerontology &                                                                      1978.67 ($\pm$0.60), 1993.61 ($\pm$0.25) \\
                           Gynecology &                                                                                                       \\
                              History &                                                                      1959.22 ($\pm$0.99), 1982.35 ($\pm$1.07) \\
                         Horticulture &                                                     1955.90 ($\pm$1.92), 1976.37 ($\pm$0.73), 2003.65 ($\pm$0.19) \\
           Human-computer interaction &                                                                                       1977.78 ($\pm$0.38) \\
                           Humanities &                                                                                       1997.26 ($\pm$0.29) \\
                            Hydrology &                                    1938.24 ($\pm$0.72), 1962.76 ($\pm$0.17), 1980.57 ($\pm$0.23), 2004.00 ($\pm$0.57) \\
                           Immunology &                                                     1965.86 ($\pm$0.20), 1983.49 ($\pm$0.59), 1998.36 ($\pm$0.40) \\
               Industrial engineering &                                    1952.60 ($\pm$2.69), 1974.63 ($\pm$1.00), 1987.42 ($\pm$0.53), 2009.46 ($\pm$0.25) \\
              Industrial organization &                                                                                       1975.20 ($\pm$0.40) \\
                Information retrieval &                                                                      1966.62 ($\pm$1.63), 1991.77 ($\pm$0.48) \\
                  Inorganic chemistry &                   1939.83 ($\pm$0.59), 1956.73 ($\pm$0.50), 1964.49 ($\pm$0.53), 1979.35 ($\pm$0.99), 2006.49 ($\pm$0.18) \\
              Intensive care medicine &                                                     1949.77 ($\pm$1.41), 1981.60 ($\pm$0.30), 1995.74 ($\pm$0.28) \\
                    Internal medicine &                                                                      1939.49 ($\pm$0.38), 1963.31 ($\pm$0.34) \\
              International economics &                                                                      1968.33 ($\pm$1.34), 1985.70 ($\pm$0.86) \\
                  International trade &                                                                      1972.24 ($\pm$0.60), 1991.63 ($\pm$0.78) \\
                     Internet privacy &                                                                      1987.50 ($\pm$0.98), 2000.65 ($\pm$0.24) \\
                  Keynesian economics &                                                                                       1969.63 ($\pm$0.40) \\
                 Knowledge management &                                                                      1970.68 ($\pm$0.80), 1983.35 ($\pm$0.50) \\
                     Labour economics &                                                                                       1972.42 ($\pm$0.29) \\
                                  Law &                                                                      1971.37 ($\pm$0.76), 1993.46 ($\pm$0.61) \\
                    Law and economics &                                                     1965.68 ($\pm$0.84), 1983.86 ($\pm$0.85), 1999.27 ($\pm$0.24) \\
                      Library science &                                                                                       1979.59 ($\pm$1.00) \\
                          Linguistics &                                                                                       1965.43 ($\pm$0.24) \\
                           Literature &                                                     1959.24 ($\pm$1.23), 1977.47 ($\pm$0.81), 1993.87 ($\pm$0.47) \\
                     Machine learning &                                                                      1960.56 ($\pm$1.25), 1978.34 ($\pm$0.86) \\
                       Macroeconomics &                                                                      1968.52 ($\pm$0.21), 2011.66 ($\pm$0.67) \\
                           Management &                                                                      1971.71 ($\pm$1.17), 1988.61 ($\pm$0.47) \\
                   Management science &                                                                      1960.27 ($\pm$0.65), 1971.68 ($\pm$0.34) \\
            Manufacturing engineering &                                                     1954.36 ($\pm$2.02), 1983.66 ($\pm$0.16), 2010.49 ($\pm$0.12) \\
                   Marine engineering &                                                                      1963.67 ($\pm$1.03), 2008.39 ($\pm$0.16) \\
                       Market economy &                                                                                       1976.81 ($\pm$0.42) \\
                            Marketing &                                                                      1969.26 ($\pm$0.34), 1985.70 ($\pm$0.59) \\
                    Materials science &                                                     1949.01 ($\pm$0.99), 1964.78 ($\pm$0.55), 2009.28 ($\pm$0.20) \\
        \bottomrule
        \end{tabular}        
        \begin{tablenotes}
            \item[] Continued on next page...
        \end{tablenotes}
    \end{threeparttable}
\end{table*}

\begin{table*}
    \centering
    \begin{threeparttable}
        \scriptsize
        \begin{tabular}{p{1.7in}p{4in}}
        \toprule
                                   Discipline &                                                                                      Breakpoints Year \\
        \midrule
                Mathematical analysis &                                    1948.28 ($\pm$0.19), 1964.18 ($\pm$0.51), 1978.39 ($\pm$0.26), 2004.86 ($\pm$0.48) \\
               Mathematical economics &                                                                                       1966.39 ($\pm$0.28) \\
            Mathematical optimization &                                                     1952.69 ($\pm$0.72), 1962.84 ($\pm$0.31), 1979.20 ($\pm$0.34) \\
                 Mathematical physics &                                                     1962.38 ($\pm$0.62), 1979.54 ($\pm$0.51), 1999.15 ($\pm$0.47) \\
                          Mathematics &                                                     1948.63 ($\pm$0.29), 1963.21 ($\pm$0.61), 1977.43 ($\pm$0.34) \\
                Mathematics education &                                                     1940.29 ($\pm$2.55), 1969.64 ($\pm$0.55), 1983.48 ($\pm$0.43) \\
               Mechanical engineering &                                                     1970.80 ($\pm$1.18), 1990.95 ($\pm$0.61), 2010.34 ($\pm$0.24) \\
                            Mechanics &                                                                                       1948.83 ($\pm$0.96) \\
                        Media studies &                                                     1968.42 ($\pm$1.10), 1989.53 ($\pm$0.71), 2003.78 ($\pm$0.27) \\
                    Medical education &                                    1954.33 ($\pm$1.07), 1975.58 ($\pm$0.44), 1989.92 ($\pm$0.49), 2000.58 ($\pm$0.14) \\
                    Medical emergency &                                                     1965.60 ($\pm$0.76), 1982.53 ($\pm$0.79), 1999.82 ($\pm$0.28) \\
                      Medical physics &                                                     1954.42 ($\pm$1.99), 1981.52 ($\pm$0.66), 2001.70 ($\pm$0.25) \\
                  Medicinal chemistry &                                                                      1956.37 ($\pm$0.96), 1966.88 ($\pm$0.61) \\
                             Medicine &                                                                                       1984.62 ($\pm$0.84) \\
                           Metallurgy &                                                                      1960.80 ($\pm$0.38), 2008.24 ($\pm$0.63) \\
                          Meteorology &                                                     1946.58 ($\pm$0.28), 1975.44 ($\pm$0.68), 1997.66 ($\pm$0.50) \\
                         Microbiology &                                    1935.56 ($\pm$0.59), 1951.81 ($\pm$0.61), 1969.68 ($\pm$0.32), 2002.11 ($\pm$0.33) \\
                       Microeconomics &                                                     1955.64 ($\pm$1.05), 1967.30 ($\pm$0.18), 1985.26 ($\pm$0.27) \\
                           Mineralogy &                                    1948.60 ($\pm$0.33), 1962.69 ($\pm$0.25), 1979.77 ($\pm$0.32), 1999.00 ($\pm$0.97) \\
                   Mining engineering &                                                     1946.79 ($\pm$1.22), 1977.31 ($\pm$0.77), 2010.40 ($\pm$0.13) \\
                    Molecular biology &                                                                      1955.29 ($\pm$0.28), 1980.62 ($\pm$0.39) \\
                    Molecular physics &                                                                      1967.29 ($\pm$0.50), 2009.50 ($\pm$0.46) \\
                   Monetary economics &                                                                                       1971.34 ($\pm$0.36) \\
                           Multimedia &                                                                      1969.54 ($\pm$2.45), 1987.46 ($\pm$0.47) \\
                       Nanotechnology &                                                                      1974.56 ($\pm$0.87), 1998.65 ($\pm$0.15) \\
          Natural language processing &                                    1960.59 ($\pm$0.32), 1974.20 ($\pm$0.34), 1987.90 ($\pm$0.28), 2014.90 ($\pm$0.16) \\
           Natural resource economics &                                                     1970.22 ($\pm$0.89), 1990.86 ($\pm$0.66), 2009.60 ($\pm$0.26) \\
               Neoclassical economics &                                                                      1960.34 ($\pm$1.14), 1974.15 ($\pm$0.42) \\
                         Neuroscience &                                                     1942.76 ($\pm$1.33), 1958.82 ($\pm$0.39), 2004.03 ($\pm$0.88) \\
                    Nuclear chemistry &                                                                                       1952.43 ($\pm$1.70) \\
                  Nuclear engineering &                                                     1958.44 ($\pm$2.18), 1975.70 ($\pm$0.66), 2006.65 ($\pm$0.22) \\
           Nuclear magnetic resonance &                                                                                                       \\
                     Nuclear medicine &                                                     1951.72 ($\pm$0.82), 1972.52 ($\pm$0.53), 1992.76 ($\pm$0.38) \\
                      Nuclear physics &                                                                                                       \\
                              Nursing &                                                     1966.61 ($\pm$0.45), 1980.43 ($\pm$0.36), 1993.51 ($\pm$0.21) \\
                           Obstetrics &                                                                                       1985.44 ($\pm$0.70) \\
                         Oceanography &                                                                      1954.57 ($\pm$0.49), 1974.27 ($\pm$0.49) \\
                             Oncology &                                                     1946.71 ($\pm$1.83), 1971.40 ($\pm$0.54), 2007.55 ($\pm$0.32) \\
                     Operating system &                                                                                       1973.40 ($\pm$0.54) \\
                Operations management &                                                     1961.35 ($\pm$0.60), 1980.85 ($\pm$0.41), 2004.16 ($\pm$0.43) \\
                  Operations research &                                    1954.69 ($\pm$0.63), 1969.13 ($\pm$0.43), 1987.71 ($\pm$0.32), 2005.41 ($\pm$0.14) \\
                        Ophthalmology &                                    1951.29 ($\pm$0.88), 1976.65 ($\pm$0.56), 1984.73 ($\pm$0.40), 2000.81 ($\pm$0.20) \\
                               Optics &                                                                                       1950.71 ($\pm$0.60) \\
                      Optoelectronics &                                                                                       1957.60 ($\pm$0.73) \\
                            Optometry &                                                     1929.53 ($\pm$1.86), 1980.81 ($\pm$0.35), 1999.28 ($\pm$0.24) \\
                    Organic chemistry &                                                     1956.91 ($\pm$0.41), 1969.69 ($\pm$0.94), 2002.64 ($\pm$0.29) \\
                         Orthodontics &                                                     1945.10 ($\pm$0.59), 1978.72 ($\pm$0.70), 1999.66 ($\pm$0.45) \\
                         Paleontology &                                                     1950.55 ($\pm$0.90), 1965.93 ($\pm$0.42), 2004.85 ($\pm$1.96) \\
                   Parallel computing &                                                     1962.91 ($\pm$1.15), 1971.50 ($\pm$0.48), 1998.00 ($\pm$0.78) \\
                     Particle physics &                                    1929.93 ($\pm$1.90), 1949.36 ($\pm$0.29), 1959.36 ($\pm$0.53), 2005.00 ($\pm$1.25) \\
                            Pathology &                                                                      1967.59 ($\pm$0.43), 1981.79 ($\pm$0.81) \\
                  Pattern recognition &                                    1958.78 ($\pm$0.46), 1966.74 ($\pm$0.22), 1987.33 ($\pm$0.13), 2013.50 ($\pm$0.17) \\
                             Pedagogy &                                                                      1931.03 ($\pm$0.85), 1976.23 ($\pm$0.16) \\
                           Pediatrics &                                                                      1981.16 ($\pm$0.59), 1998.33 ($\pm$0.33) \\
                Petroleum engineering &                                                     1968.35 ($\pm$0.83), 1980.57 ($\pm$0.77), 2008.68 ($\pm$0.06) \\
                            Petrology &                                                     1953.77 ($\pm$0.65), 1968.45 ($\pm$0.31), 2008.17 ($\pm$0.29) \\
                         Pharmacology &                   1957.62 ($\pm$0.44), 1964.67 ($\pm$0.36), 1973.21 ($\pm$0.23), 1985.50 ($\pm$0.15), 2005.35 ($\pm$0.26) \\
                           Philosophy &                                                                      1966.68 ($\pm$0.62), 1995.50 ($\pm$0.62) \\
                       Photochemistry &                                                                                       1964.09 ($\pm$0.67) \\
                   Physical chemistry &                                                                      1961.61 ($\pm$0.59), 1980.75 ($\pm$0.87) \\
                   Physical geography &                                                     1973.00 ($\pm$1.19), 1993.91 ($\pm$1.11), 2015.00 ($\pm$0.36) \\
 Physical medicine and rehabilitation &                                                     1966.81 ($\pm$0.90), 1980.27 ($\pm$0.32), 1997.62 ($\pm$0.26) \\
        \bottomrule
        \end{tabular}        
        \begin{tablenotes}
            \item[] Continued on next page...
        \end{tablenotes}
    \end{threeparttable}
\end{table*}

\begin{table*}
    \centering
    \begin{threeparttable}
        \scriptsize
        \begin{tabular}{p{1.7in}p{4in}}
        \toprule
                                   Discipline &                                                                                      Breakpoints Year \\
        \midrule
                     Physical therapy &                                                     1955.68 ($\pm$1.33), 1978.51 ($\pm$0.34), 1991.83 ($\pm$0.44) \\
                              Physics &                                                     1953.58 ($\pm$0.21), 1976.28 ($\pm$0.30), 2004.26 ($\pm$0.75) \\
                           Physiology &                                                                                       1997.27 ($\pm$2.06) \\
                    Political economy &                                    1936.71 ($\pm$2.49), 1964.65 ($\pm$0.39), 1981.54 ($\pm$1.15), 2000.23 ($\pm$0.20) \\
                    Political science &                                                                      1966.62 ($\pm$1.14), 1992.78 ($\pm$0.70) \\
                    Polymer chemistry &                                                     1935.67 ($\pm$1.84), 1963.45 ($\pm$0.58), 1995.59 ($\pm$0.29) \\
                      Polymer science &                                                     1934.00 ($\pm$1.66), 1969.23 ($\pm$0.30), 2006.29 ($\pm$0.59) \\
                   Positive economics &                                                                                       1969.61 ($\pm$0.31) \\
                  Process engineering &                                                                      1980.38 ($\pm$1.05), 2008.74 ($\pm$0.21) \\
                   Process management &                                                                      1971.56 ($\pm$0.90), 1993.48 ($\pm$0.24) \\
                 Programming language &                                                                      1959.95 ($\pm$0.19), 1995.54 ($\pm$0.57) \\
                           Psychiatry &                                                                      1962.58 ($\pm$0.32), 1977.54 ($\pm$0.36) \\
                       Psychoanalysis &                                                     1939.33 ($\pm$1.92), 1950.53 ($\pm$1.13), 1983.68 ($\pm$0.74) \\
                           Psychology &                                                                      1960.54 ($\pm$0.53), 1980.66 ($\pm$0.84) \\
                      Psychotherapist &                                                                                       1951.71 ($\pm$0.51) \\
                Public administration &                                                                      1971.62 ($\pm$0.44), 1992.41 ($\pm$0.59) \\
                     Public economics &                                                                                       1971.46 ($\pm$0.30) \\
                     Public relations &                                                     1965.73 ($\pm$0.66), 1974.57 ($\pm$0.36), 1992.62 ($\pm$0.31) \\
              Pulp and paper industry &                                                                      1983.32 ($\pm$1.36), 2009.54 ($\pm$0.28) \\
                     Pure mathematics &                                    1949.47 ($\pm$0.25), 1964.29 ($\pm$1.08), 1976.41 ($\pm$0.34), 2007.61 ($\pm$0.27) \\
              Quantum electrodynamics &                                                                      1952.36 ($\pm$0.24), 1972.36 ($\pm$0.62) \\
                    Quantum mechanics &                                                                                       1961.34 ($\pm$0.52) \\
                       Radiochemistry &                                    1954.14 ($\pm$0.57), 1970.39 ($\pm$0.61), 1988.30 ($\pm$0.59), 2008.29 ($\pm$0.89) \\
                            Radiology &                                                     1949.60 ($\pm$0.89), 1969.41 ($\pm$0.29), 1999.11 ($\pm$0.69) \\
                  Real-time computing &                                                     1962.26 ($\pm$0.28), 1986.95 ($\pm$0.23), 1998.61 ($\pm$0.35) \\
                     Regional science &                                    1956.92 ($\pm$1.38), 1964.59 ($\pm$1.11), 1977.78 ($\pm$0.86), 2002.76 ($\pm$0.20) \\
              Reliability engineering &                                                     1954.39 ($\pm$0.62), 1972.82 ($\pm$0.30), 2009.09 ($\pm$0.32) \\
                    Religious studies &                                                     1950.43 ($\pm$1.87), 1969.37 ($\pm$0.84), 1999.26 ($\pm$0.30) \\
                       Remote sensing &  1937.61 ($\pm$3.34), 1949.68 ($\pm$0.51), 1965.60 ($\pm$0.45), 1980.31 ($\pm$0.24), 1988.52 ($\pm$0.35), 2009.51 ($\pm$0.26) \\
          Risk analysis (engineering) &                                    1967.14 ($\pm$1.37), 1982.35 ($\pm$0.47), 1999.69 ($\pm$0.39), 2009.41 ($\pm$0.20) \\
                           Seismology &                                                     1952.85 ($\pm$0.90), 1962.70 ($\pm$0.41), 1981.78 ($\pm$0.42) \\
                           Simulation &                                                     1957.67 ($\pm$0.35), 1988.40 ($\pm$0.24), 2003.27 ($\pm$0.40) \\
                    Social psychology &                                                     1958.79 ($\pm$0.28), 1980.79 ($\pm$0.49), 2005.00 ($\pm$1.29) \\
                       Social science &                                                                      1948.46 ($\pm$0.81), 1968.33 ($\pm$0.16) \\
                       Socioeconomics &                                                     1970.40 ($\pm$0.37), 1992.40 ($\pm$0.70), 2001.38 ($\pm$0.54) \\
                            Sociology &                                                     1948.87 ($\pm$1.14), 1968.30 ($\pm$0.24), 1989.03 ($\pm$0.87) \\
                 Software engineering &                                                                                       1971.90 ($\pm$0.37) \\
                         Soil science &                                    1935.43 ($\pm$0.66), 1951.08 ($\pm$0.28), 1980.37 ($\pm$0.46), 2006.42 ($\pm$0.26) \\
                   Speech recognition &                                                                      1962.36 ($\pm$1.06), 1975.56 ($\pm$0.46) \\
                  Statistical physics &                                                                      1960.38 ($\pm$0.56), 1973.31 ($\pm$0.28) \\
                           Statistics &                                                                      1943.52 ($\pm$0.29), 1987.47 ($\pm$0.83) \\
                      Stereochemistry &                                                                      1950.32 ($\pm$0.30), 1981.62 ($\pm$1.48) \\
               Structural engineering &                                                     1949.45 ($\pm$2.08), 1967.28 ($\pm$0.29), 2011.32 ($\pm$0.26) \\
                              Surgery &                                                     1948.60 ($\pm$1.20), 1977.48 ($\pm$0.72), 1991.79 ($\pm$0.71) \\
                  Systems engineering &                                                                                       1982.20 ($\pm$0.24) \\
                   Telecommunications &                                                                                       1989.52 ($\pm$0.56) \\
                             Theology &                                                     1950.57 ($\pm$2.02), 1967.48 ($\pm$1.31), 1996.84 ($\pm$0.40) \\
         Theoretical computer science &                                                                                       1958.28 ($\pm$0.28) \\
                  Theoretical physics &                                                     1953.63 ($\pm$1.15), 1967.50 ($\pm$0.78), 1997.37 ($\pm$0.17) \\
                       Thermodynamics &                                                                                       1951.60 ($\pm$0.85) \\
                             Topology &                                    1947.59 ($\pm$0.20), 1963.40 ($\pm$0.59), 1977.11 ($\pm$0.26), 2000.83 ($\pm$0.75) \\
                           Toxicology &                                                     1940.55 ($\pm$1.02), 1969.36 ($\pm$0.33), 2004.69 ($\pm$0.54) \\
                 Traditional medicine &                                                                      1983.81 ($\pm$0.77), 2003.35 ($\pm$0.26) \\
                Transport engineering &                                                     1960.86 ($\pm$0.45), 1995.59 ($\pm$0.24), 2007.56 ($\pm$0.27) \\
                              Urology &                                                     1945.62 ($\pm$0.45), 1969.31 ($\pm$0.56), 1993.86 ($\pm$0.37) \\
                  Veterinary medicine &                   1945.30 ($\pm$1.61), 1961.09 ($\pm$1.58), 1979.72 ($\pm$0.51), 1997.29 ($\pm$0.43), 2004.88 ($\pm$0.35) \\
                             Virology &                                                                                       1960.37 ($\pm$0.20) \\
                          Visual arts &                                                                      1967.39 ($\pm$0.19), 2002.69 ($\pm$0.21) \\
                     Waste management &                                    1963.80 ($\pm$0.99), 1979.47 ($\pm$0.47), 1996.40 ($\pm$0.37), 2007.64 ($\pm$0.10) \\
            Water resource management &                                    1933.31 ($\pm$1.16), 1982.00 ($\pm$1.85), 1996.58 ($\pm$0.36), 2006.71 ($\pm$0.15) \\
                    Welfare economics &                                                                      1967.47 ($\pm$0.32), 2012.24 ($\pm$0.55) \\
                       World Wide Web &                                                                                       1990.71 ($\pm$0.43) \\
                              Zoology &                                                                      1968.58 ($\pm$0.45), 1997.38 ($\pm$0.38) \\
        \bottomrule
        \end{tabular}        
        \begin{tablenotes}
            \item[] 
        \end{tablenotes}
    \end{threeparttable}
\end{table*}

\begin{figure}[H]
    \centering
    \caption{The growth of KQI with breakpoints across disciplines, sorted in alphabetical order.}
    
    \includegraphics[height=4cm]{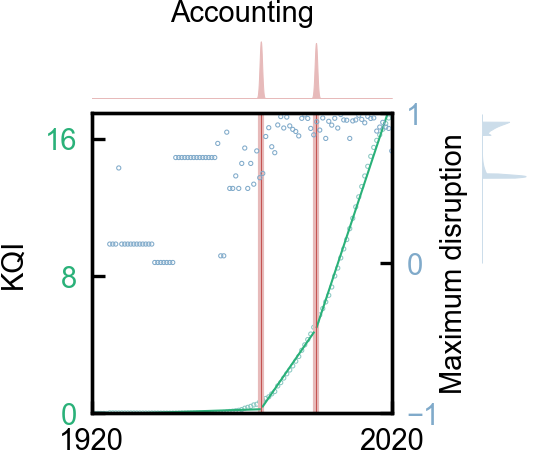}
    \includegraphics[height=4cm]{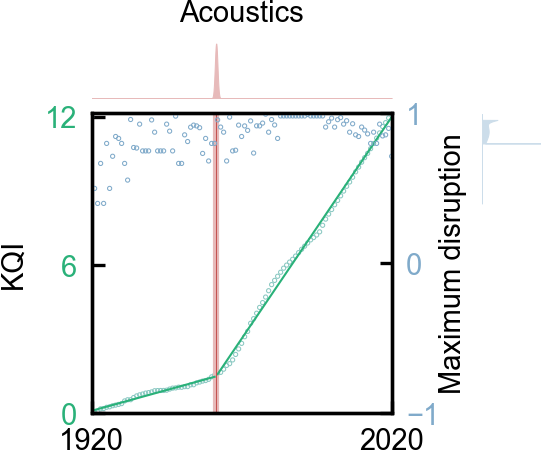}
    \includegraphics[height=4cm]{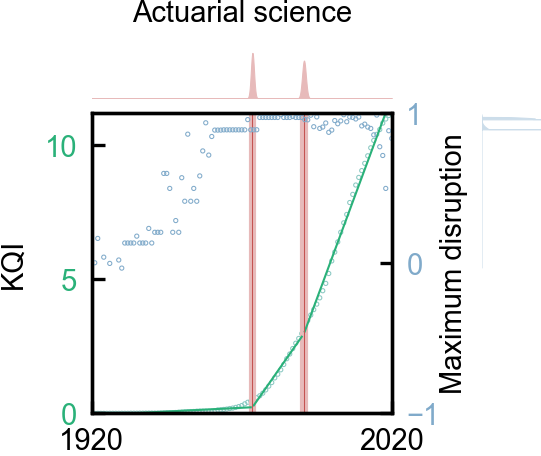}
    \includegraphics[height=4cm]{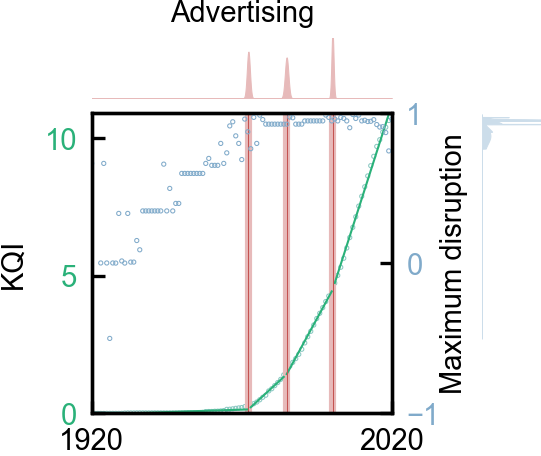}
    \includegraphics[height=4cm]{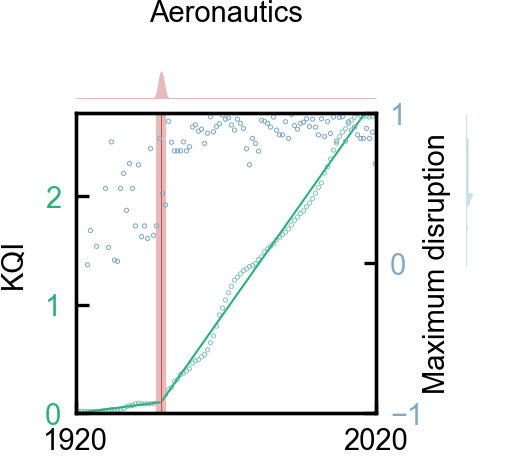}
    \includegraphics[height=4cm]{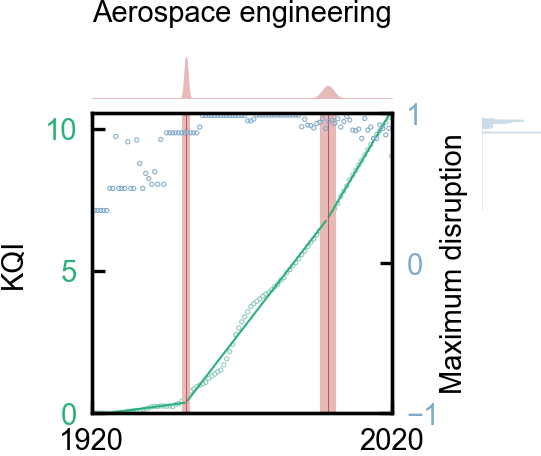}
    \includegraphics[height=4cm]{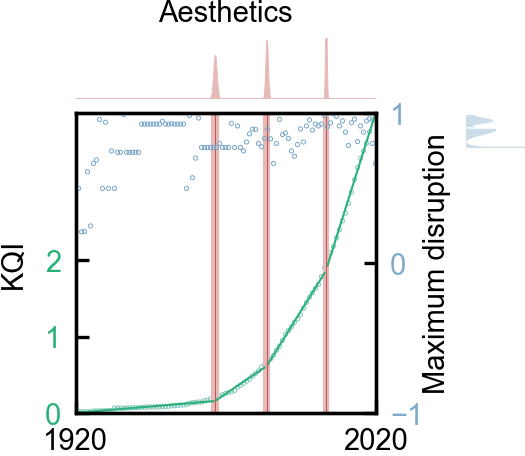}
    \includegraphics[height=4cm]{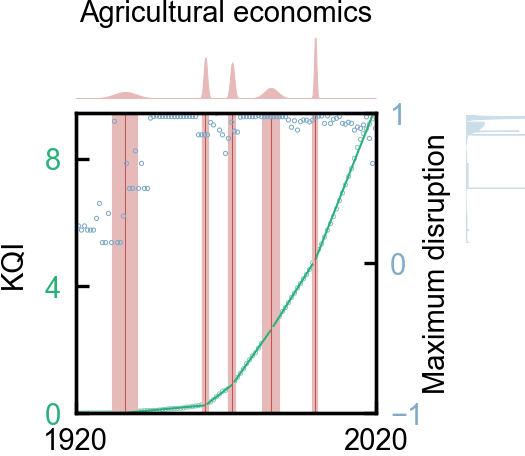}
    \includegraphics[height=4cm]{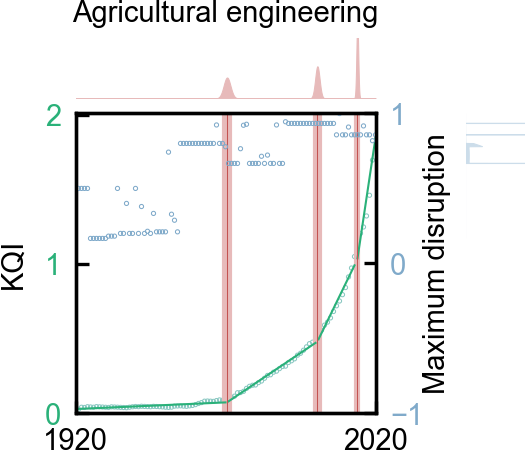}
    \includegraphics[height=4cm]{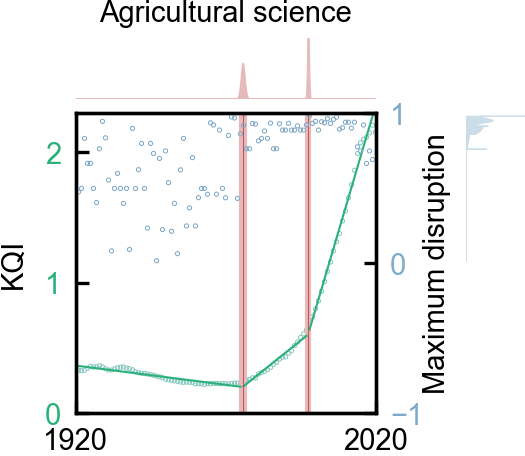}
    \includegraphics[height=4cm]{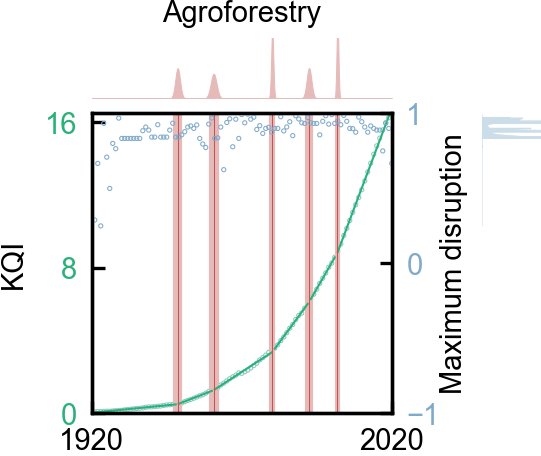}
    \includegraphics[height=4cm]{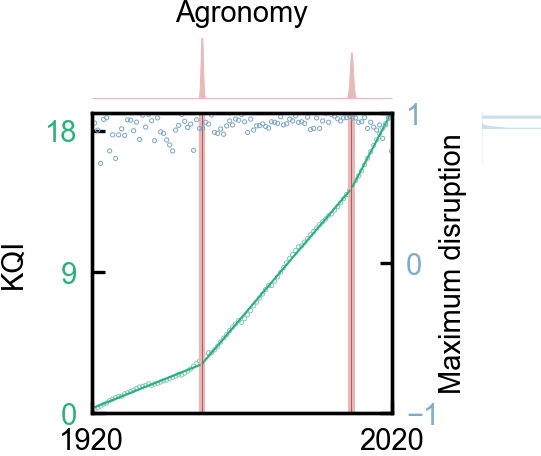}
    \includegraphics[height=4cm]{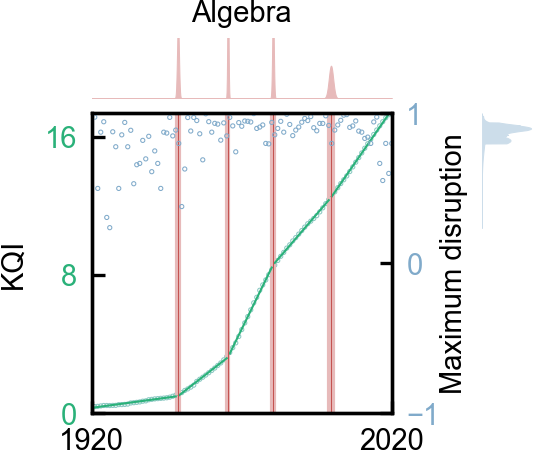}
    \includegraphics[height=4cm]{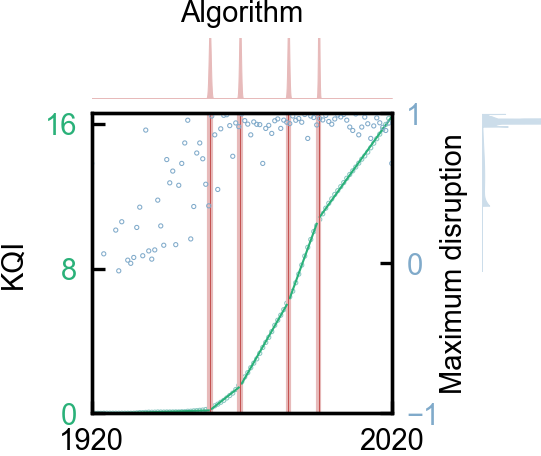}
    \includegraphics[height=4cm]{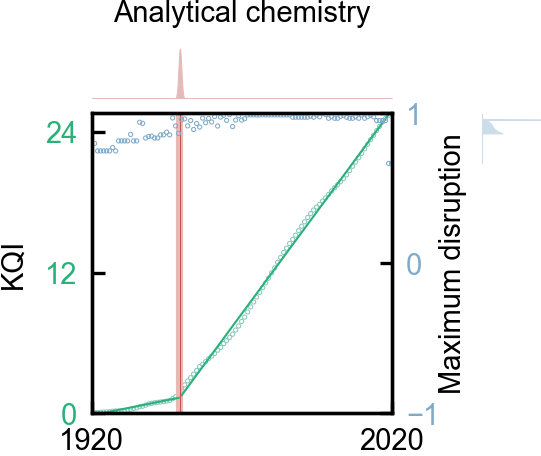}
\end{figure}
\begin{figure}[H]
    \centering
    \includegraphics[height=4cm]{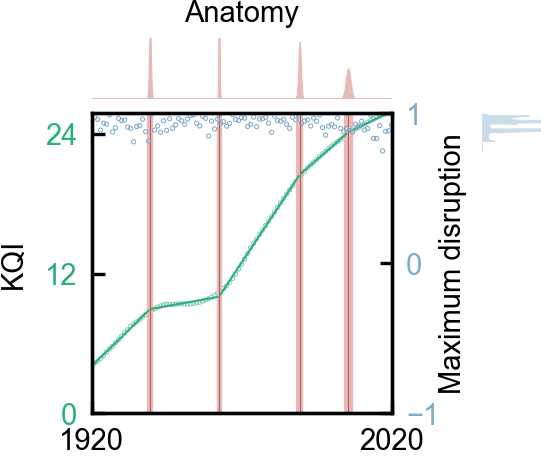}
    \includegraphics[height=4cm]{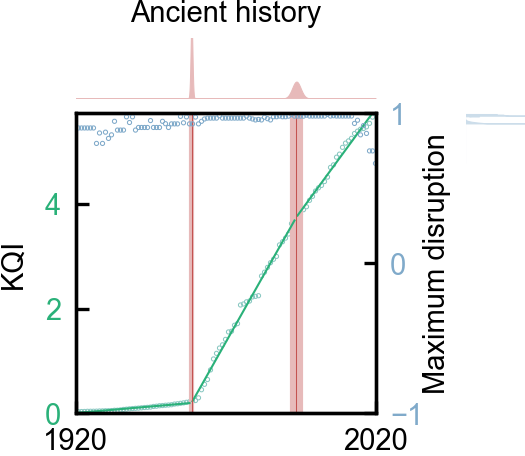}
    \includegraphics[height=4cm]{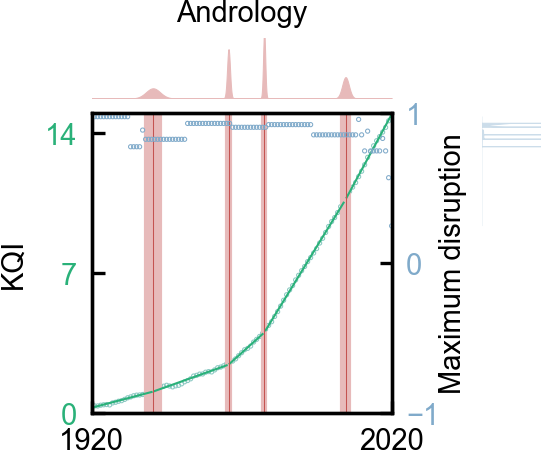}
    \includegraphics[height=4cm]{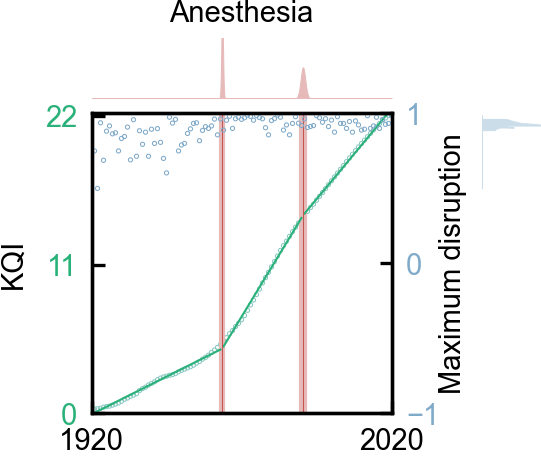}
    \includegraphics[height=4cm]{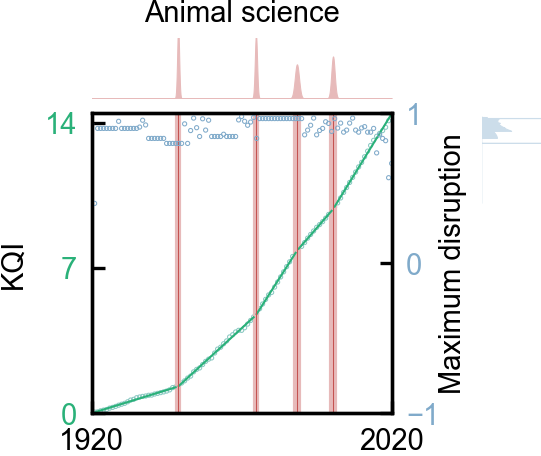}
    \includegraphics[height=4cm]{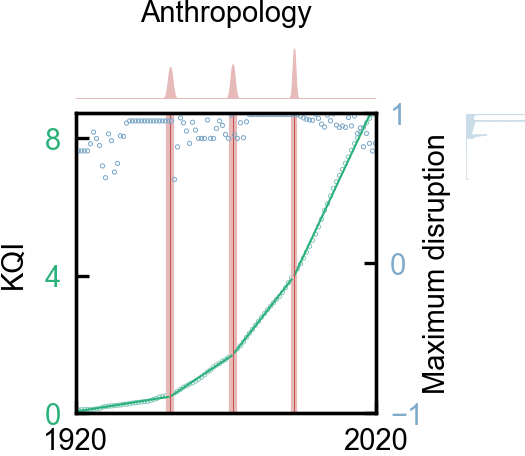}
    \includegraphics[height=4cm]{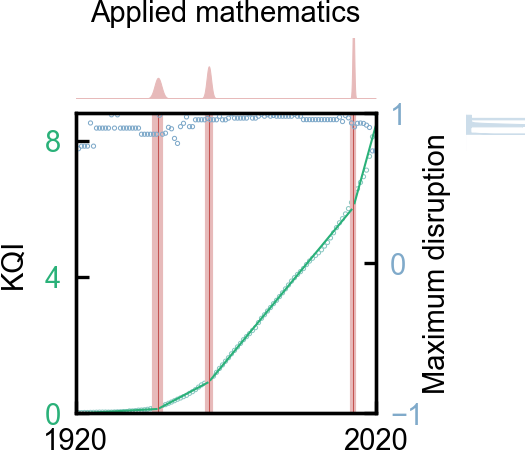}
    \includegraphics[height=4cm]{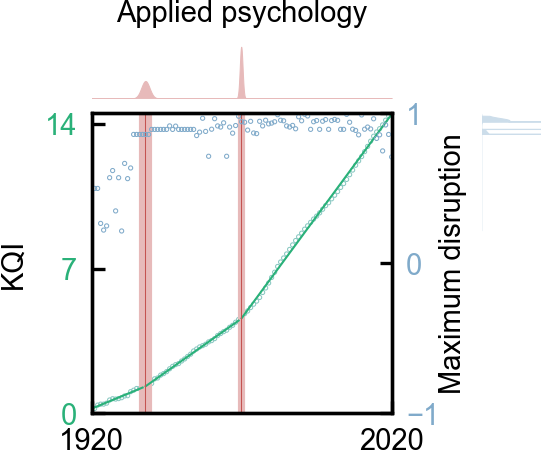}
    \includegraphics[height=4cm]{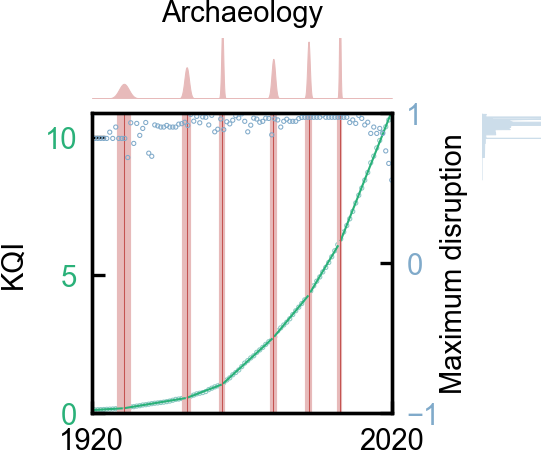}
    \includegraphics[height=4cm]{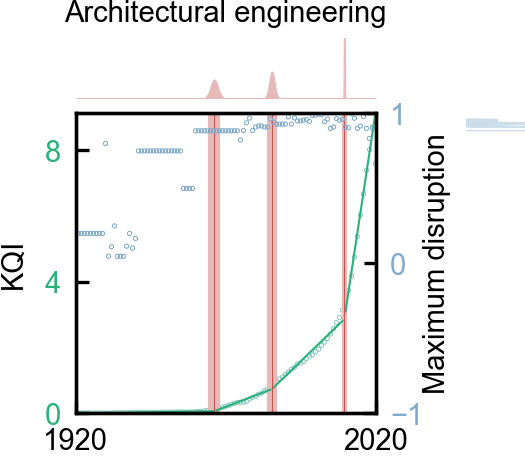}
    \includegraphics[height=4cm]{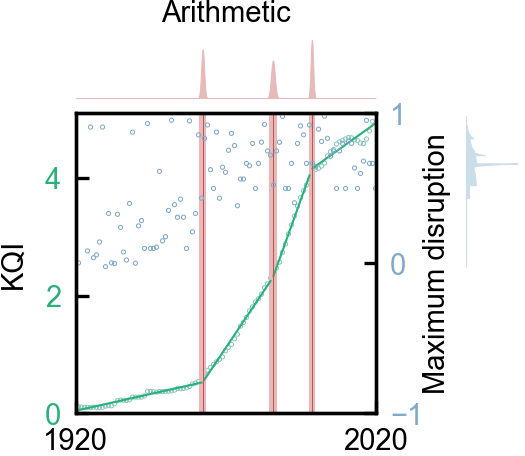}
    \includegraphics[height=4cm]{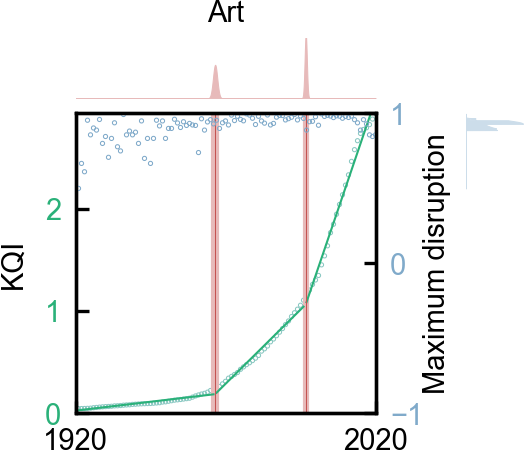}
    \includegraphics[height=4cm]{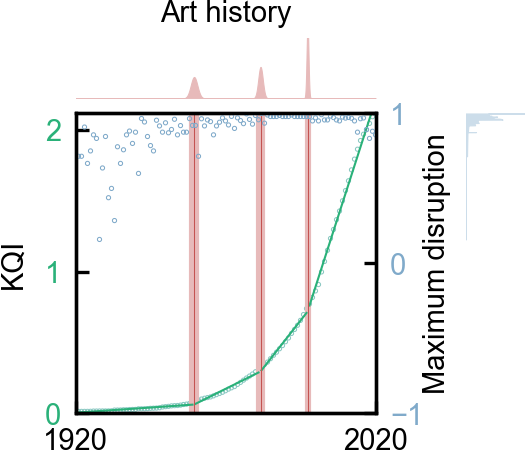}
    \includegraphics[height=4cm]{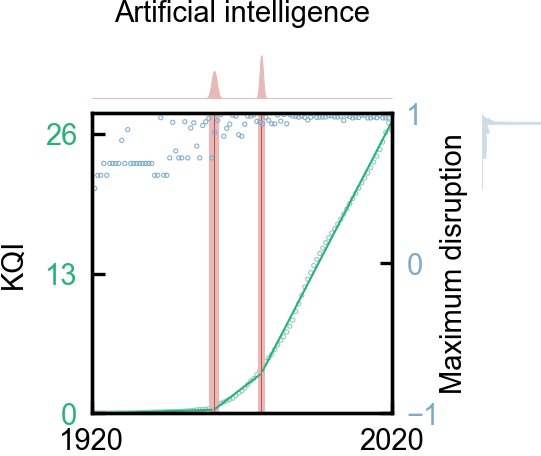}
    \includegraphics[height=4cm]{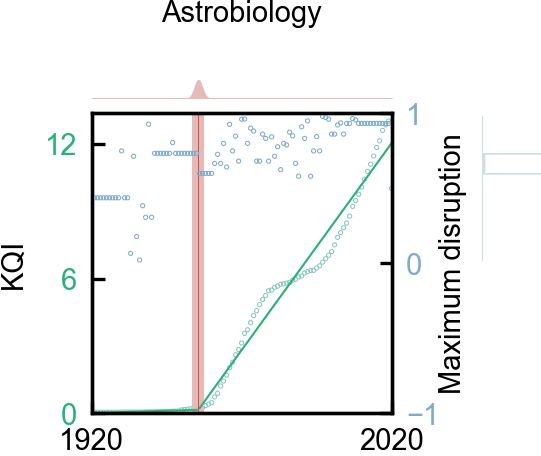}
\end{figure}
\begin{figure}[H]
    \centering
    \includegraphics[height=4cm]{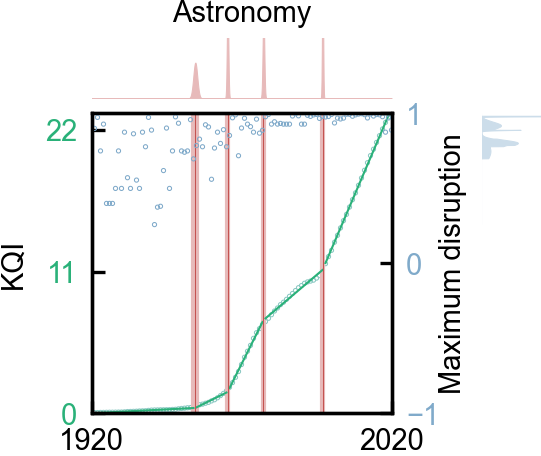}
    \includegraphics[height=4cm]{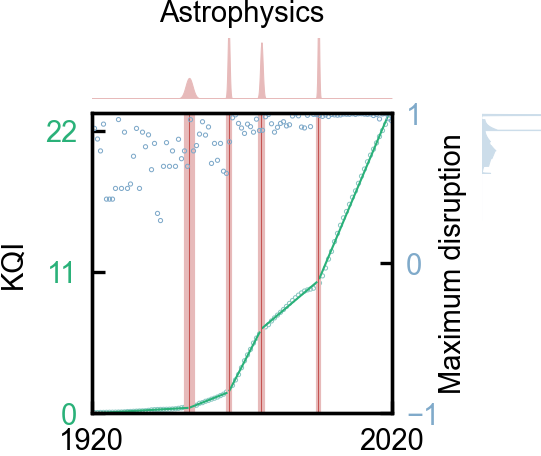}
    \includegraphics[height=4cm]{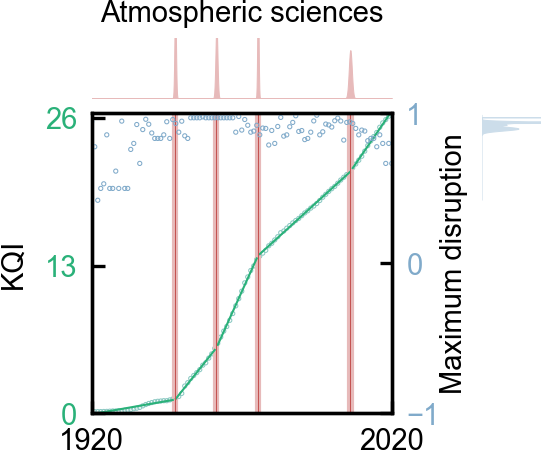}
    \includegraphics[height=4cm]{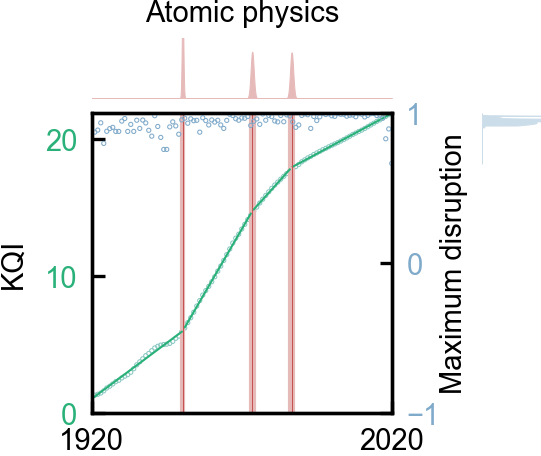}
    \includegraphics[height=4cm]{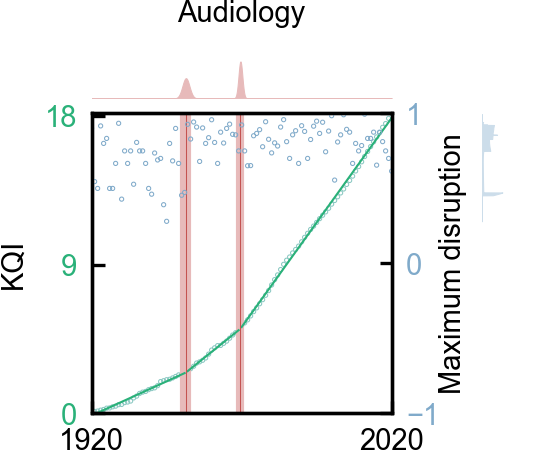}
    \includegraphics[height=4cm]{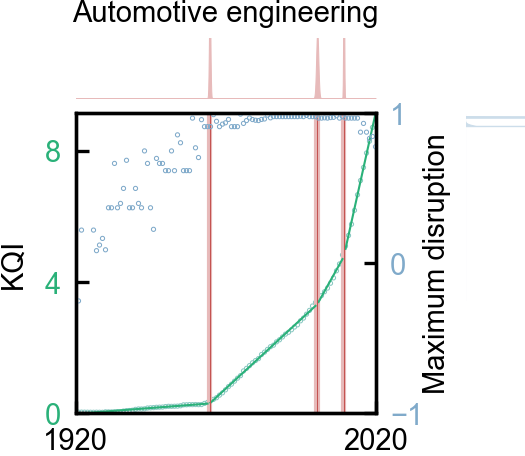}
    \includegraphics[height=4cm]{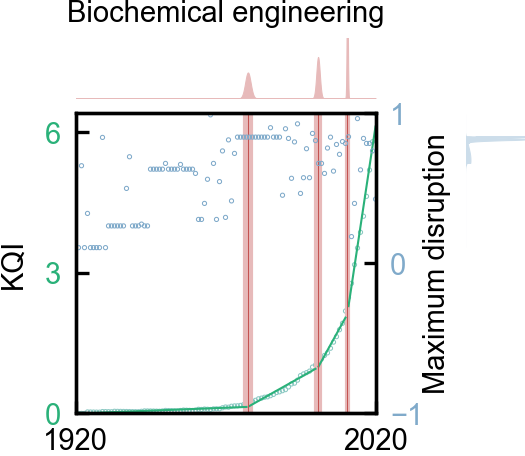}
    \includegraphics[height=4cm]{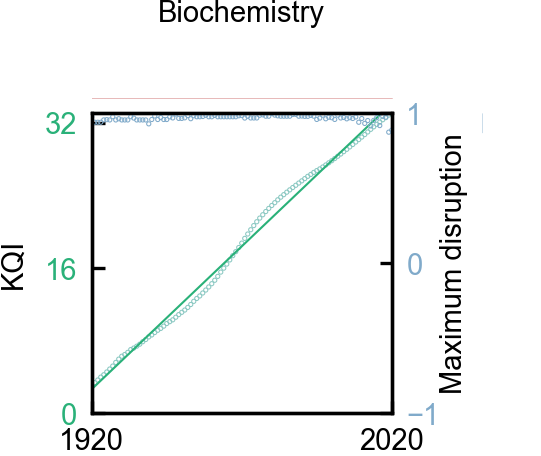}
    \includegraphics[height=4cm]{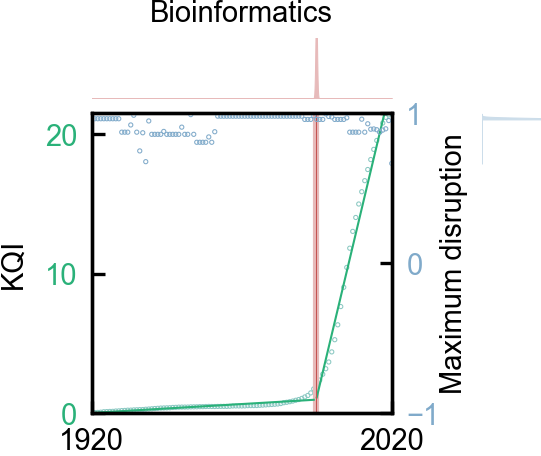}
    \includegraphics[height=4cm]{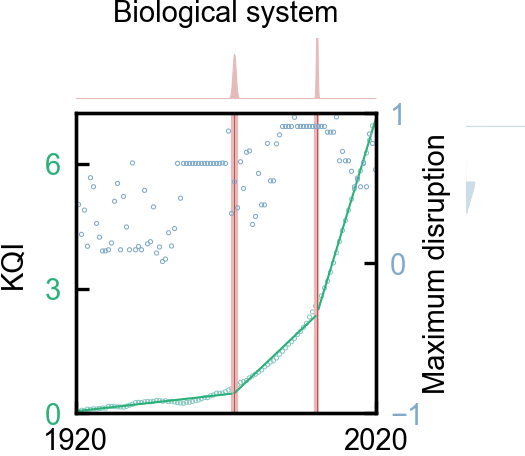}
    \includegraphics[height=4cm]{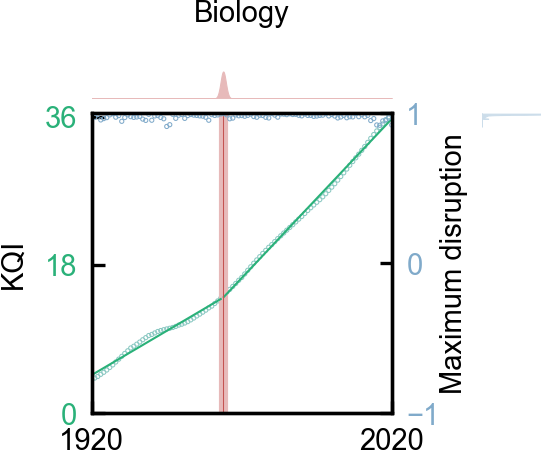}
    \includegraphics[height=4cm]{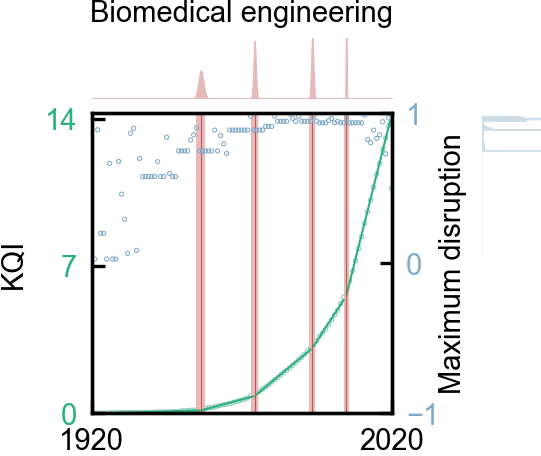}
    \includegraphics[height=4cm]{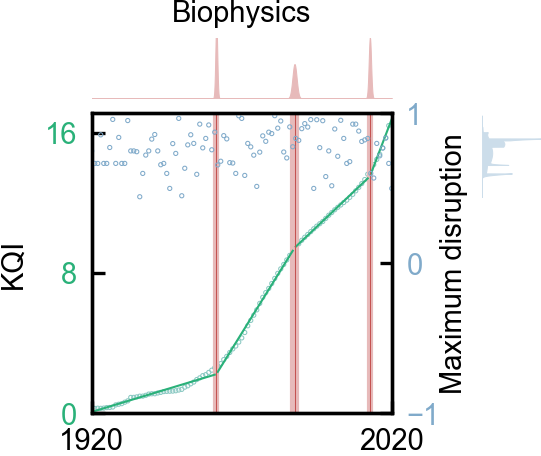}
    \includegraphics[height=4cm]{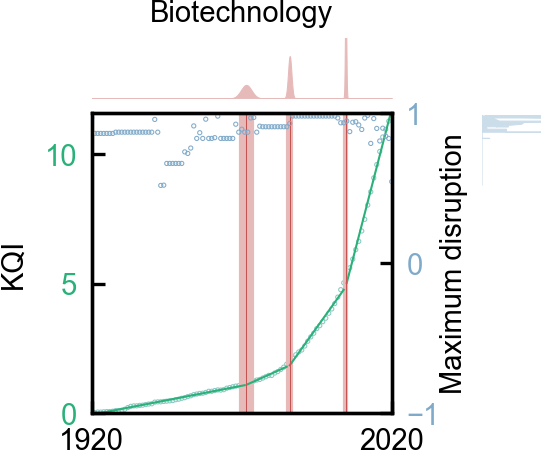}
    \includegraphics[height=4cm]{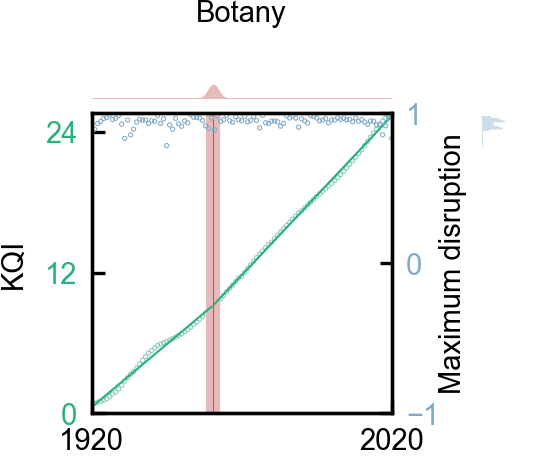}
\end{figure}
\begin{figure}[H]
    \centering
    \includegraphics[height=4cm]{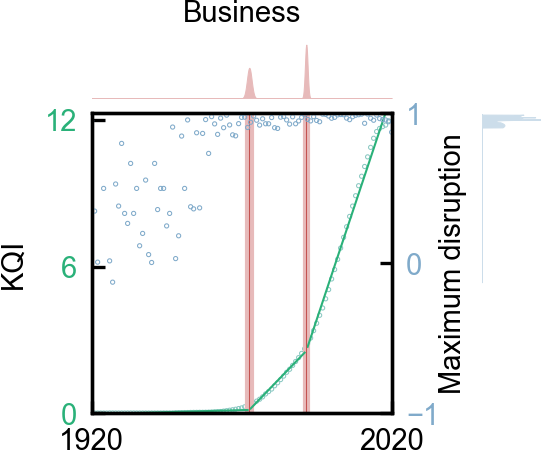}
    \includegraphics[height=4cm]{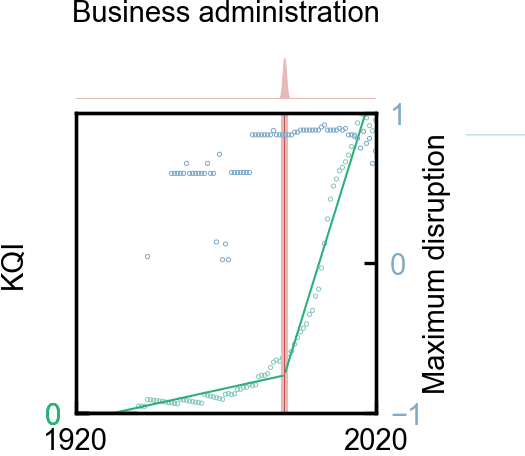}
    \includegraphics[height=4cm]{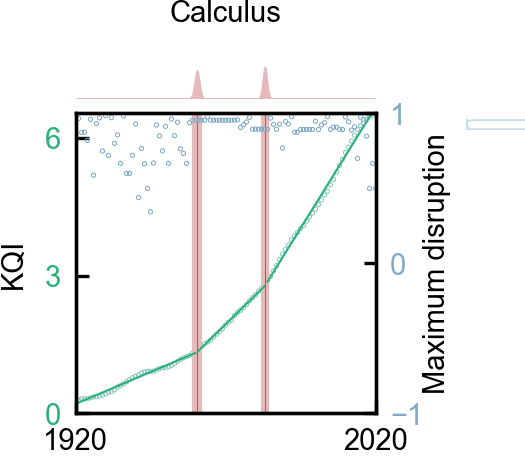}
    \includegraphics[height=4cm]{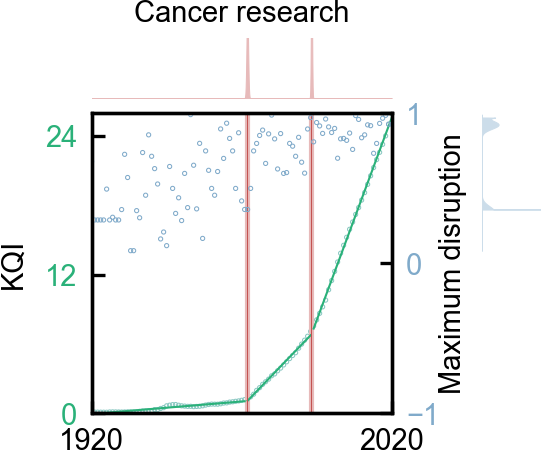}
    \includegraphics[height=4cm]{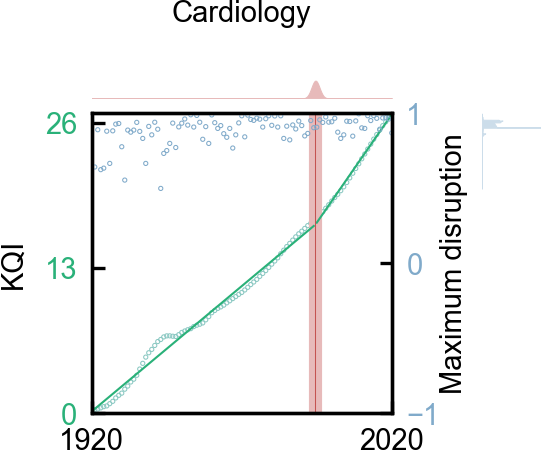}
    \includegraphics[height=4cm]{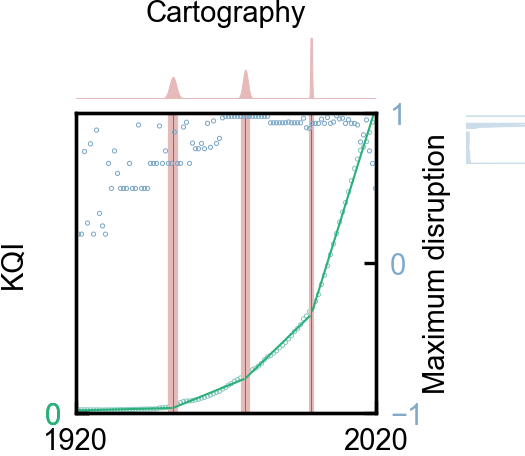}
    \includegraphics[height=4cm]{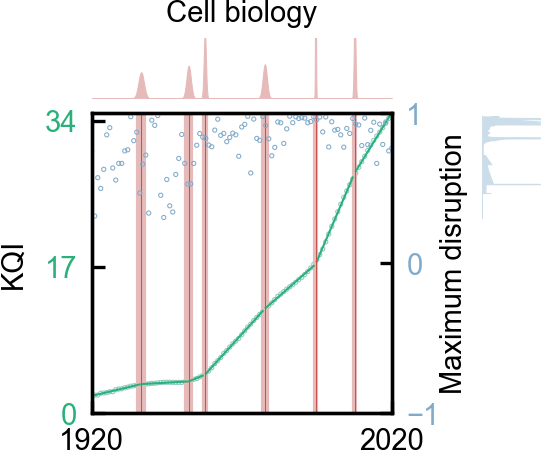}
    \includegraphics[height=4cm]{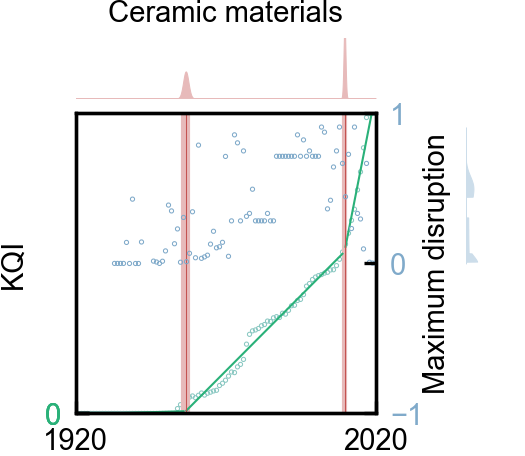}
    \includegraphics[height=4cm]{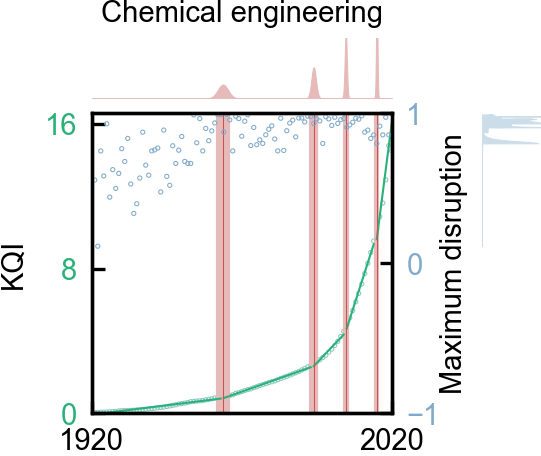}
    \includegraphics[height=4cm]{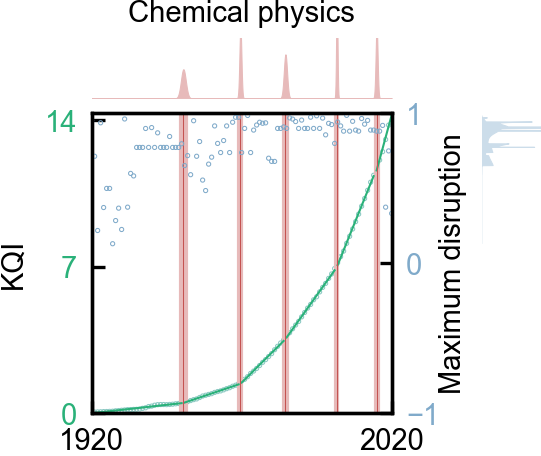}
    \includegraphics[height=4cm]{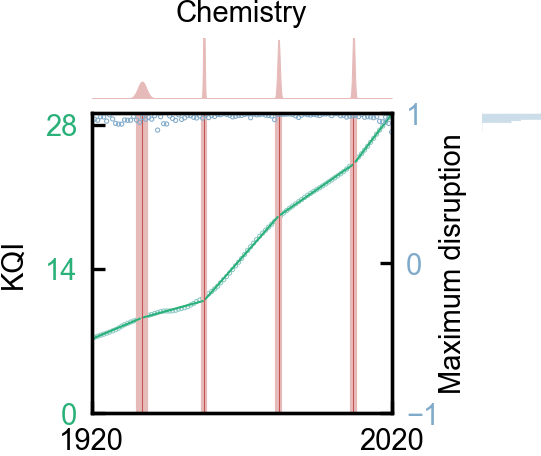}
    \includegraphics[height=4cm]{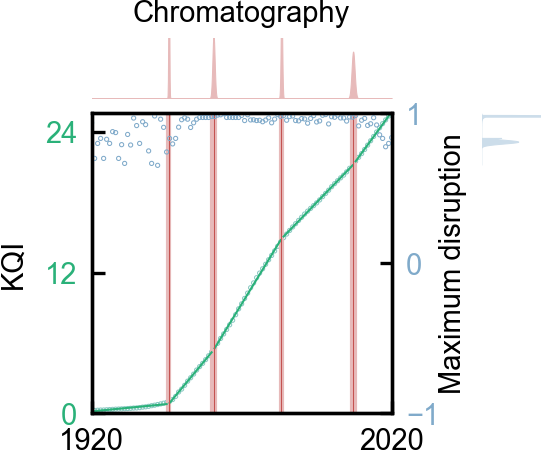}
    \includegraphics[height=4cm]{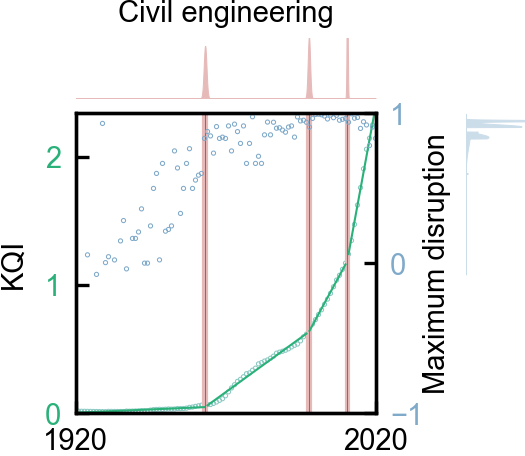}
    \includegraphics[height=4cm]{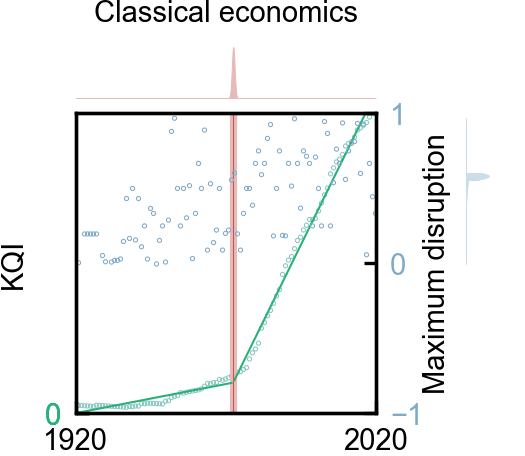}
    \includegraphics[height=4cm]{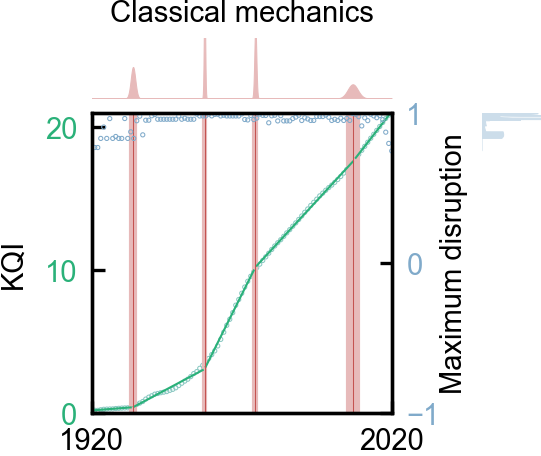}
\end{figure}
\begin{figure}[H]
    \centering
    \includegraphics[height=4cm]{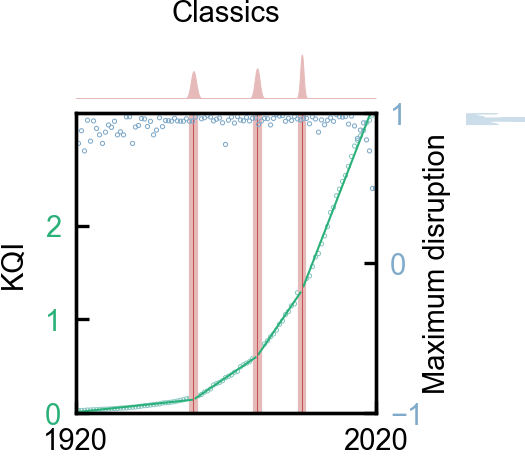}
    \includegraphics[height=4cm]{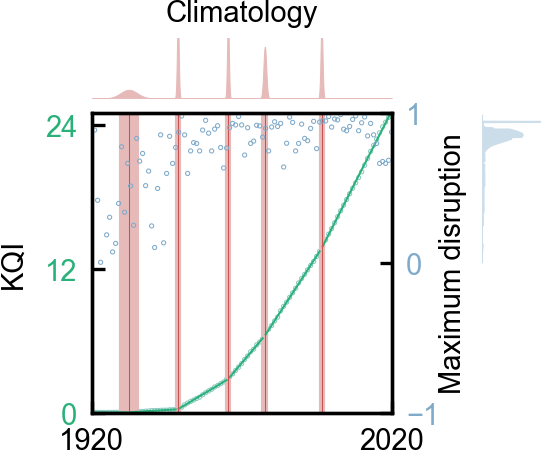}
    \includegraphics[height=4cm]{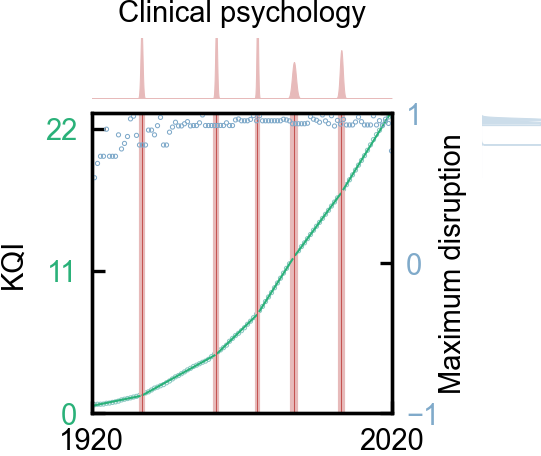}
    \includegraphics[height=4cm]{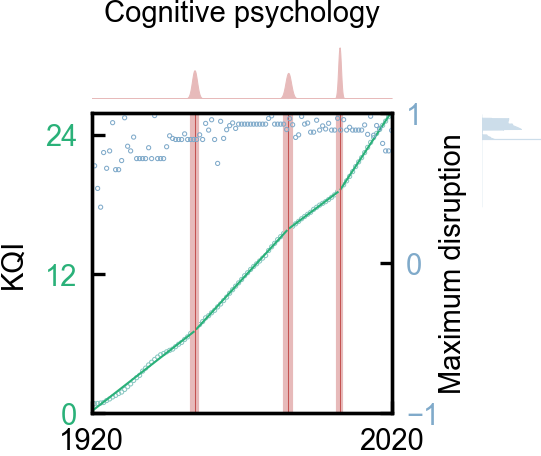}
    \includegraphics[height=4cm]{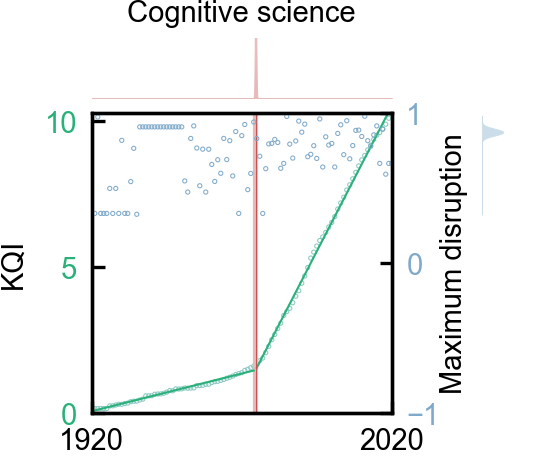}
    \includegraphics[height=4cm]{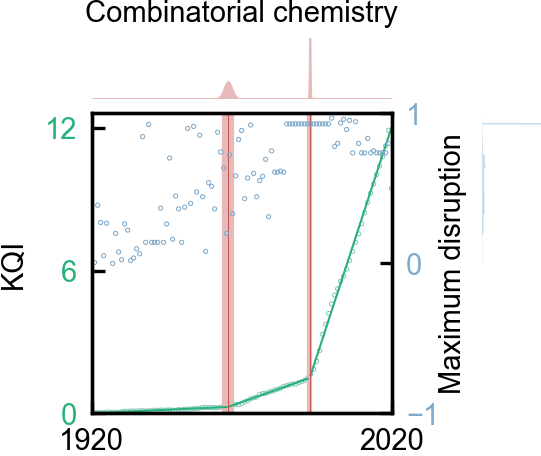}
    \includegraphics[height=4cm]{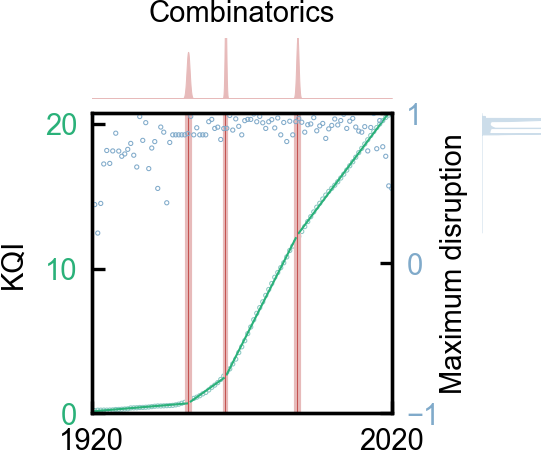}
    \includegraphics[height=4cm]{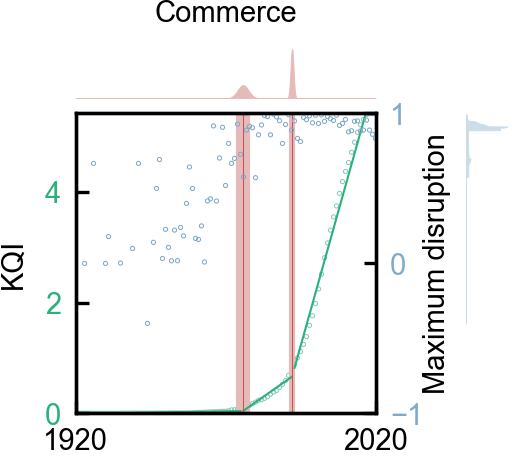}
    \includegraphics[height=4cm]{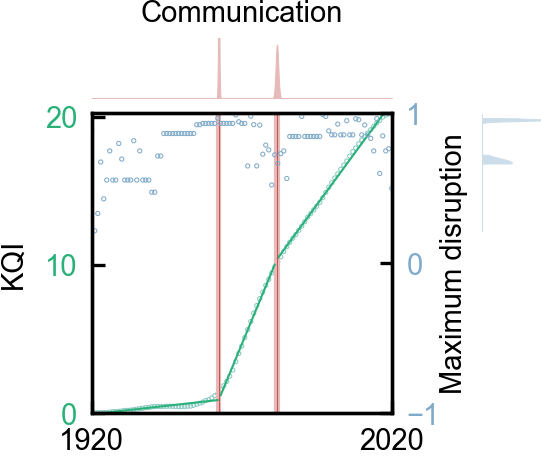}
    \includegraphics[height=4cm]{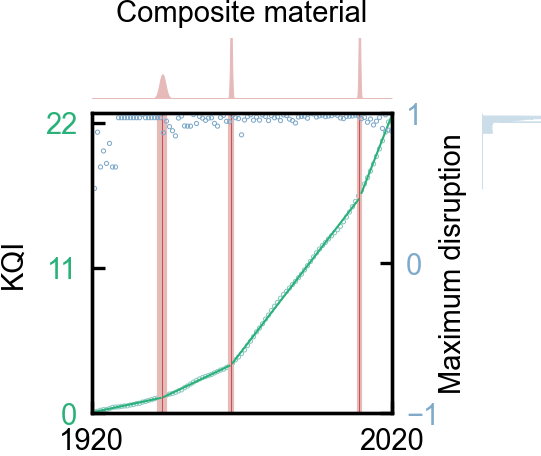}
    \includegraphics[height=4cm]{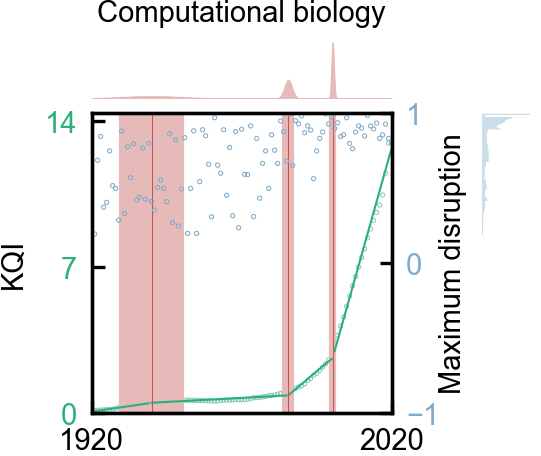}
    \includegraphics[height=4cm]{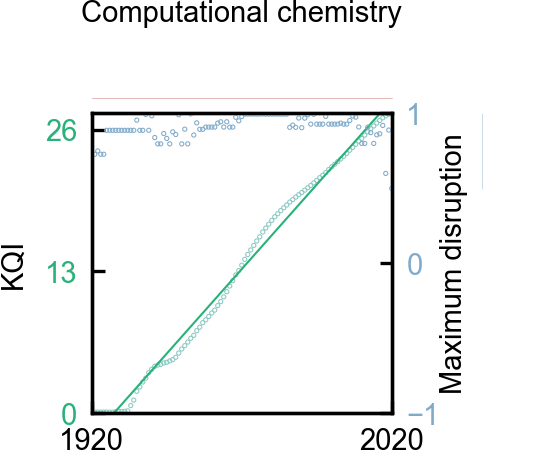}
    \includegraphics[height=4cm]{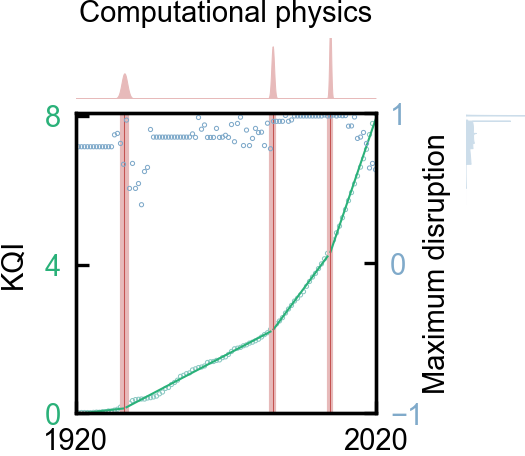}
    \includegraphics[height=4cm]{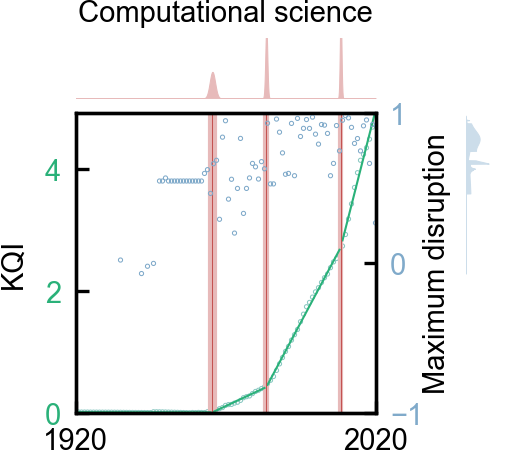}
    \includegraphics[height=4cm]{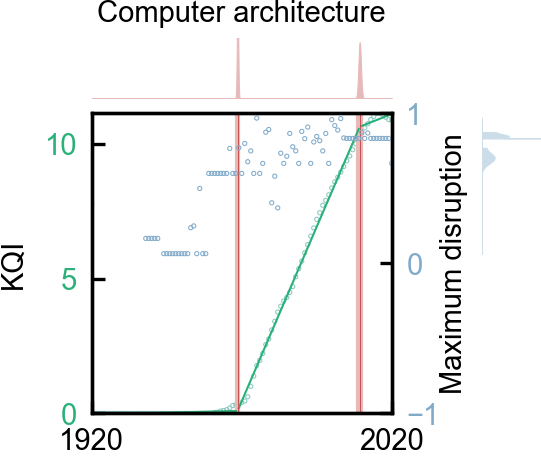}
\end{figure}
\begin{figure}[H]
    \centering
    \includegraphics[height=4cm]{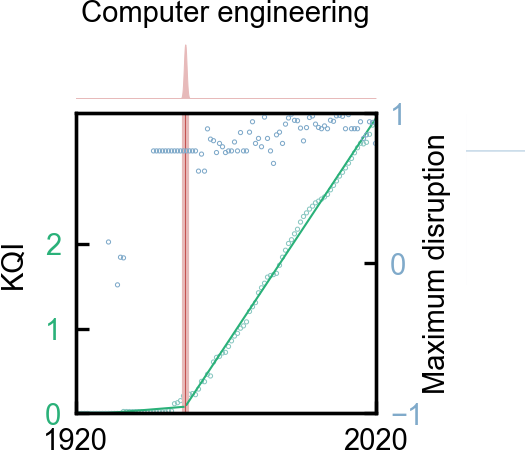}
    \includegraphics[height=4cm]{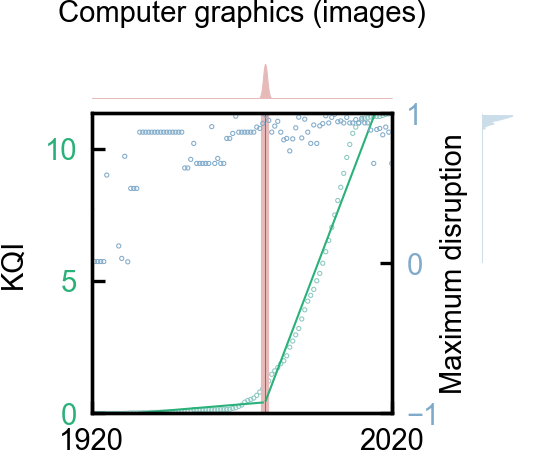}
    \includegraphics[height=4cm]{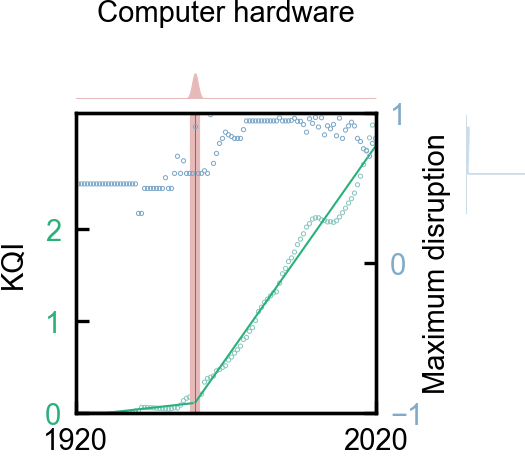}
    \includegraphics[height=4cm]{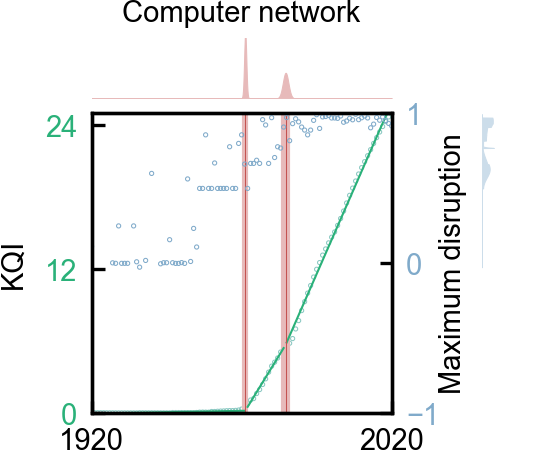}
    \includegraphics[height=4cm]{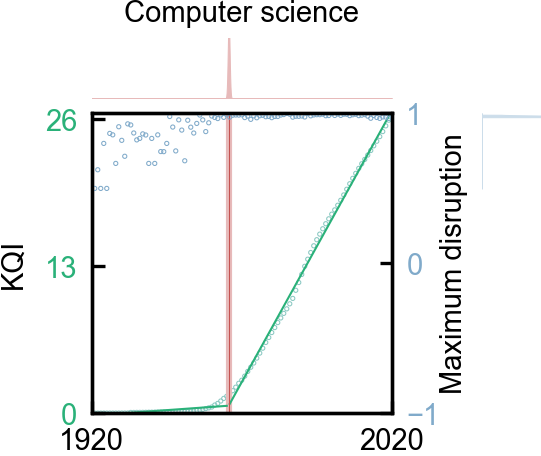}
    \includegraphics[height=4cm]{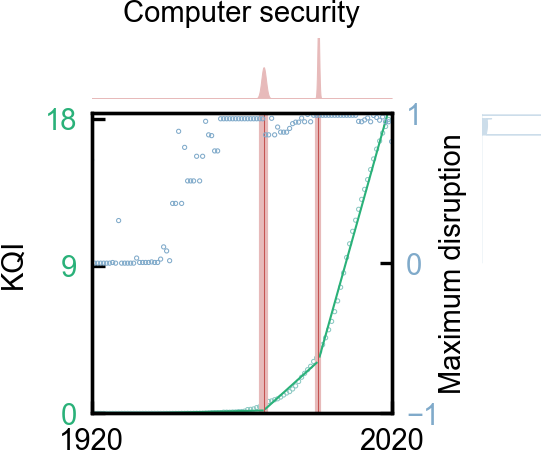}
    \includegraphics[height=4cm]{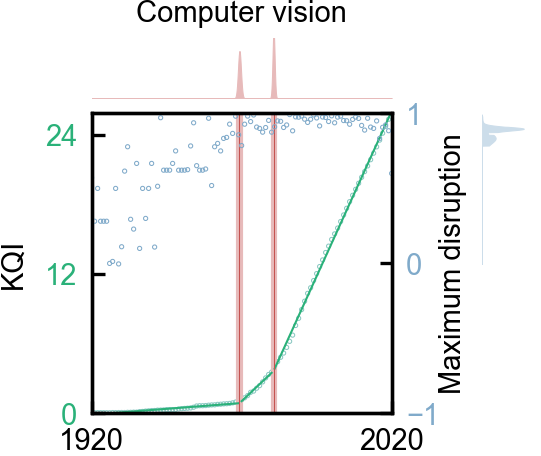}
    \includegraphics[height=4cm]{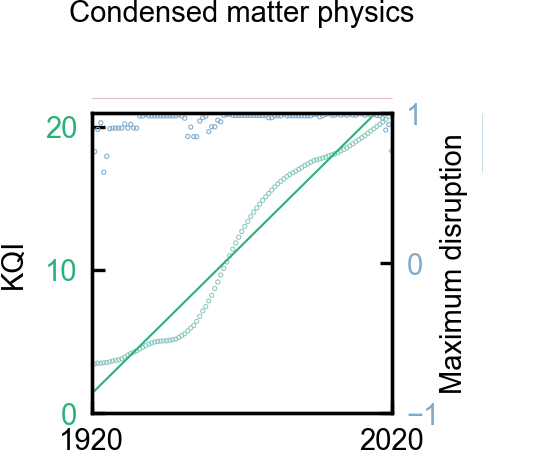}
    \includegraphics[height=4cm]{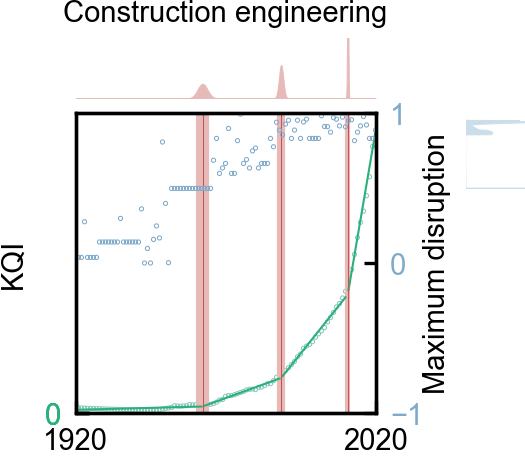}
    \includegraphics[height=4cm]{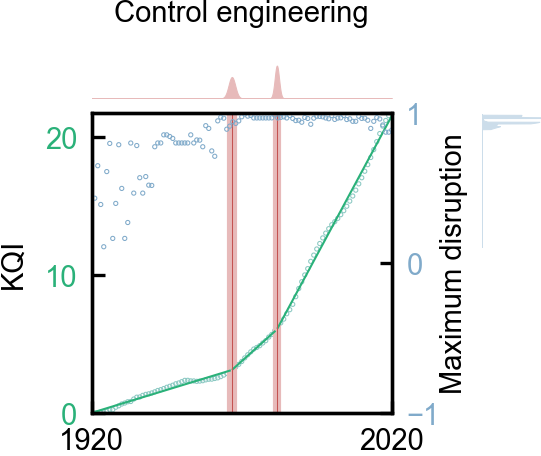}
    \includegraphics[height=4cm]{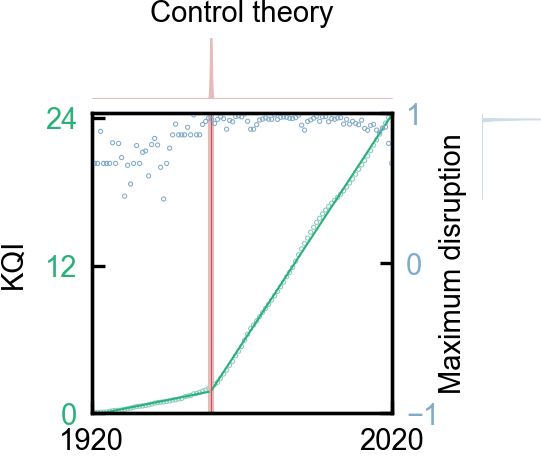}
    \includegraphics[height=4cm]{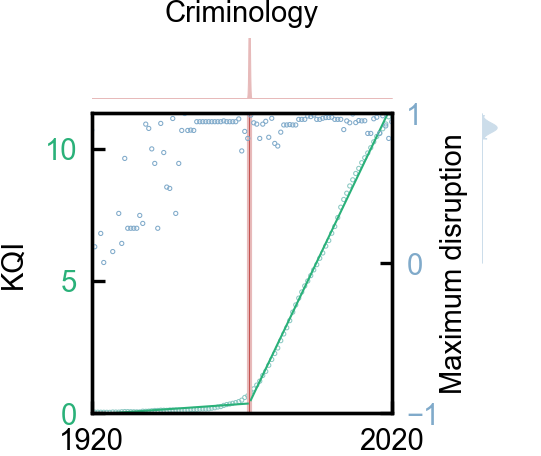}
    \includegraphics[height=4cm]{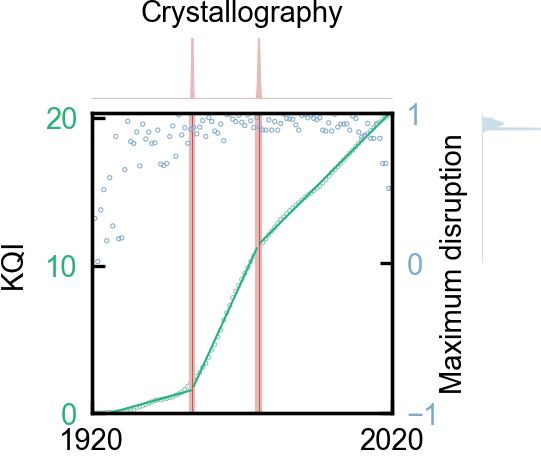}
    \includegraphics[height=4cm]{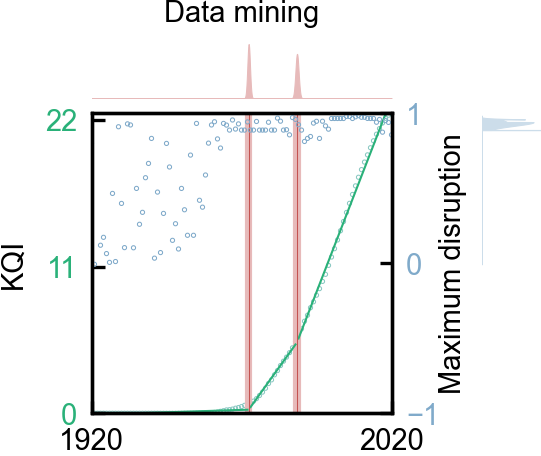}
    \includegraphics[height=4cm]{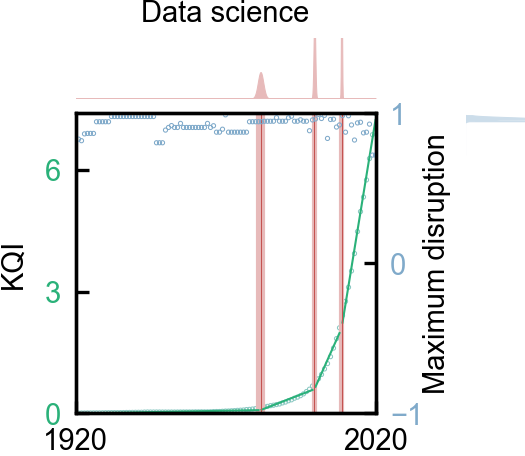}
\end{figure}
\begin{figure}[H]
    \centering
    \includegraphics[height=4cm]{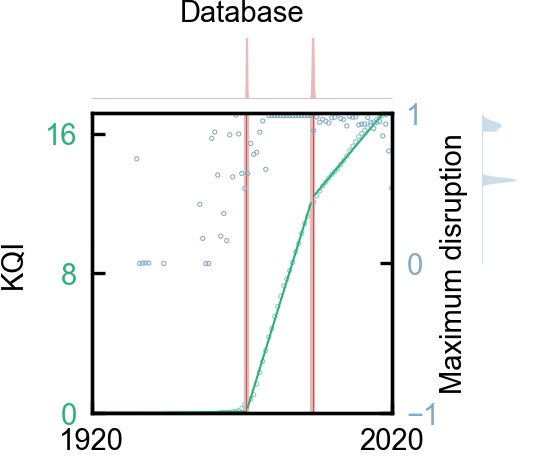}
    \includegraphics[height=4cm]{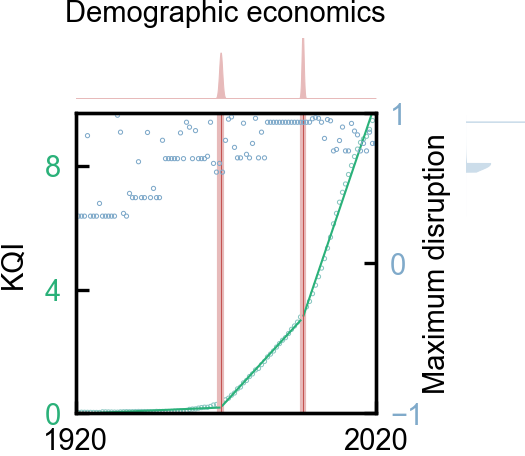}
    \includegraphics[height=4cm]{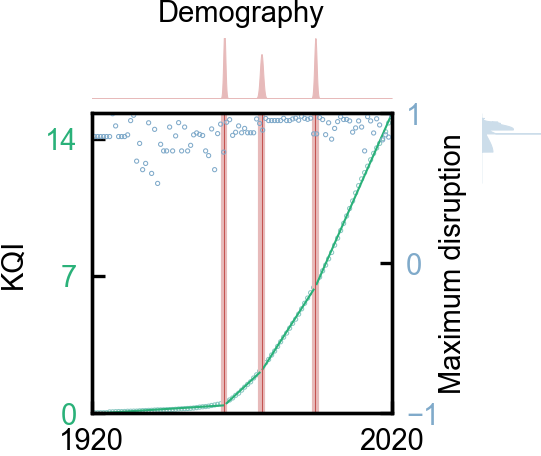}
    \includegraphics[height=4cm]{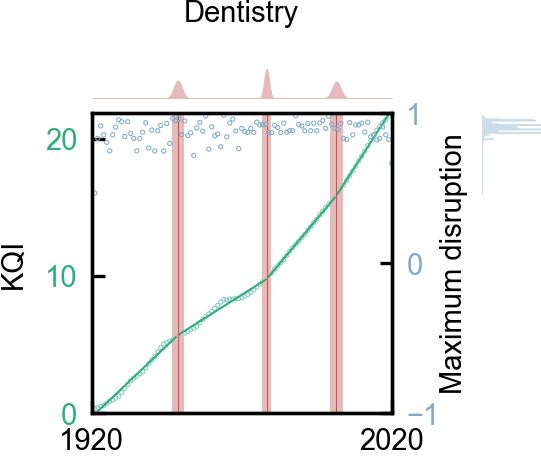}
    \includegraphics[height=4cm]{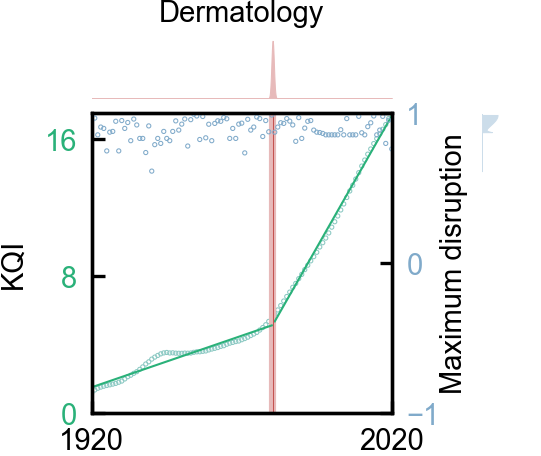}
    \includegraphics[height=4cm]{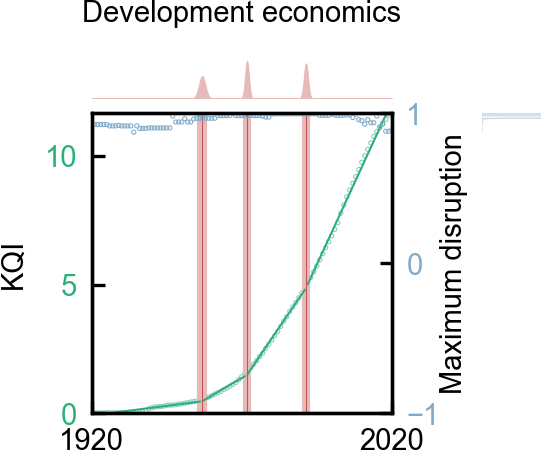}
    \includegraphics[height=4cm]{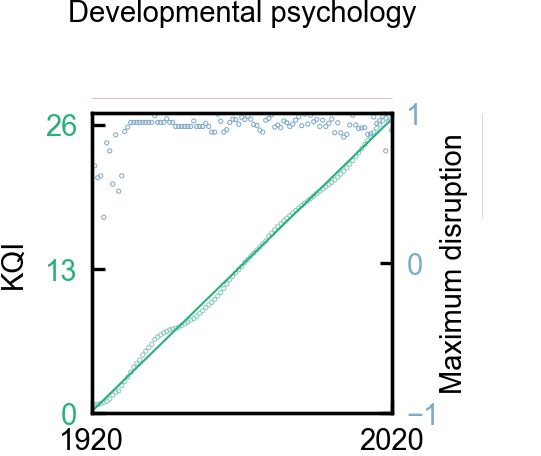}
    \includegraphics[height=4cm]{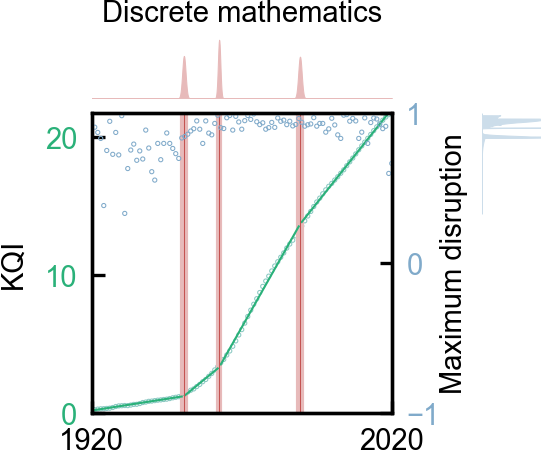}
    \includegraphics[height=4cm]{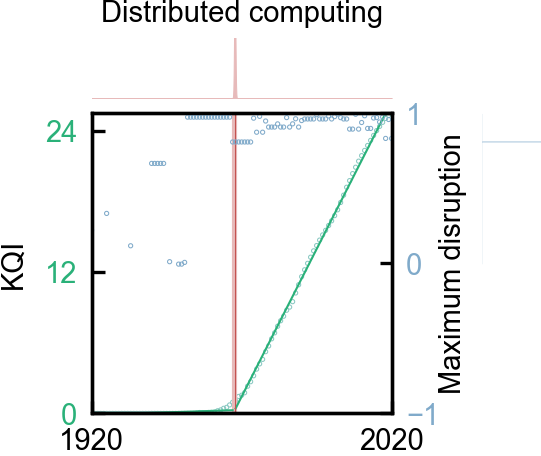}
    \includegraphics[height=4cm]{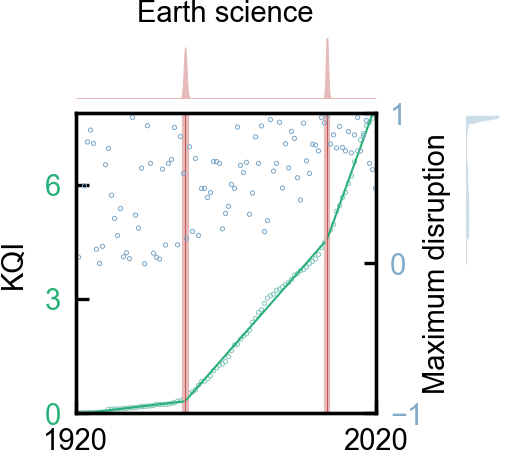}
    \includegraphics[height=4cm]{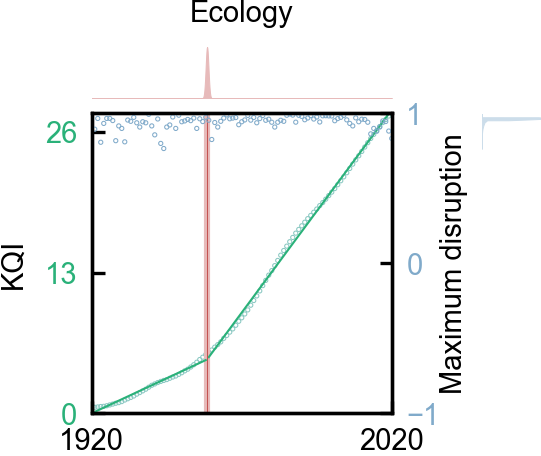}
    \includegraphics[height=4cm]{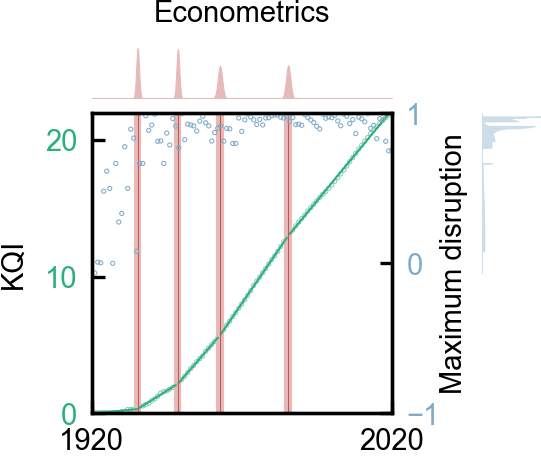}
    \includegraphics[height=4cm]{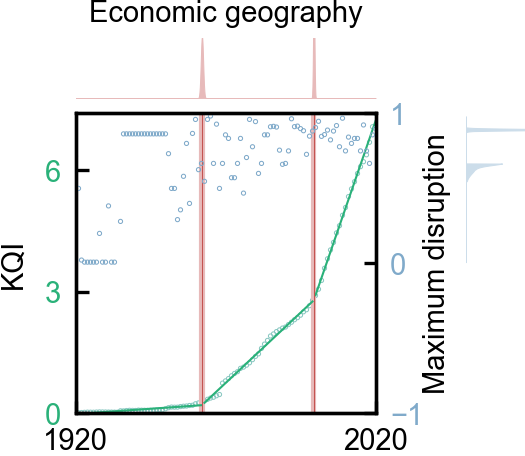}
    \includegraphics[height=4cm]{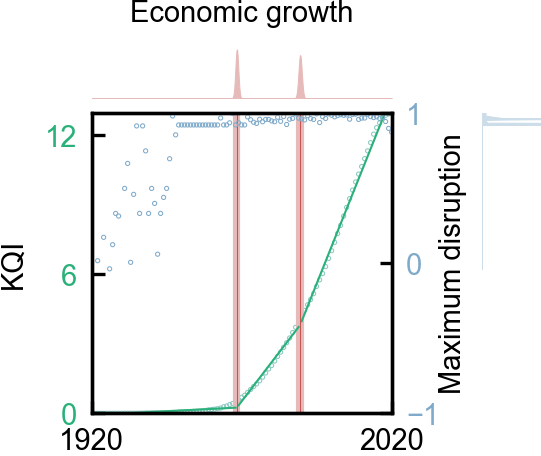}
    \includegraphics[height=4cm]{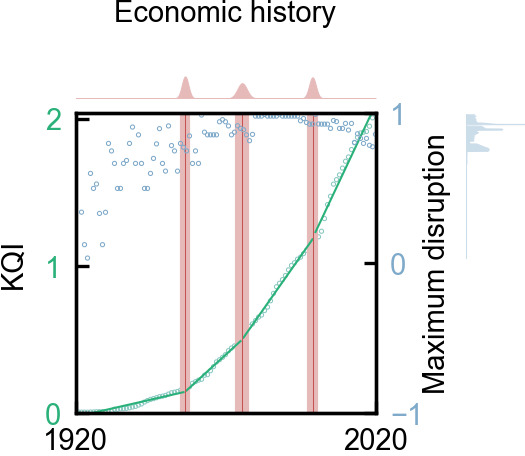}
\end{figure}
\begin{figure}[H]
    \centering
    \includegraphics[height=4cm]{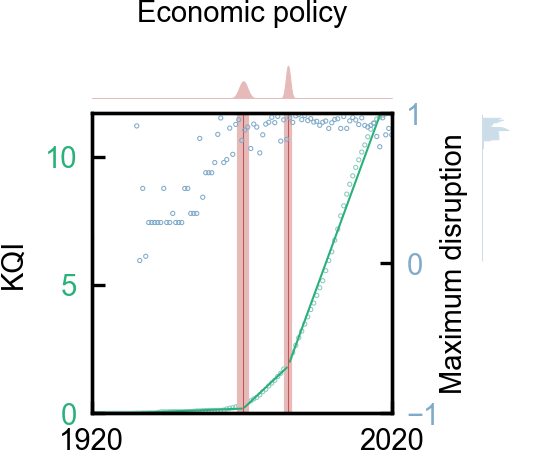}
    \includegraphics[height=4cm]{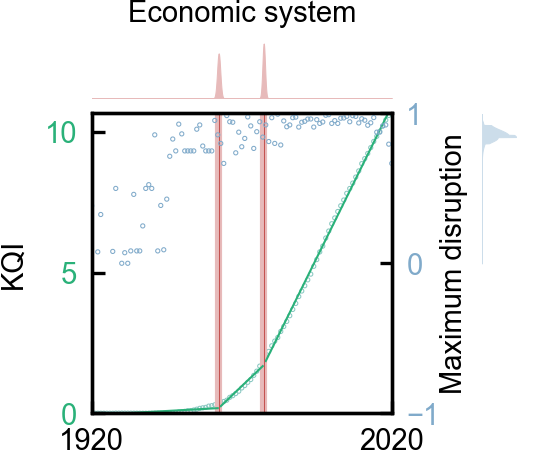}
    \includegraphics[height=4cm]{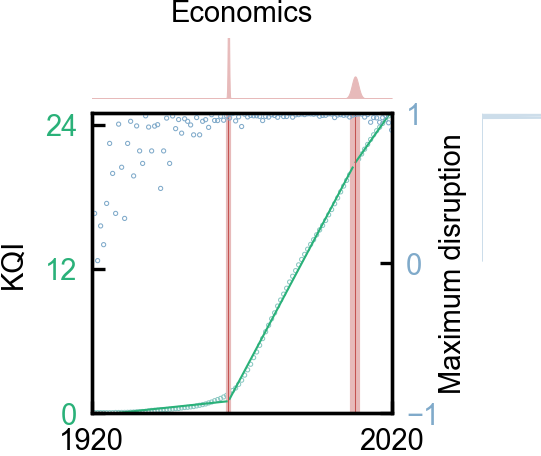}
    \includegraphics[height=4cm]{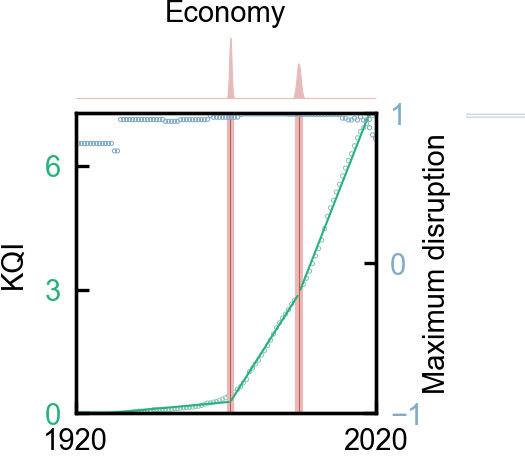}
    \includegraphics[height=4cm]{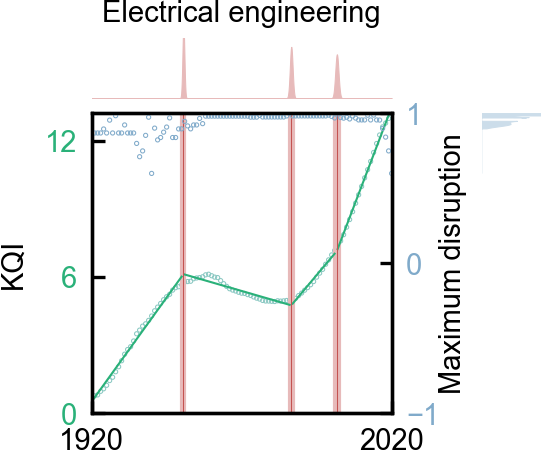}
    \includegraphics[height=4cm]{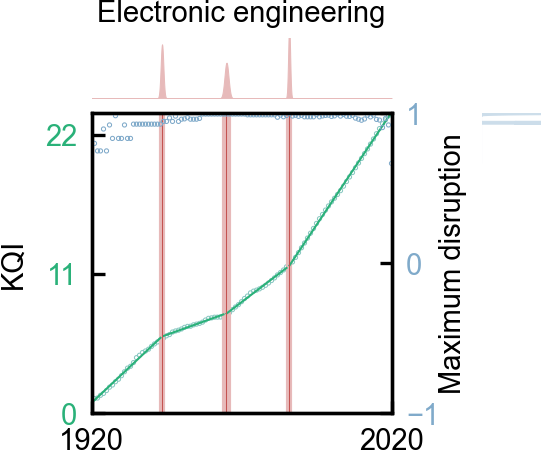}
    \includegraphics[height=4cm]{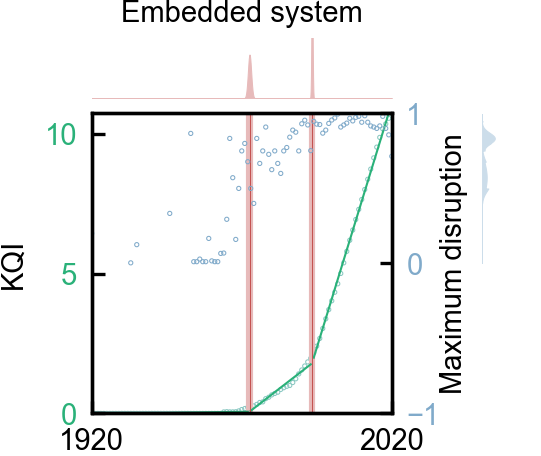}
    \includegraphics[height=4cm]{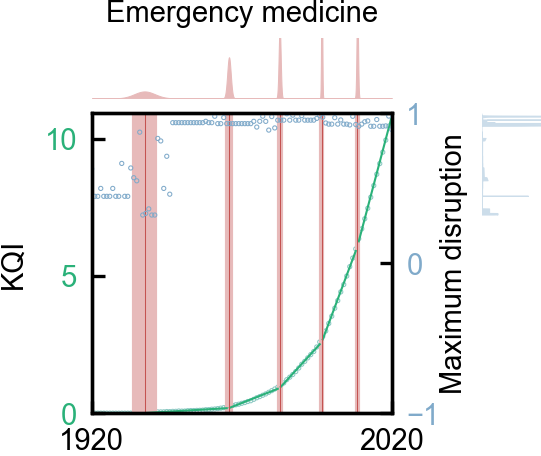}
    \includegraphics[height=4cm]{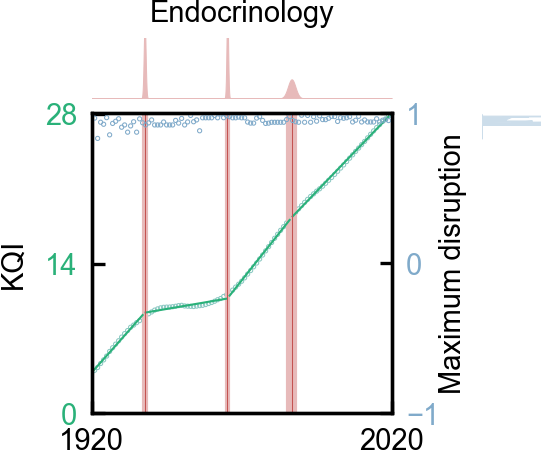}
    \includegraphics[height=4cm]{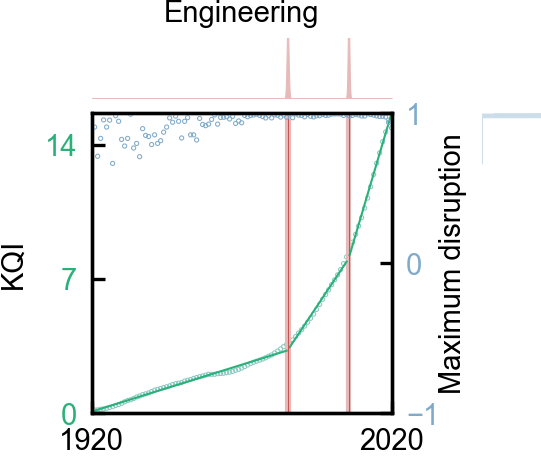}
    \includegraphics[height=4cm]{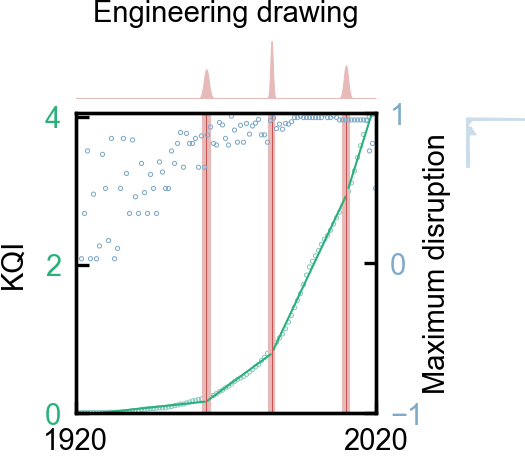}
    \includegraphics[height=4cm]{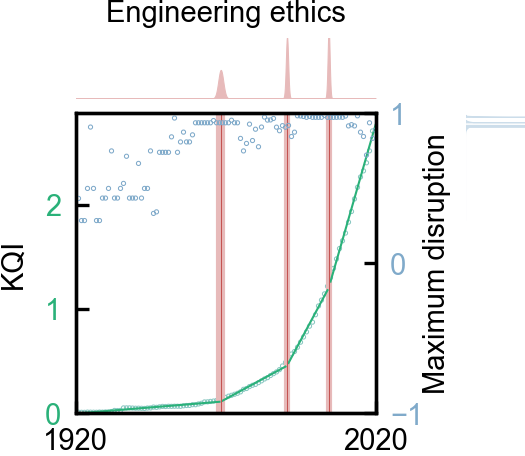}
    \includegraphics[height=4cm]{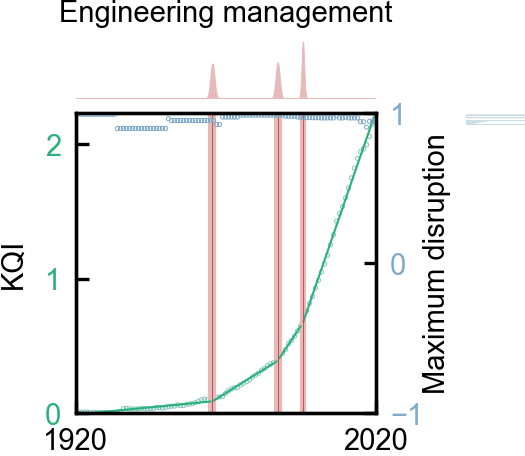}
    \includegraphics[height=4cm]{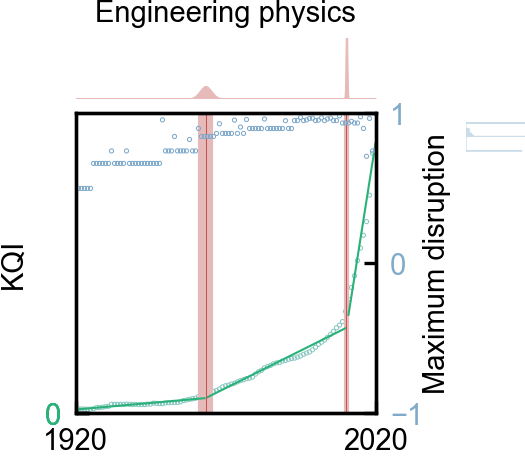}
    \includegraphics[height=4cm]{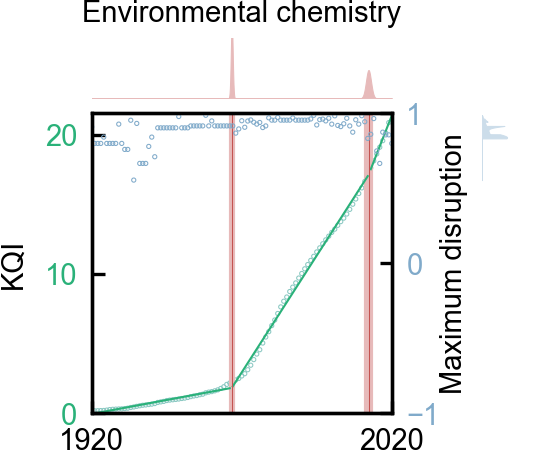}
\end{figure}
\begin{figure}[H]
    \centering
    \includegraphics[height=4cm]{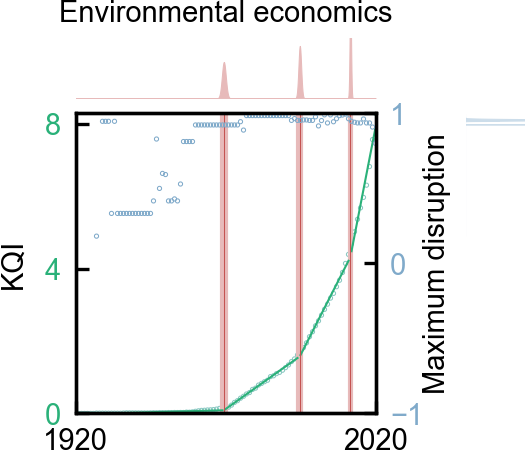}
    \includegraphics[height=4cm]{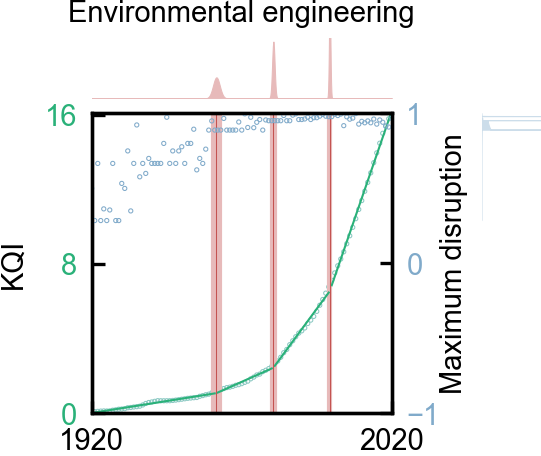}
    \includegraphics[height=4cm]{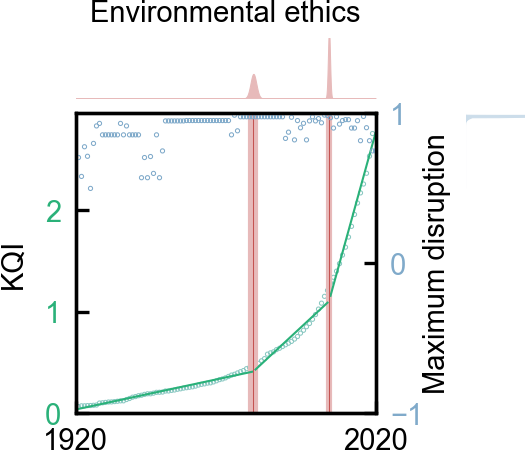}
    \includegraphics[height=4cm]{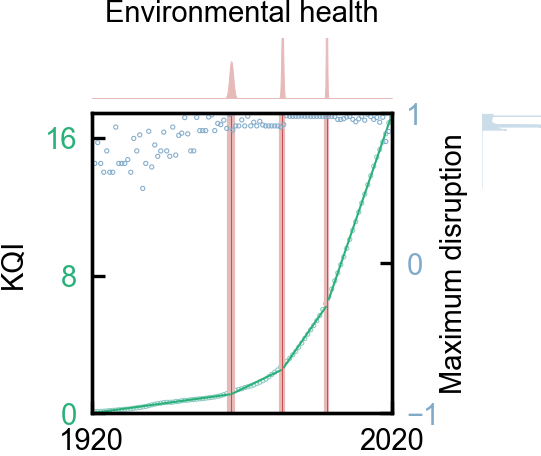}
    \includegraphics[height=4cm]{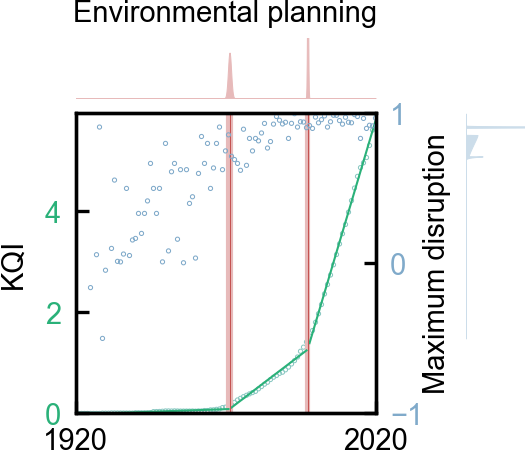}
    \includegraphics[height=4cm]{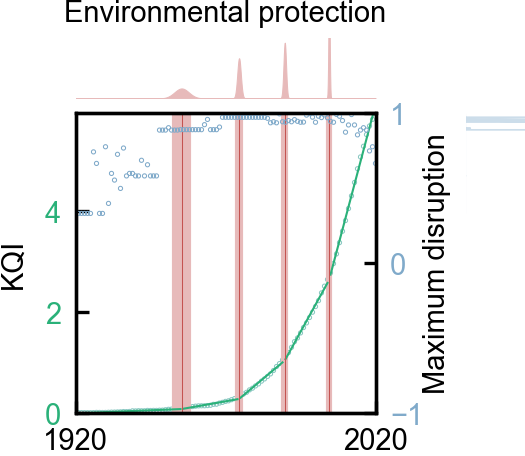}
    \includegraphics[height=4cm]{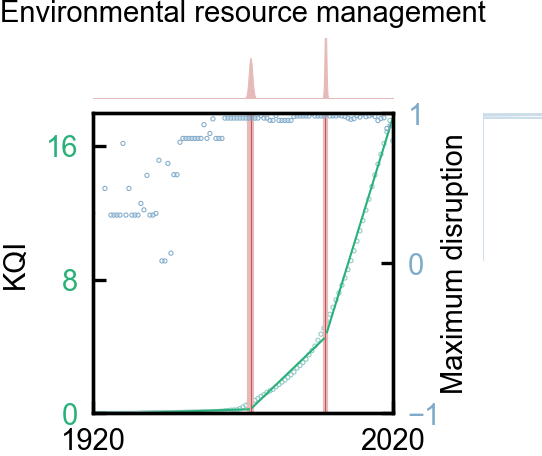}
    \includegraphics[height=4cm]{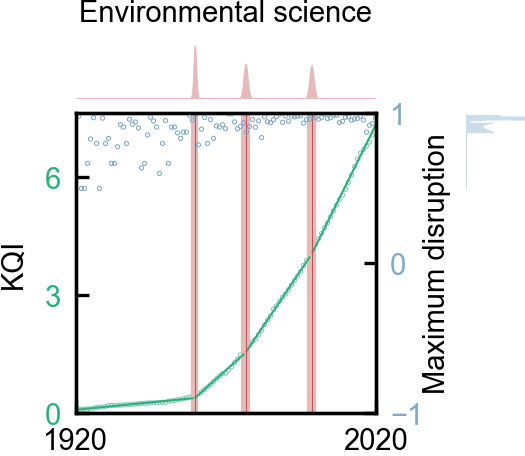}
    \includegraphics[height=4cm]{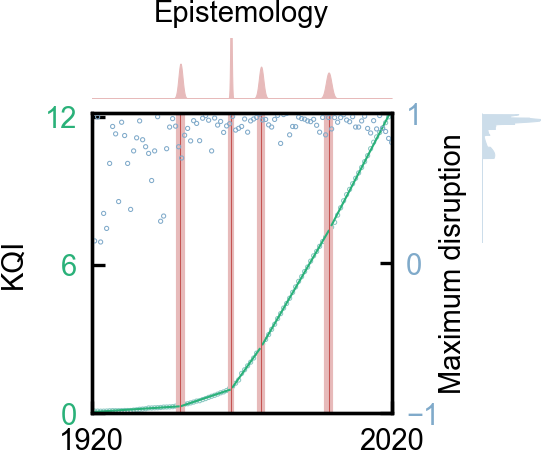}
    \includegraphics[height=4cm]{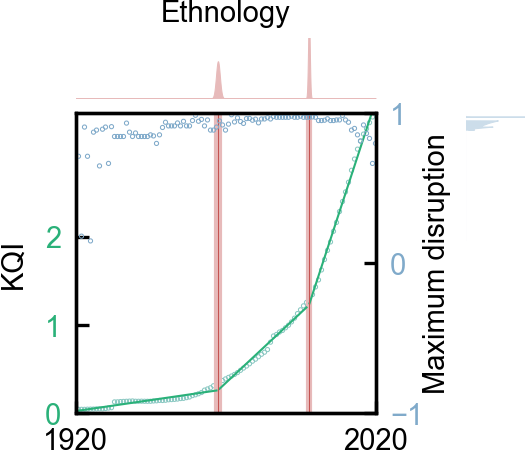}
    \includegraphics[height=4cm]{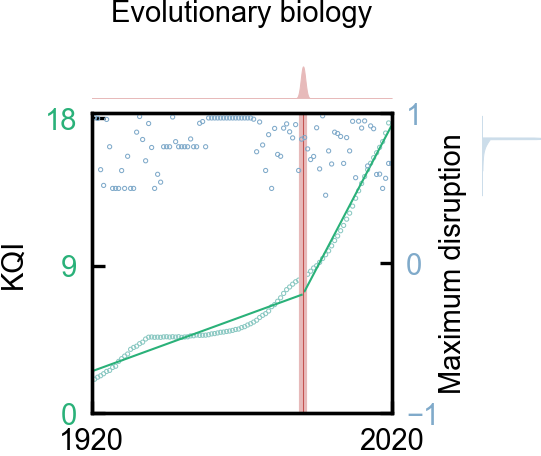}
    \includegraphics[height=4cm]{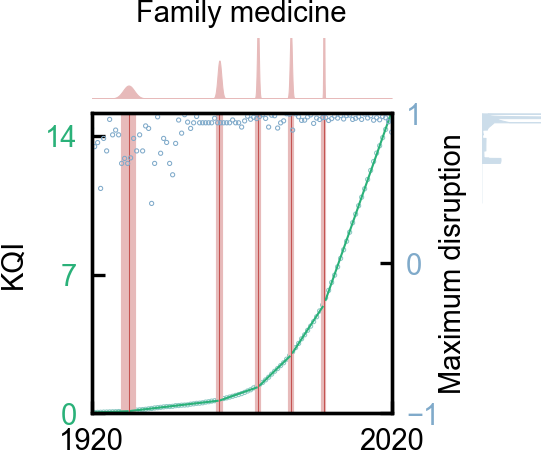}
    \includegraphics[height=4cm]{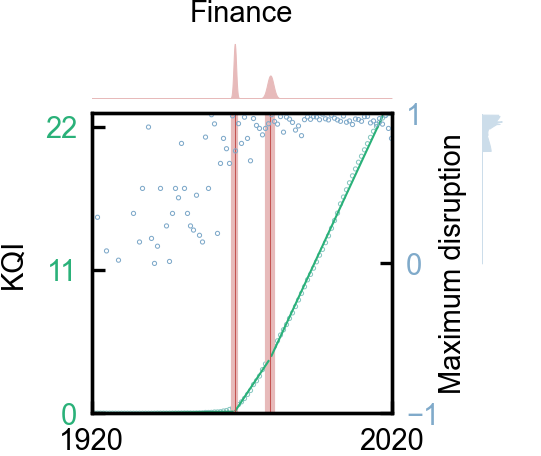}
    \includegraphics[height=4cm]{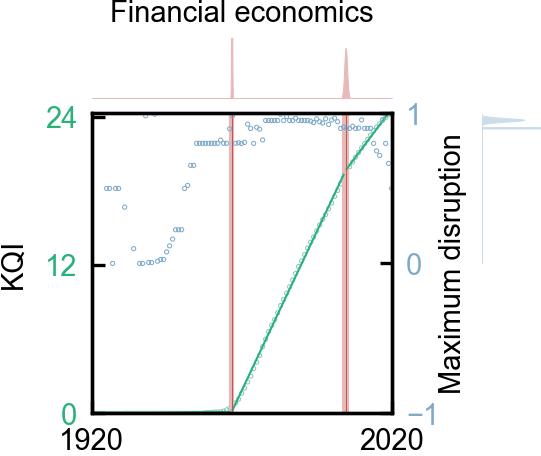}
    \includegraphics[height=4cm]{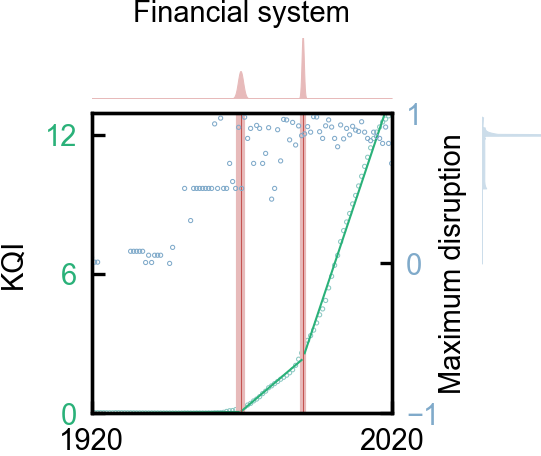}
\end{figure}
\begin{figure}[H]
    \centering
    \includegraphics[height=4cm]{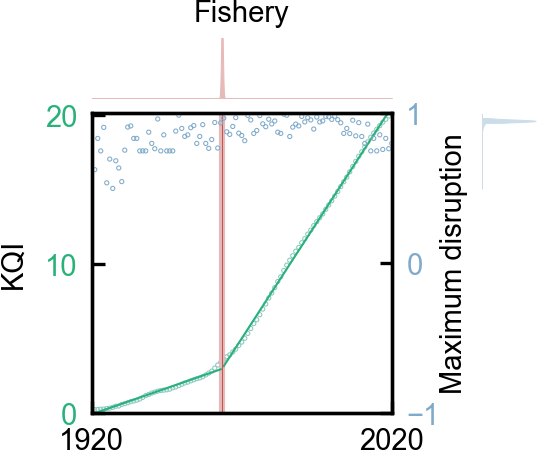}
    \includegraphics[height=4cm]{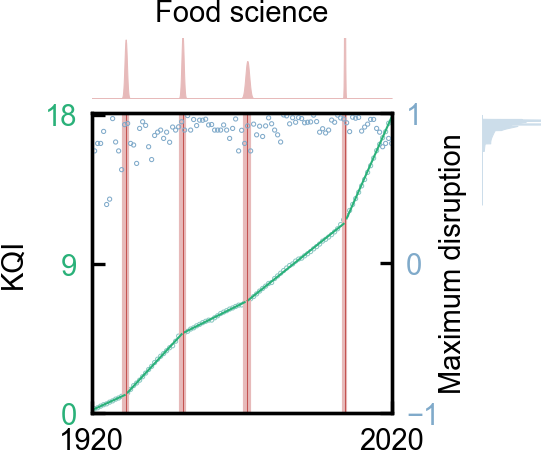}
    \includegraphics[height=4cm]{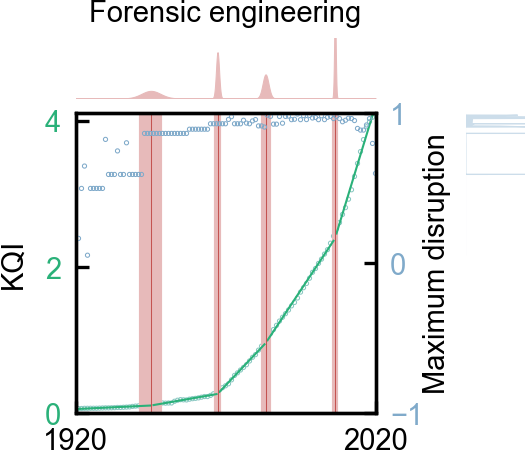}
    \includegraphics[height=4cm]{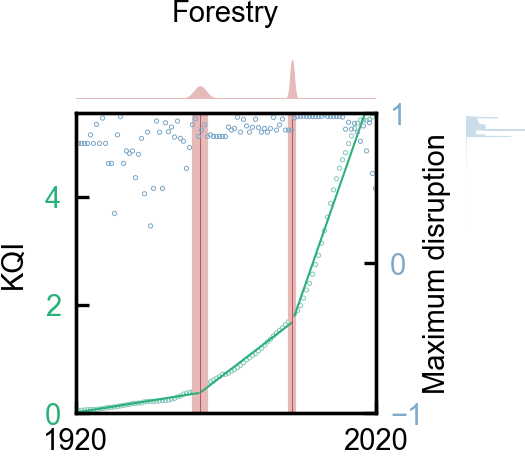}
    \includegraphics[height=4cm]{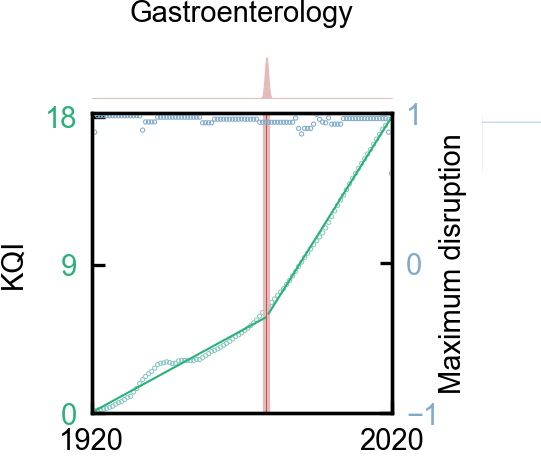}
    \includegraphics[height=4cm]{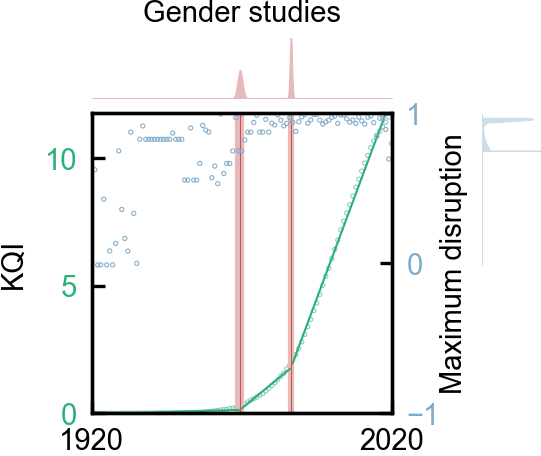}
    \includegraphics[height=4cm]{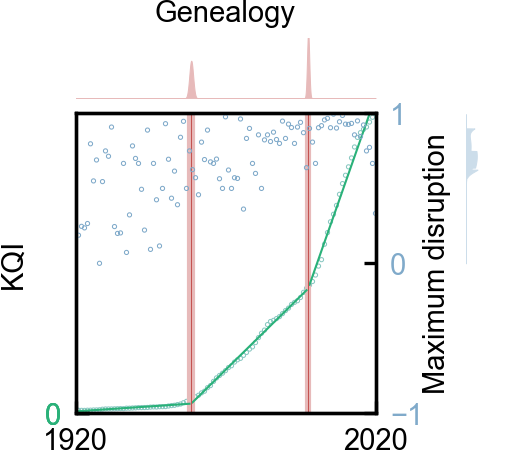}
    \includegraphics[height=4cm]{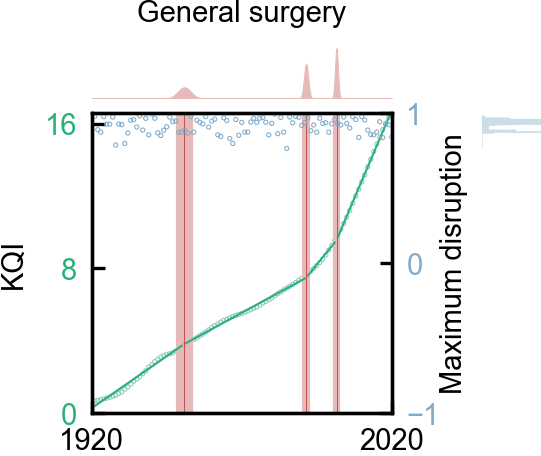}
    \includegraphics[height=4cm]{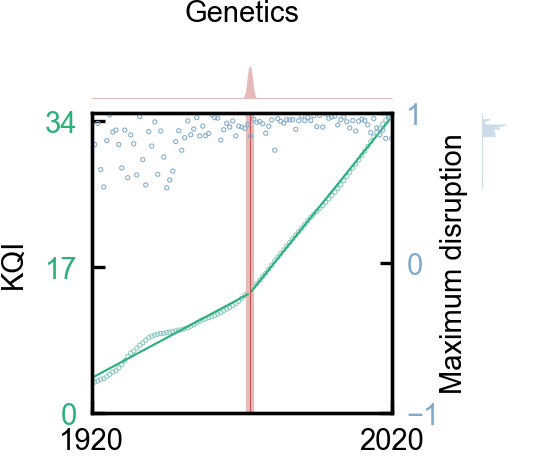}
    \includegraphics[height=4cm]{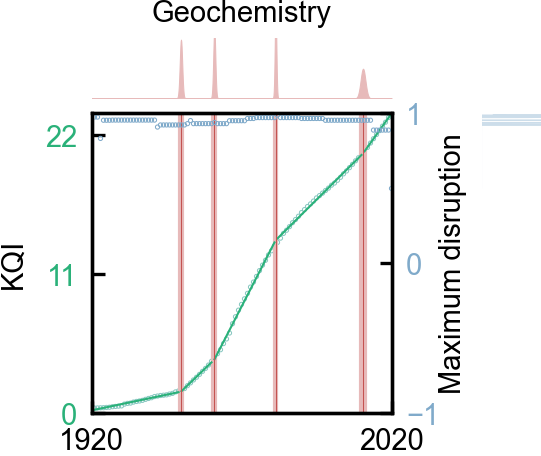}
    \includegraphics[height=4cm]{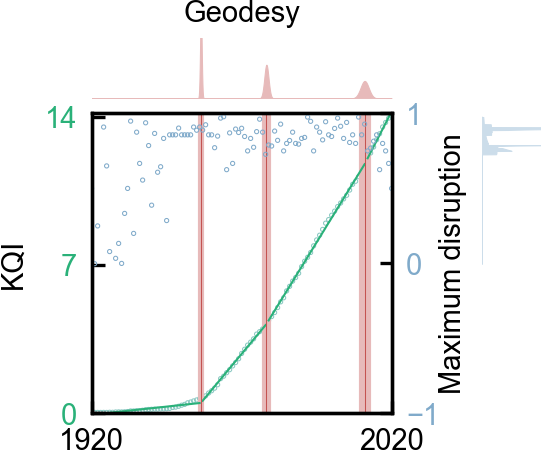}
    \includegraphics[height=4cm]{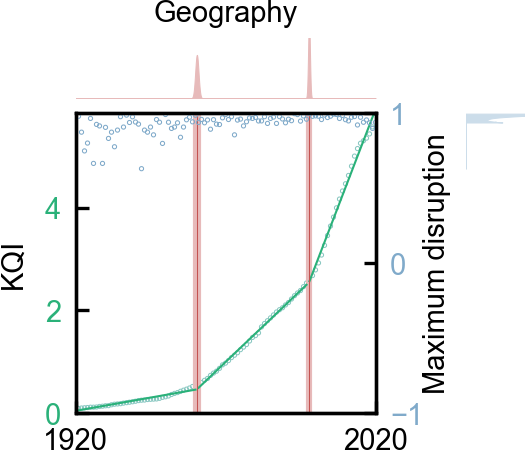}
    \includegraphics[height=4cm]{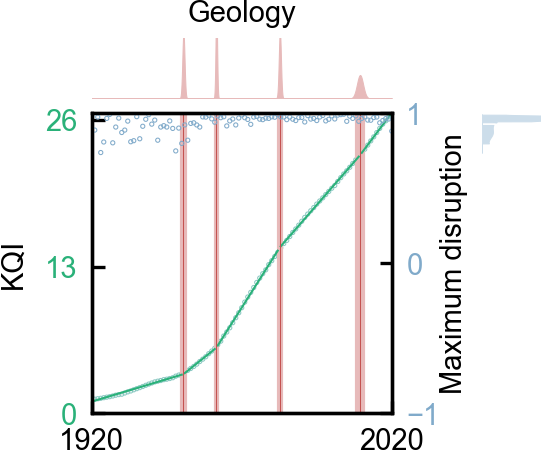}
    \includegraphics[height=4cm]{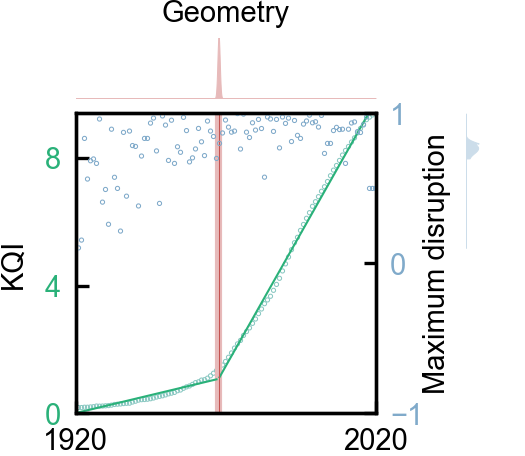}
    \includegraphics[height=4cm]{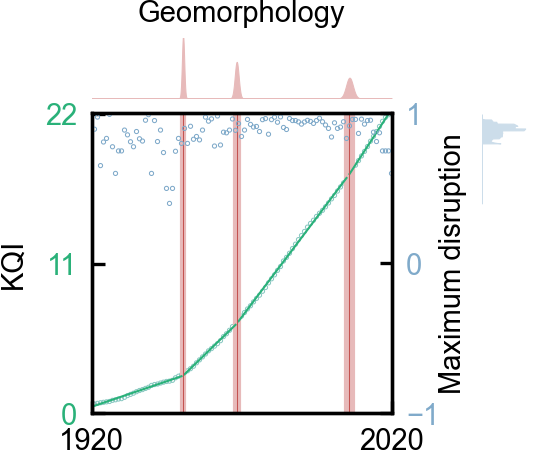}
\end{figure}
\begin{figure}[H]
    \centering
    \includegraphics[height=4cm]{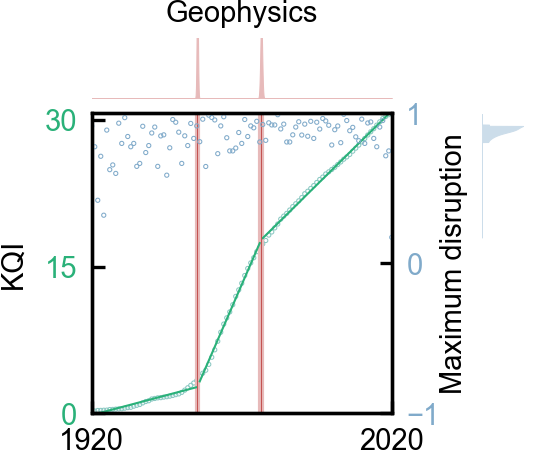}
    \includegraphics[height=4cm]{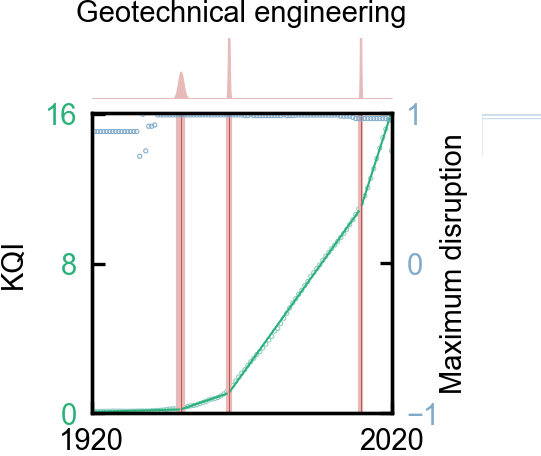}
    \includegraphics[height=4cm]{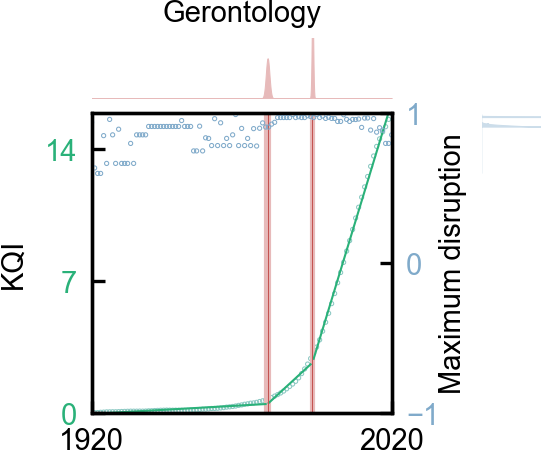}
    \includegraphics[height=4cm]{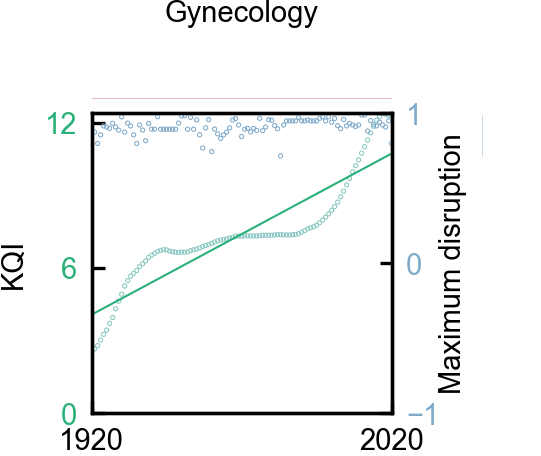}
    \includegraphics[height=4cm]{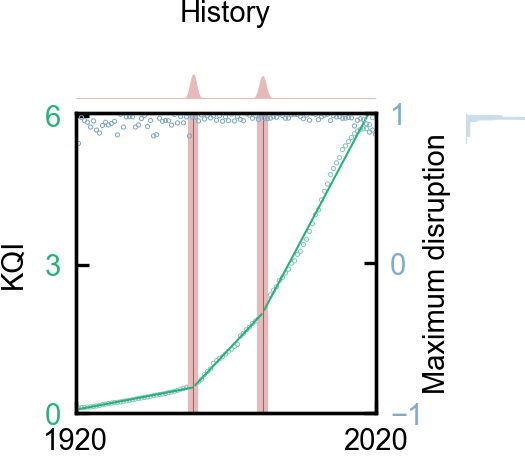}
    \includegraphics[height=4cm]{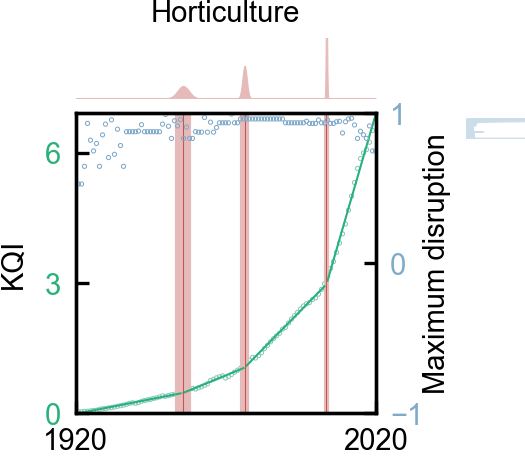}
    \includegraphics[height=4cm]{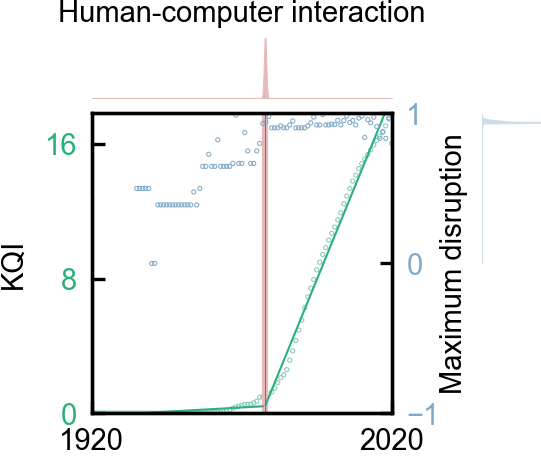}
    \includegraphics[height=4cm]{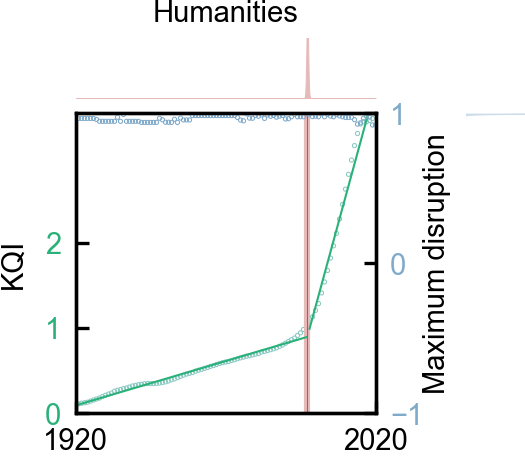}
    \includegraphics[height=4cm]{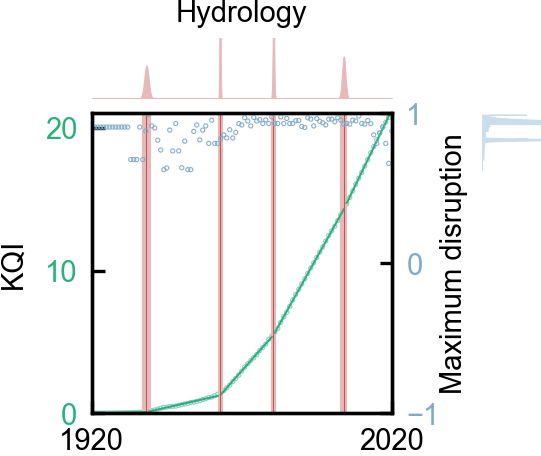}
    \includegraphics[height=4cm]{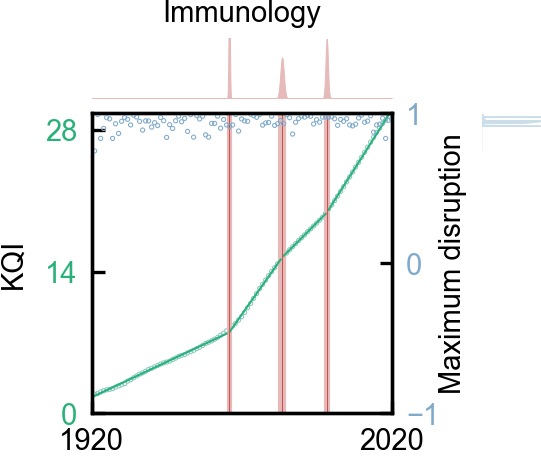}
    \includegraphics[height=4cm]{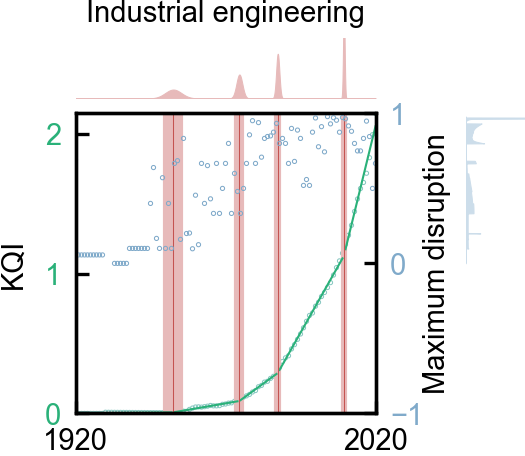}
    \includegraphics[height=4cm]{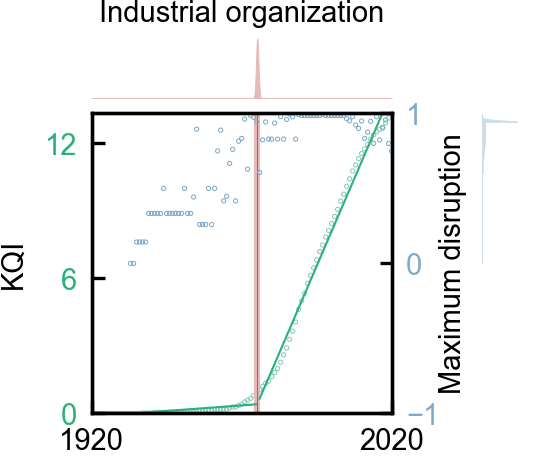}
    \includegraphics[height=4cm]{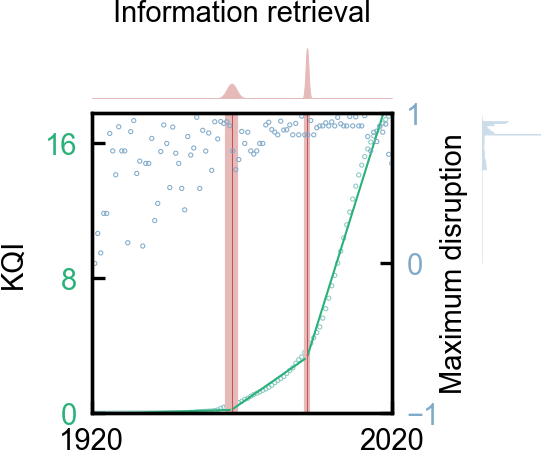}
    \includegraphics[height=4cm]{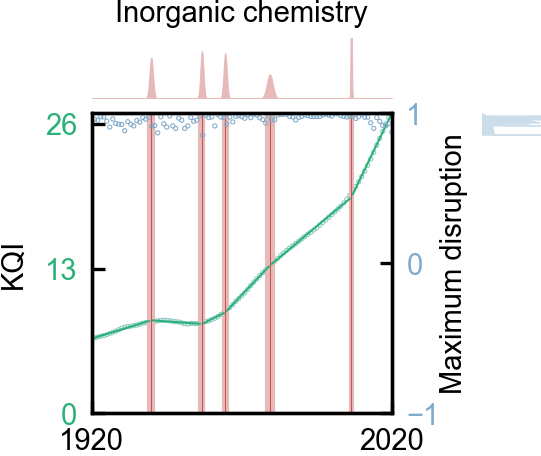}
    \includegraphics[height=4cm]{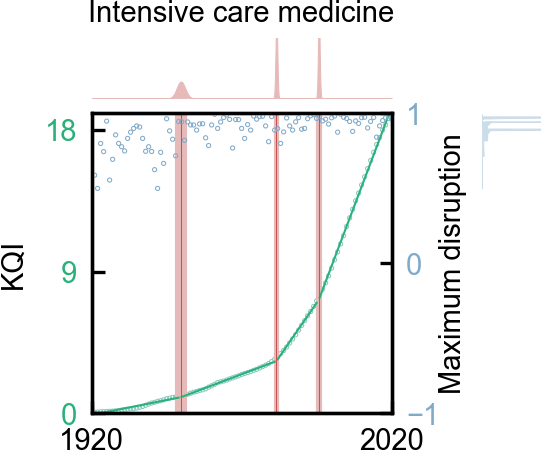}
\end{figure}
\begin{figure}[H]
    \centering
    \includegraphics[height=4cm]{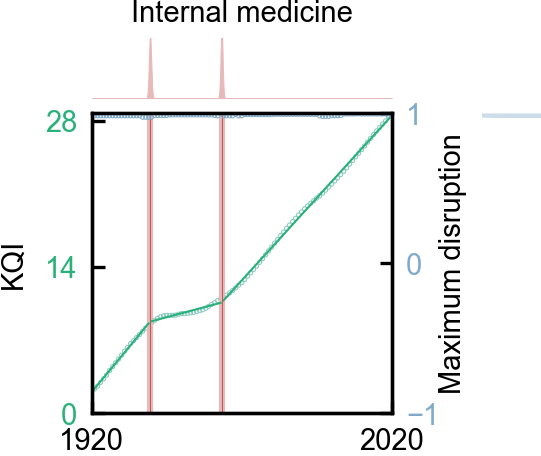}
    \includegraphics[height=4cm]{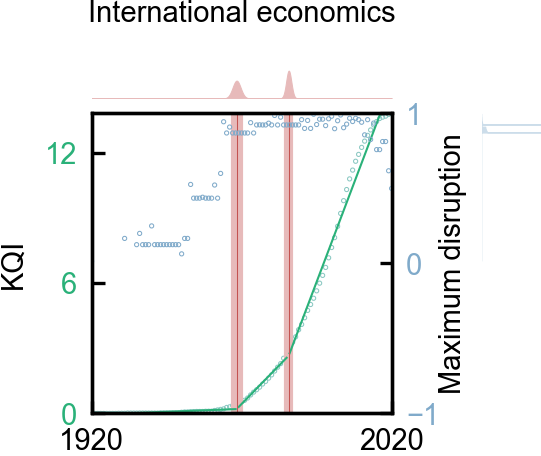}
    \includegraphics[height=4cm]{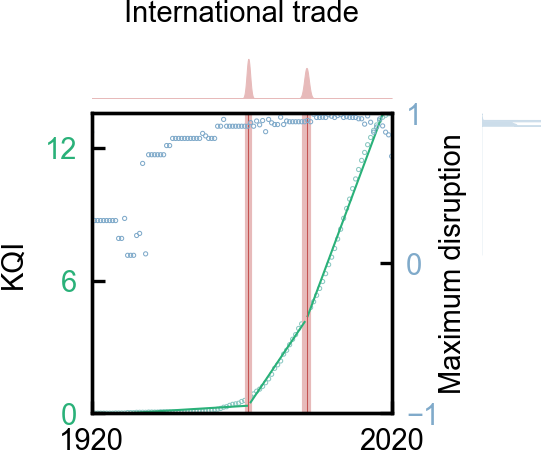}
    \includegraphics[height=4cm]{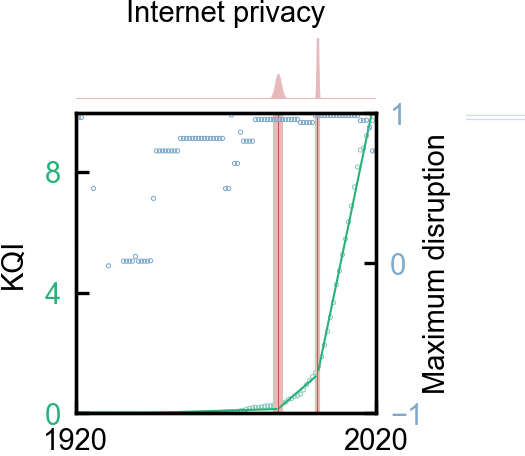}
    \includegraphics[height=4cm]{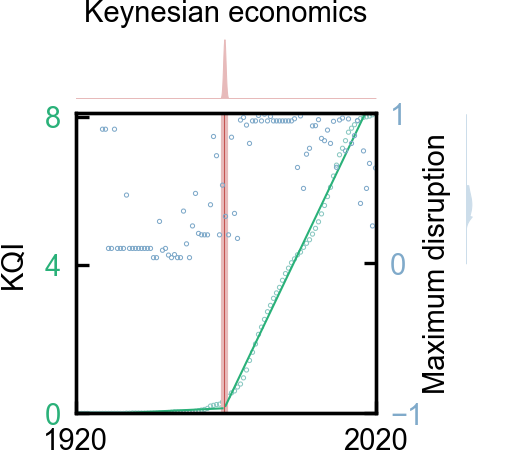}
    \includegraphics[height=4cm]{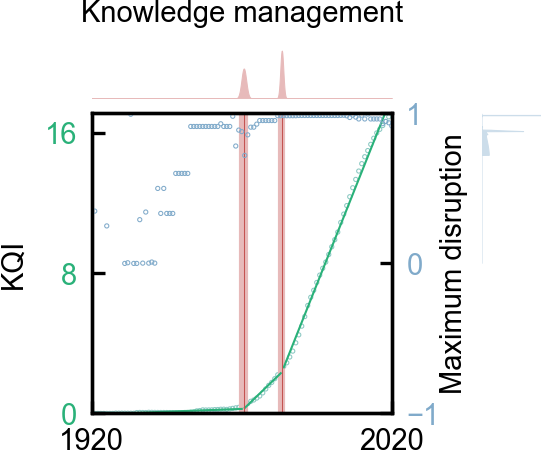}
    \includegraphics[height=4cm]{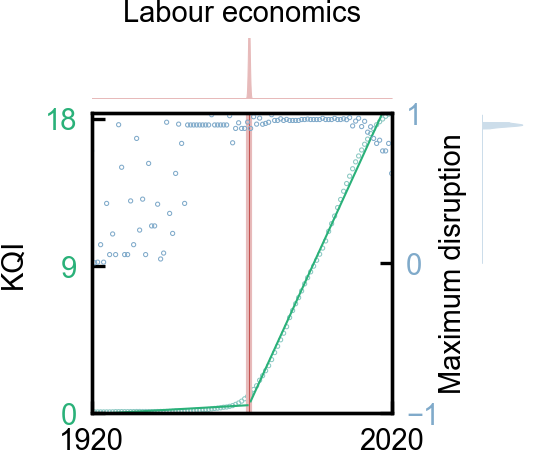}
    \includegraphics[height=4cm]{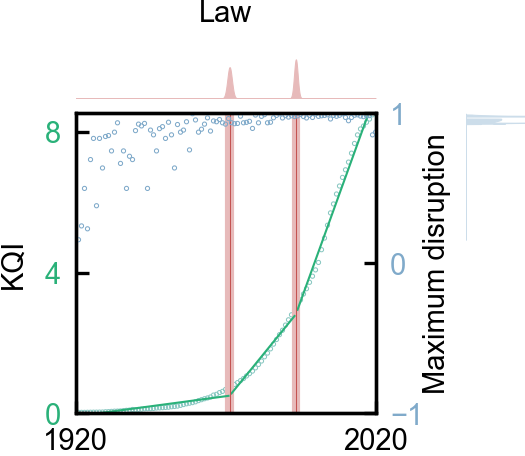}
    \includegraphics[height=4cm]{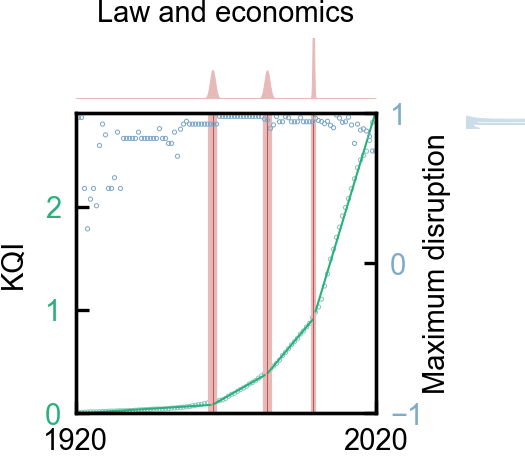}
    \includegraphics[height=4cm]{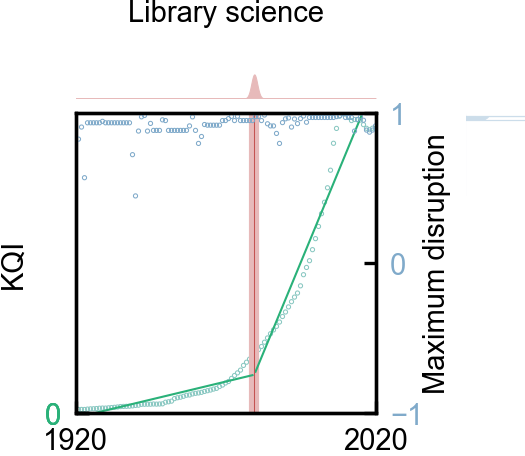}
    \includegraphics[height=4cm]{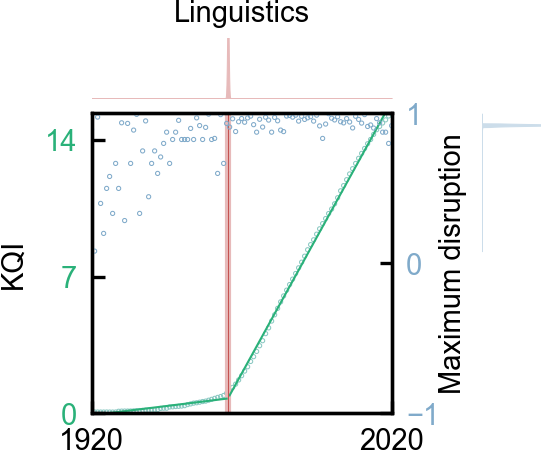}
    \includegraphics[height=4cm]{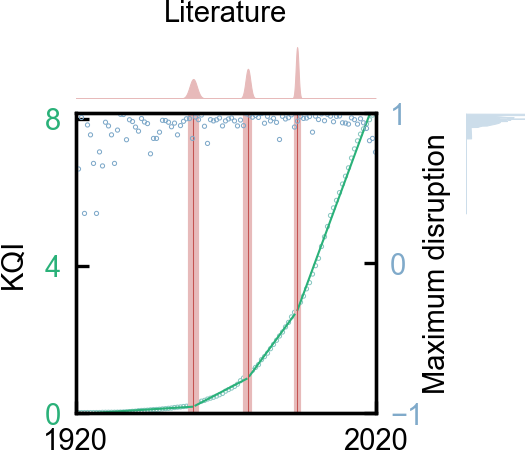}
    \includegraphics[height=4cm]{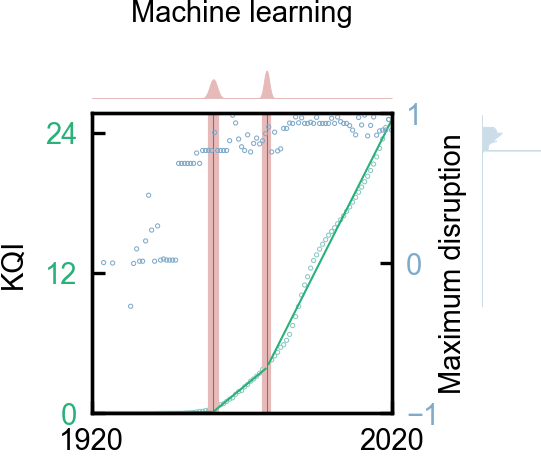}
    \includegraphics[height=4cm]{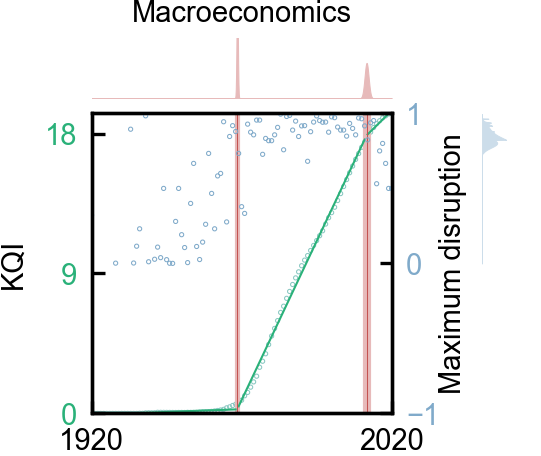}
    \includegraphics[height=4cm]{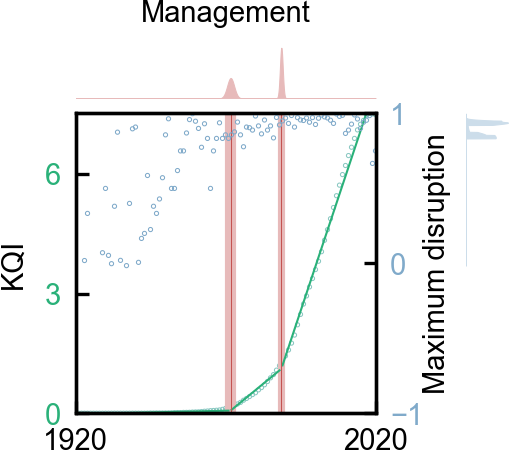}
\end{figure}
\begin{figure}[H]
    \centering
    \includegraphics[height=4cm]{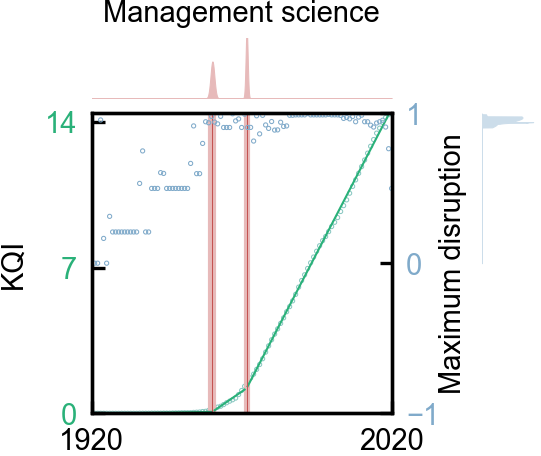}
    \includegraphics[height=4cm]{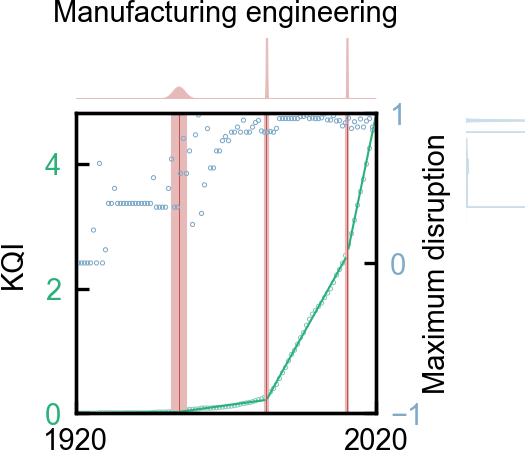}
    \includegraphics[height=4cm]{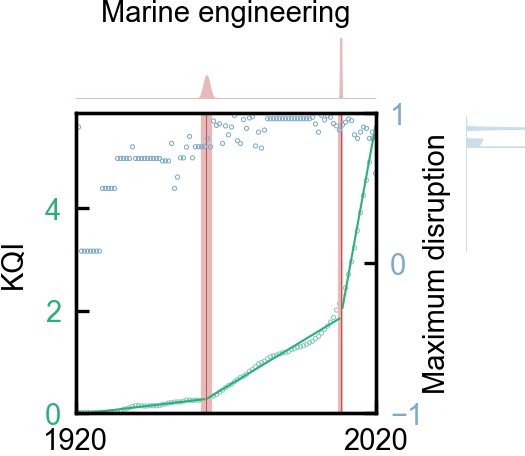}
    \includegraphics[height=4cm]{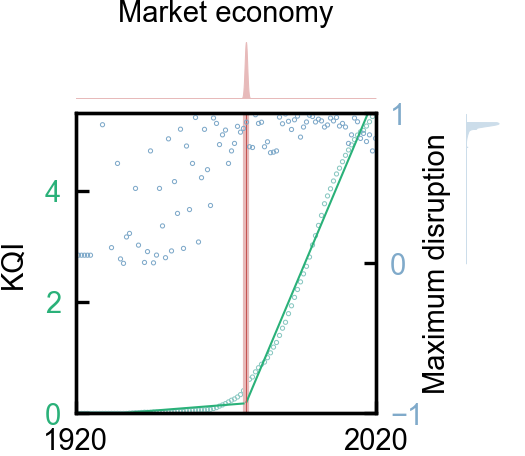}
    \includegraphics[height=4cm]{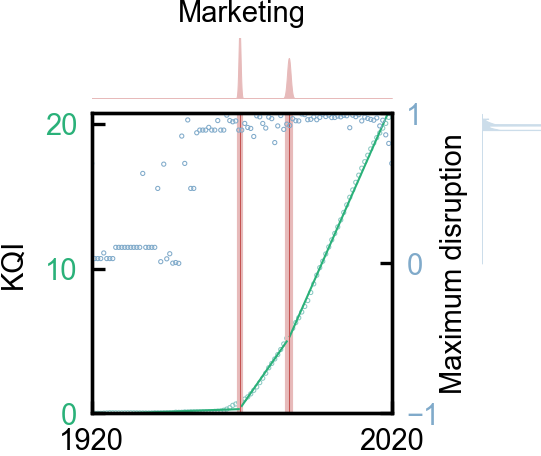}
    \includegraphics[height=4cm]{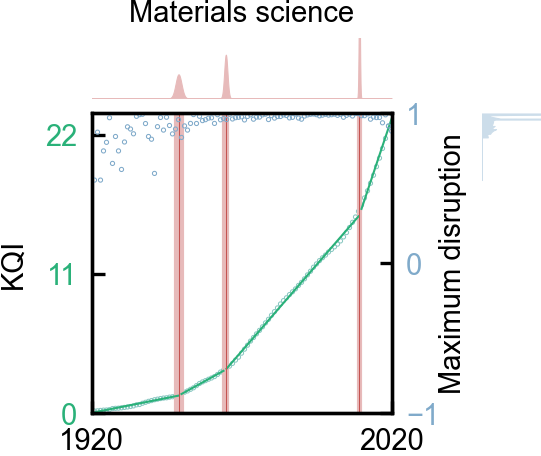}
    \includegraphics[height=4cm]{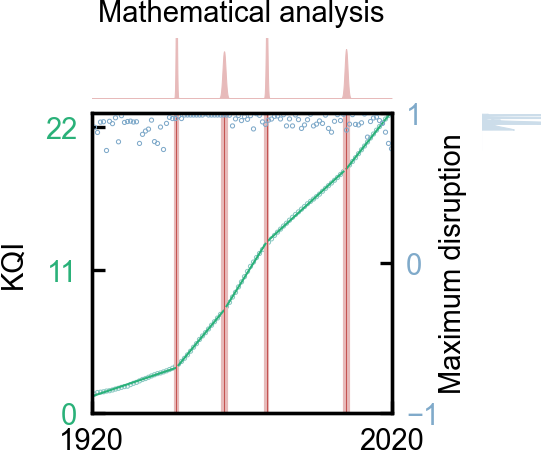}
    \includegraphics[height=4cm]{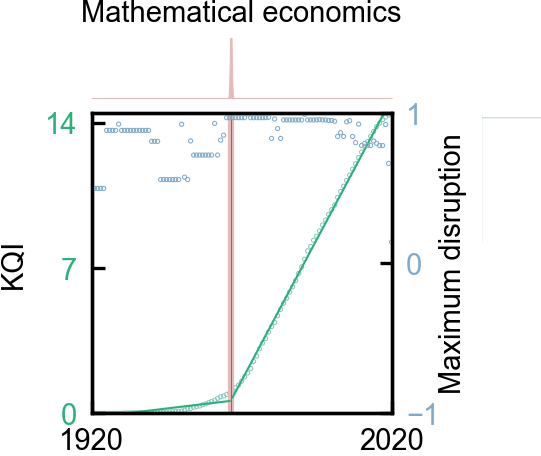}
    \includegraphics[height=4cm]{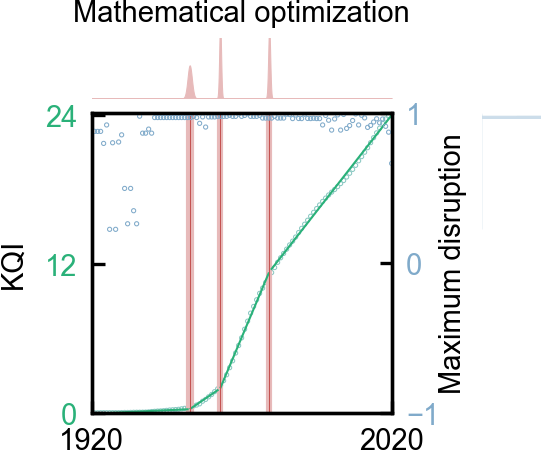}
    \includegraphics[height=4cm]{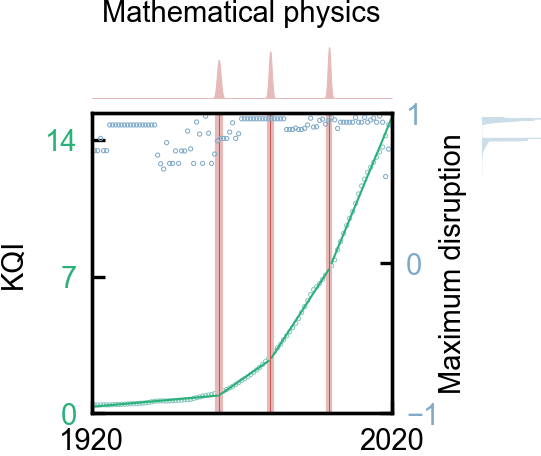}
    \includegraphics[height=4cm]{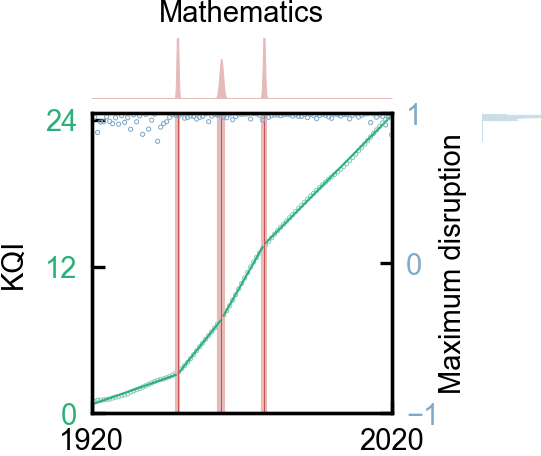}
    \includegraphics[height=4cm]{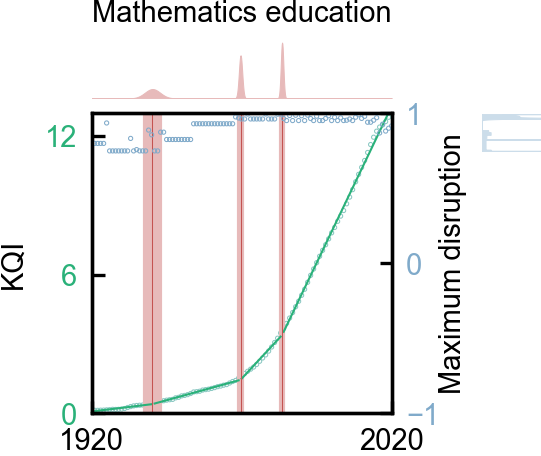}
    \includegraphics[height=4cm]{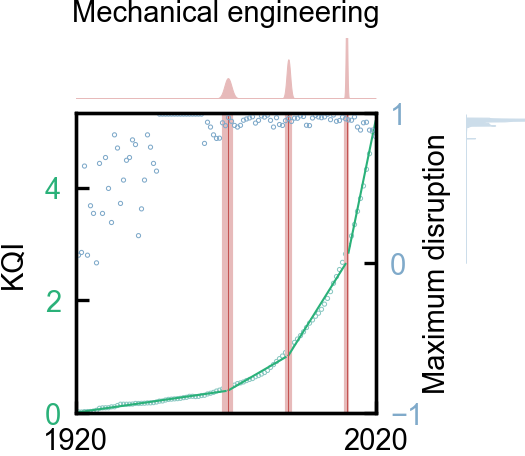}
    \includegraphics[height=4cm]{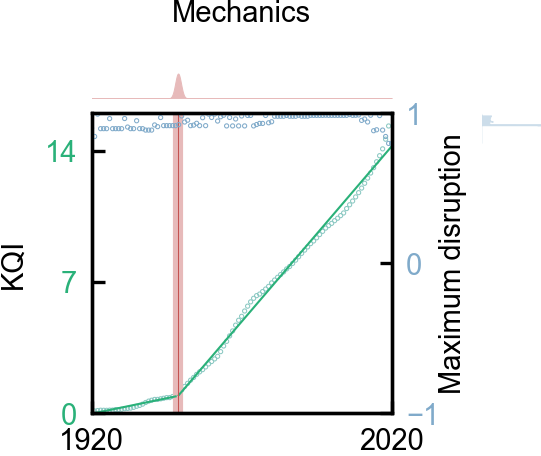}
    \includegraphics[height=4cm]{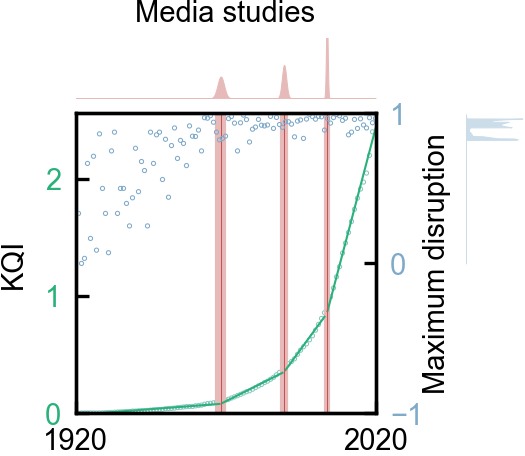}
\end{figure}
\begin{figure}[H]
    \centering
    \includegraphics[height=4cm]{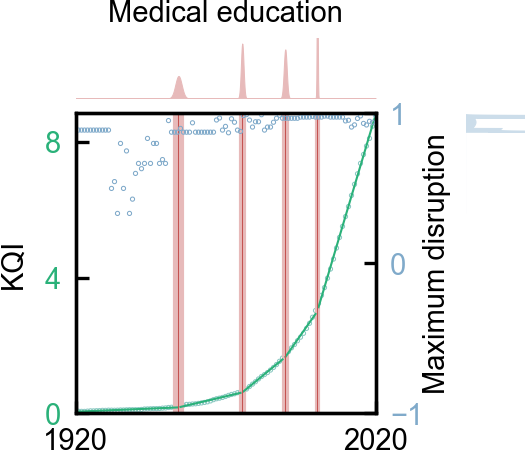}
    \includegraphics[height=4cm]{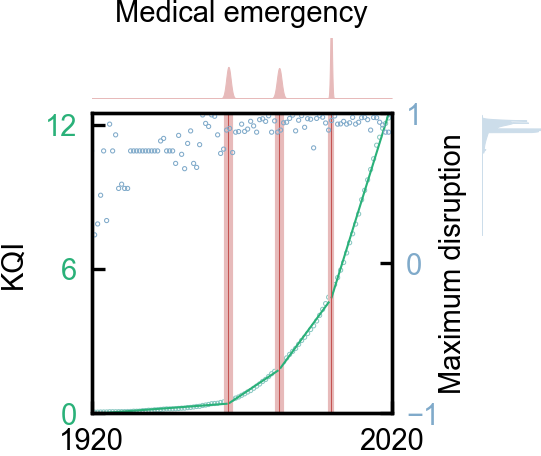}
    \includegraphics[height=4cm]{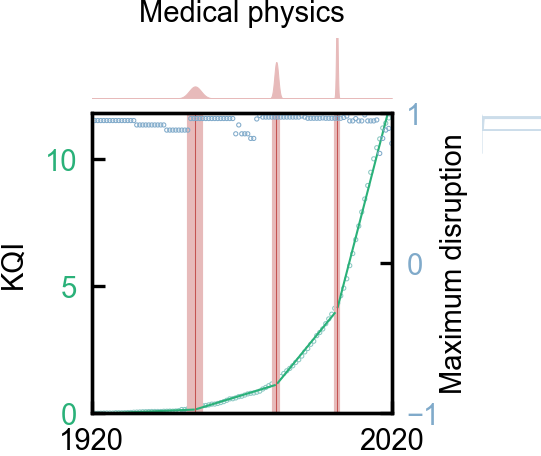}
    \includegraphics[height=4cm]{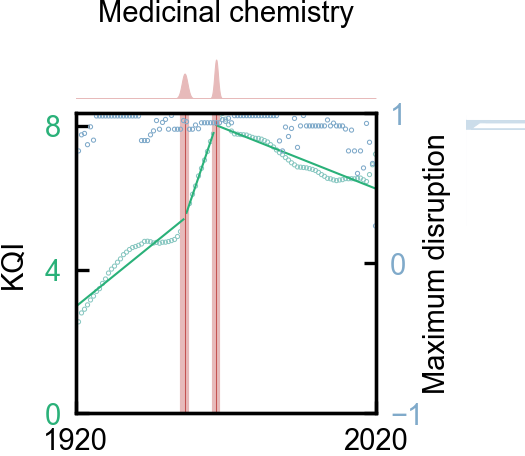}
    \includegraphics[height=4cm]{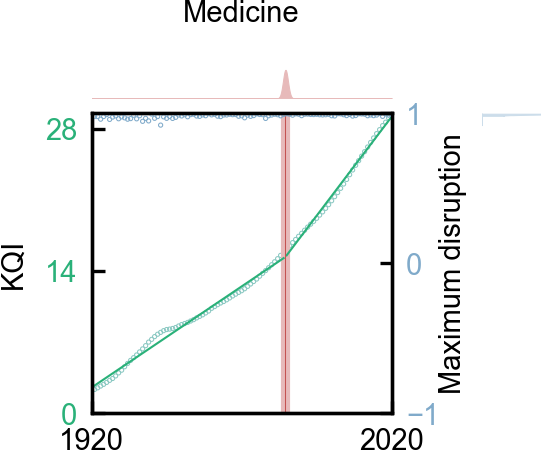}
    \includegraphics[height=4cm]{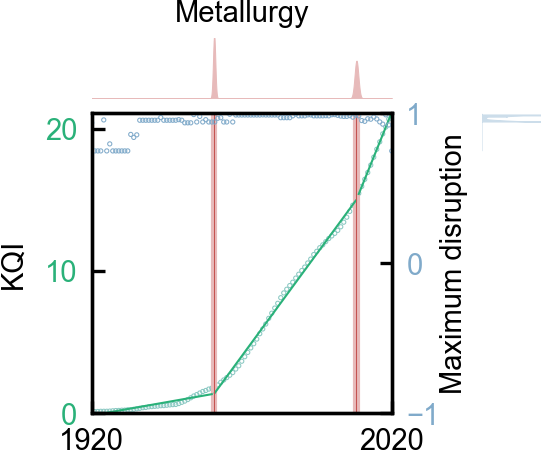}
    \includegraphics[height=4cm]{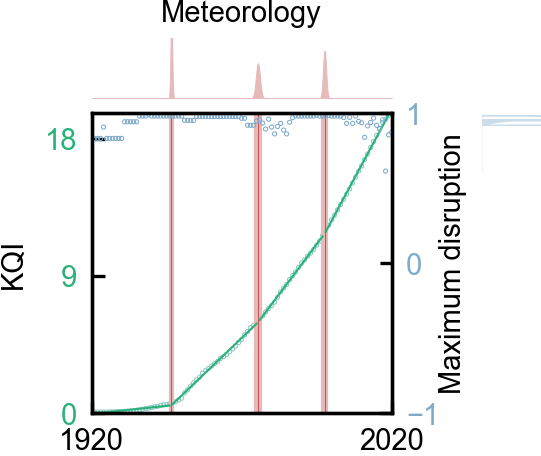}
    \includegraphics[height=4cm]{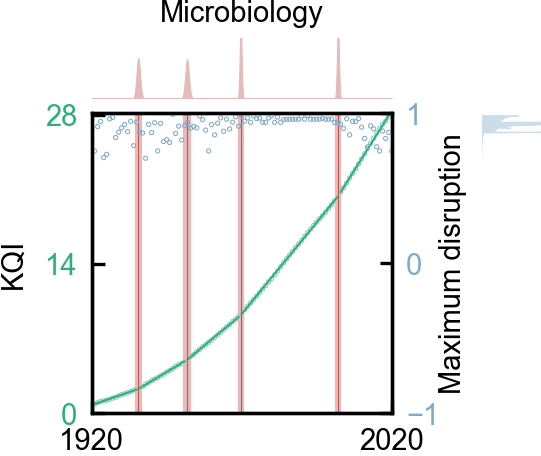}
    \includegraphics[height=4cm]{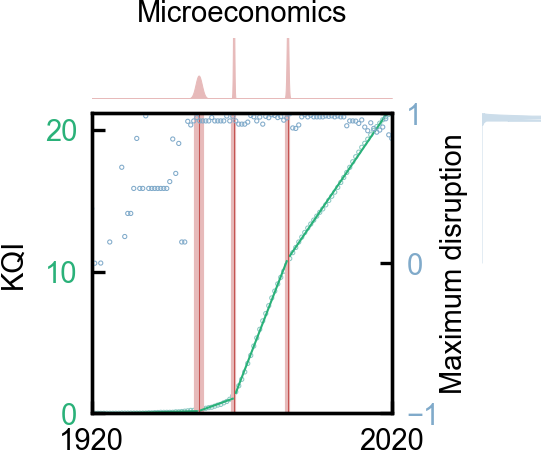}
    \includegraphics[height=4cm]{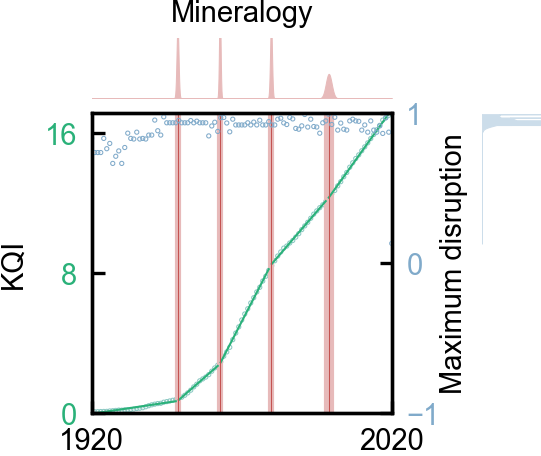}
    \includegraphics[height=4cm]{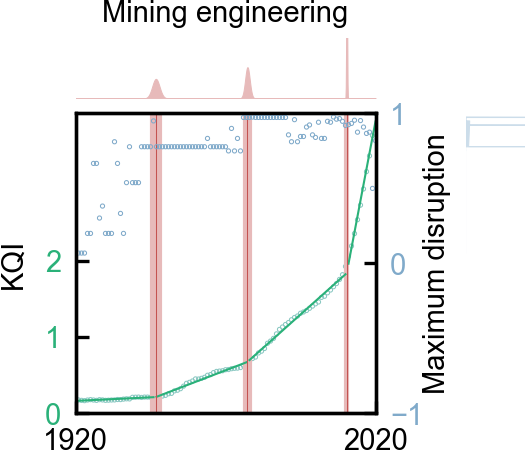}
    \includegraphics[height=4cm]{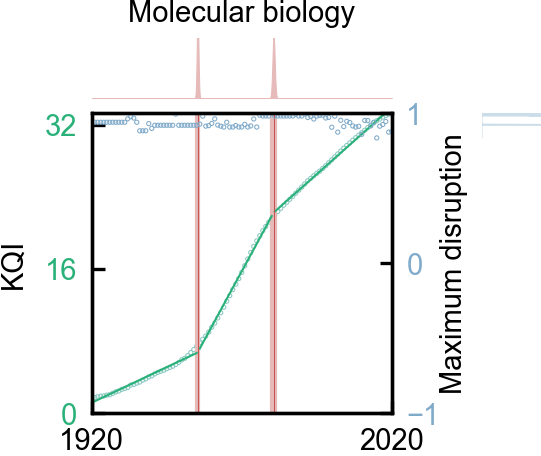}
    \includegraphics[height=4cm]{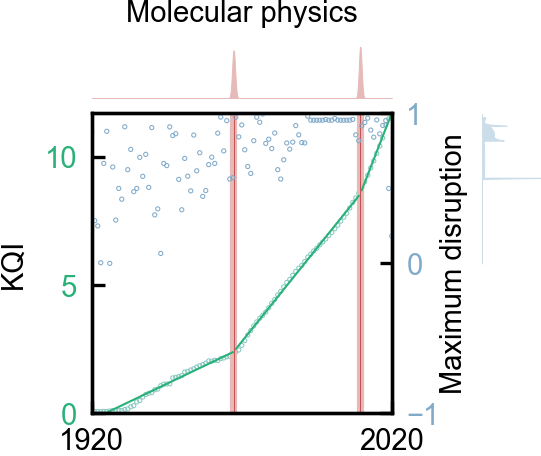}
    \includegraphics[height=4cm]{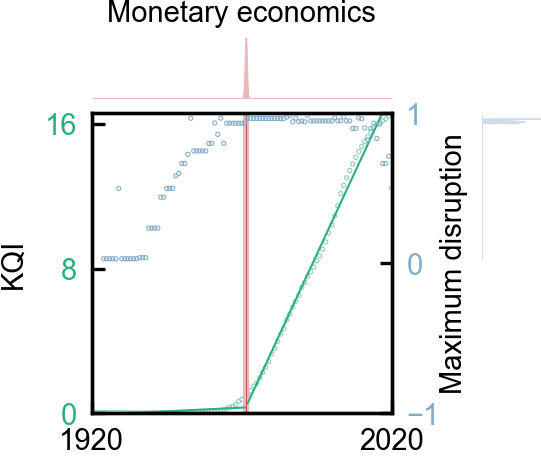}
    \includegraphics[height=4cm]{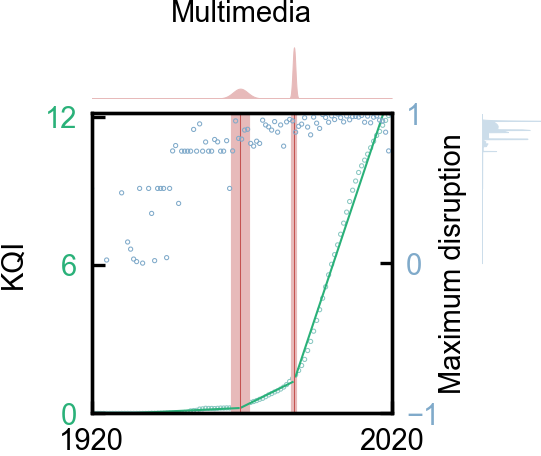}
\end{figure}
\begin{figure}[H]
    \centering
    \includegraphics[height=4cm]{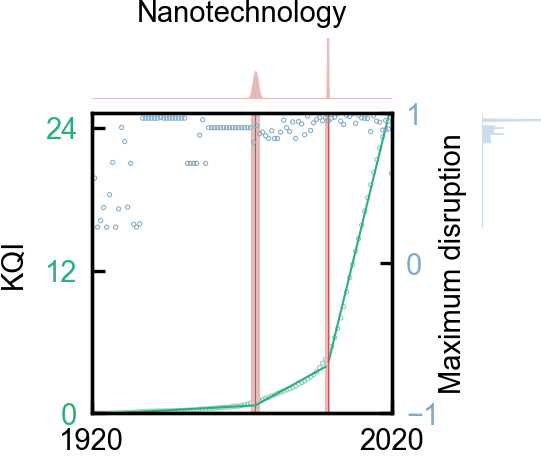}
    \includegraphics[height=4cm]{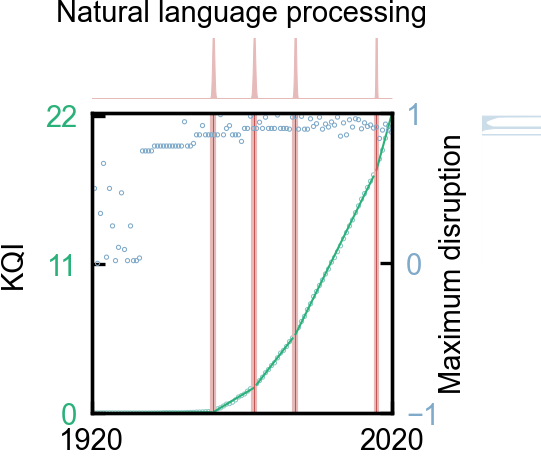}
    \includegraphics[height=4cm]{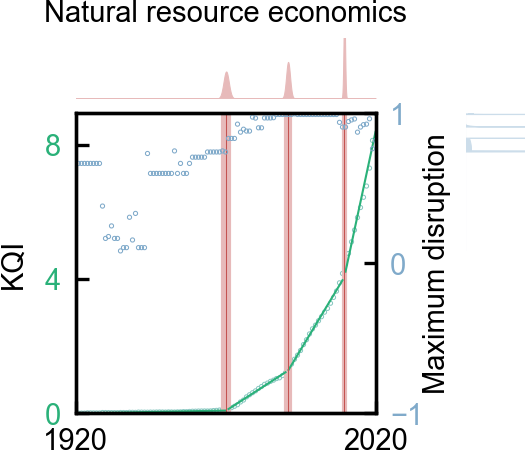}
    \includegraphics[height=4cm]{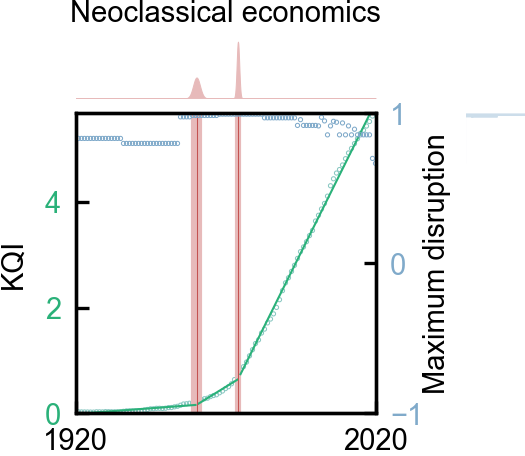}
    \includegraphics[height=4cm]{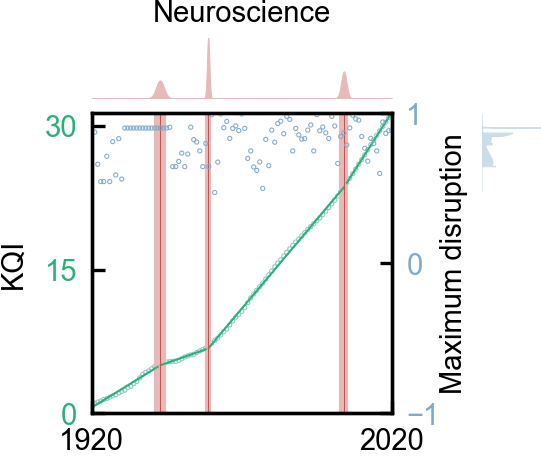}
    \includegraphics[height=4cm]{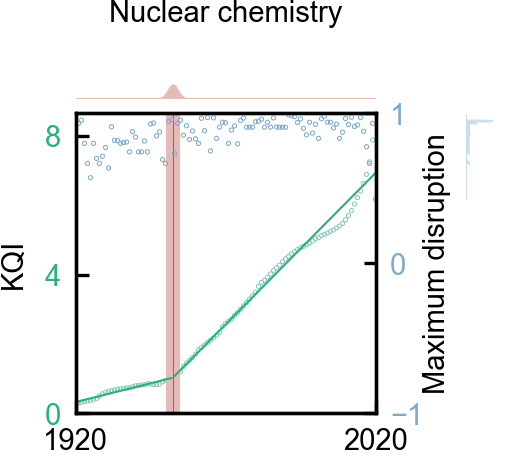}
    \includegraphics[height=4cm]{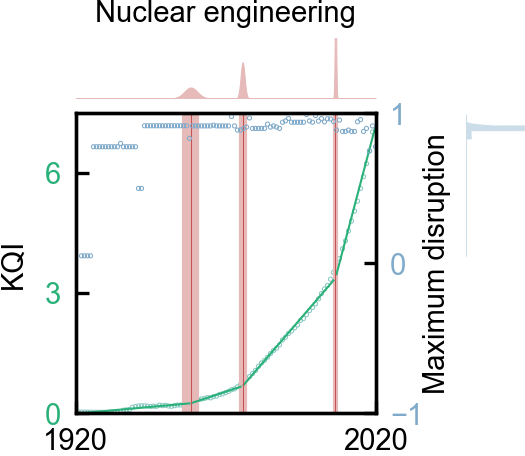}
    \includegraphics[height=4cm]{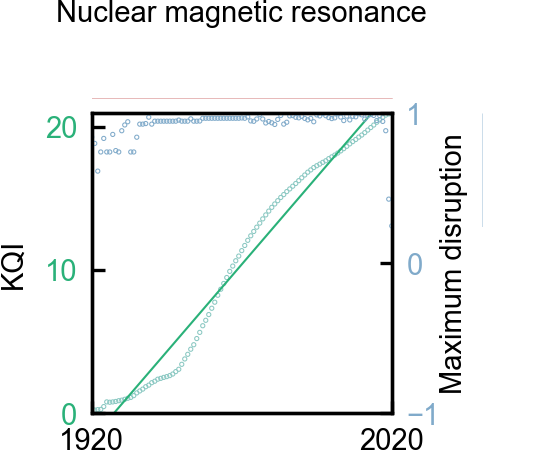}
    \includegraphics[height=4cm]{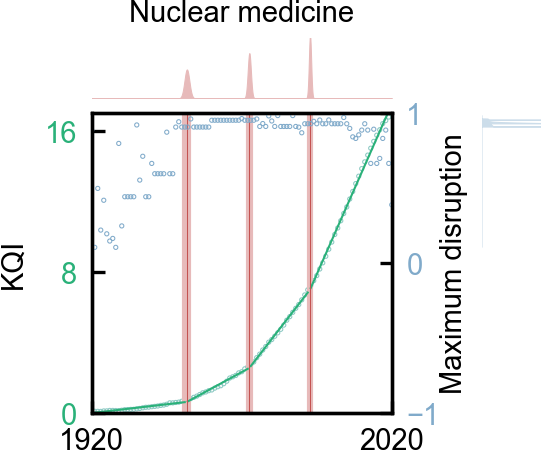}
    \includegraphics[height=4cm]{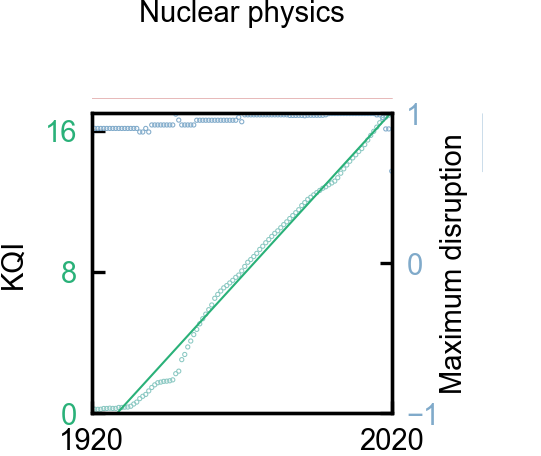}
    \includegraphics[height=4cm]{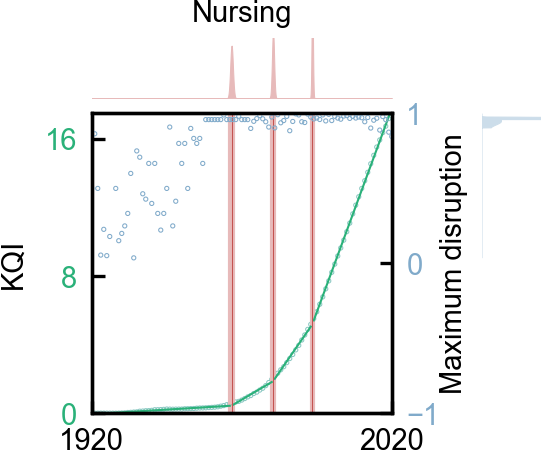}
    \includegraphics[height=4cm]{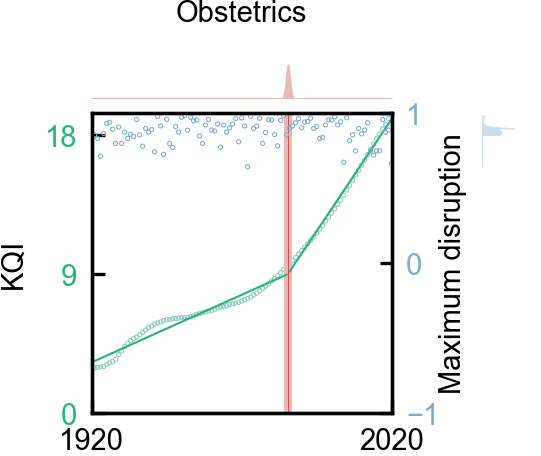}
    \includegraphics[height=4cm]{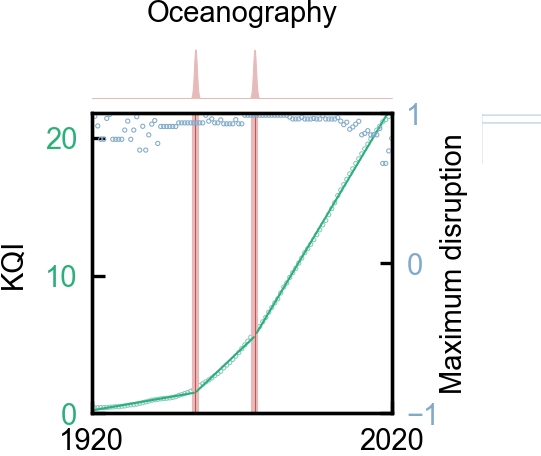}
    \includegraphics[height=4cm]{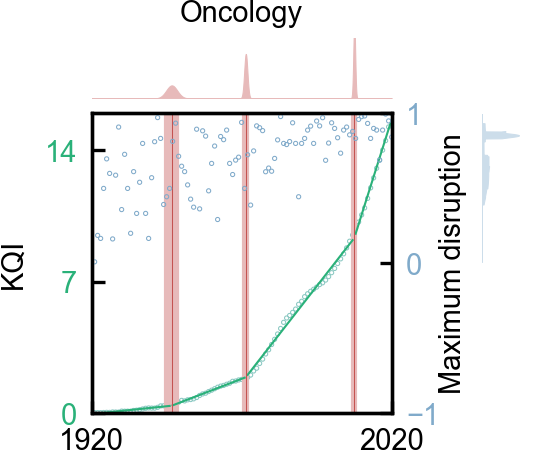}
    \includegraphics[height=4cm]{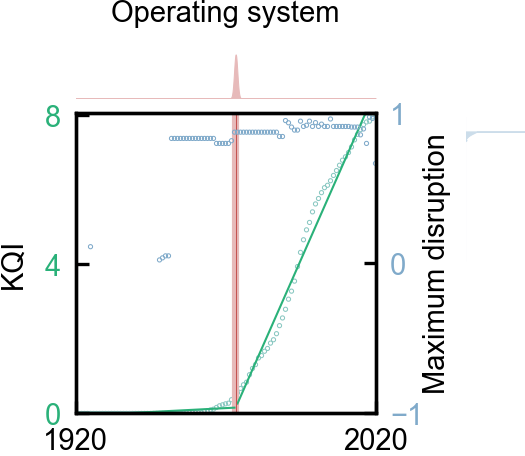}
\end{figure}
\begin{figure}[H]
    \centering
    \includegraphics[height=4cm]{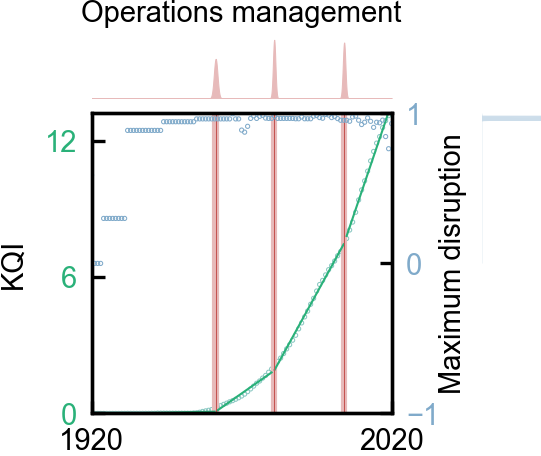}
    \includegraphics[height=4cm]{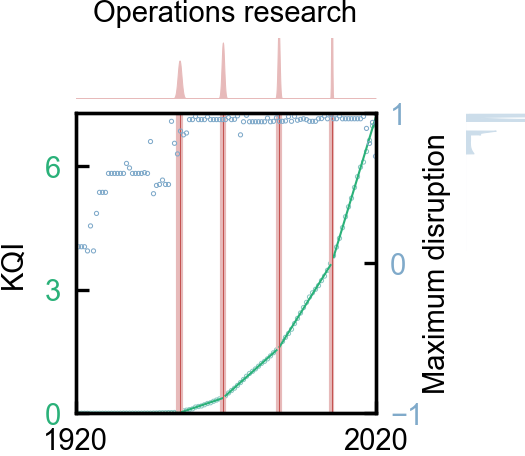}
    \includegraphics[height=4cm]{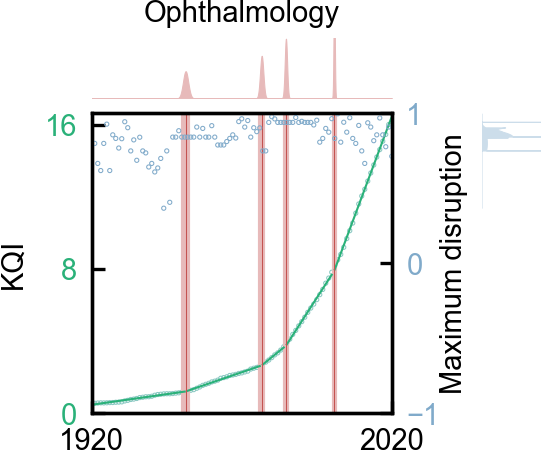}
    \includegraphics[height=4cm]{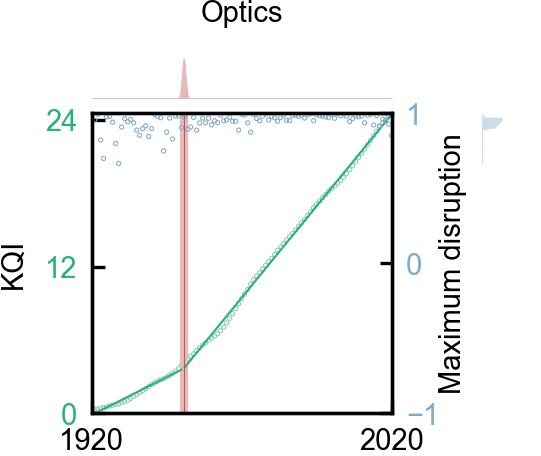}
    \includegraphics[height=4cm]{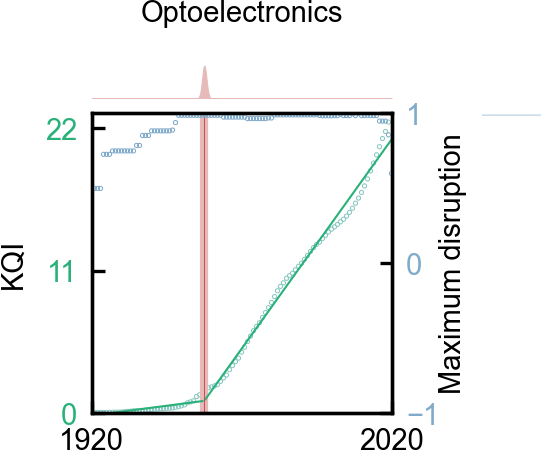}
    \includegraphics[height=4cm]{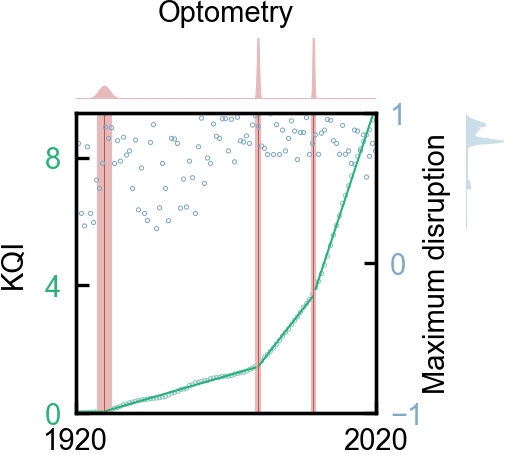}
    \includegraphics[height=4cm]{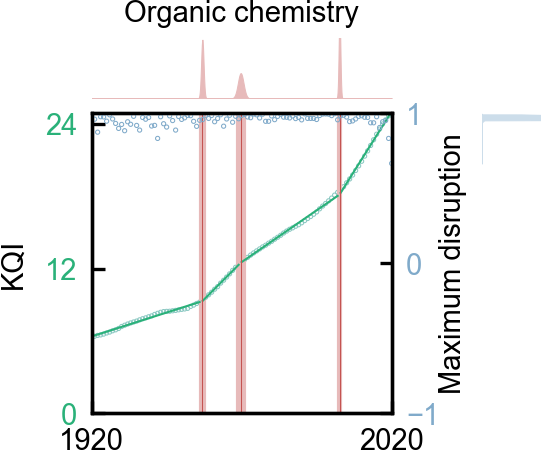}
    \includegraphics[height=4cm]{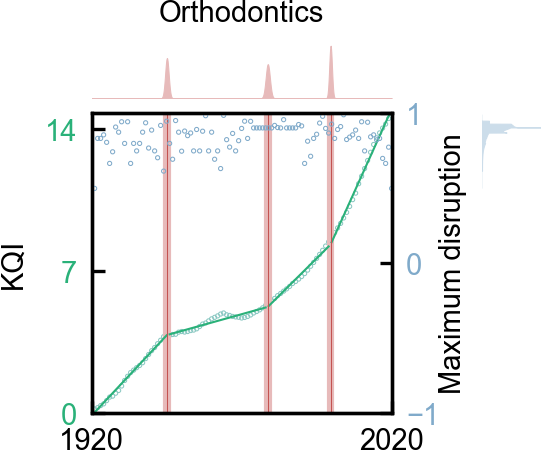}
    \includegraphics[height=4cm]{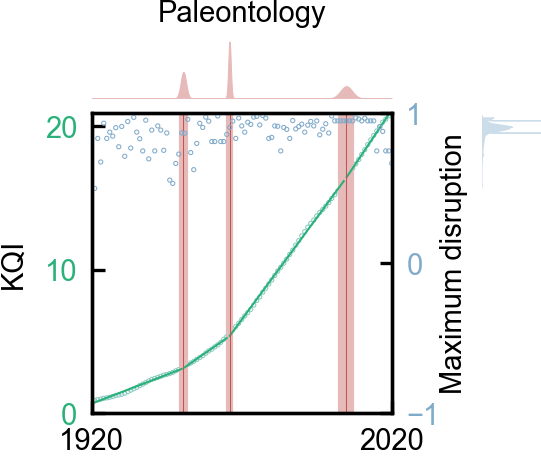}
    \includegraphics[height=4cm]{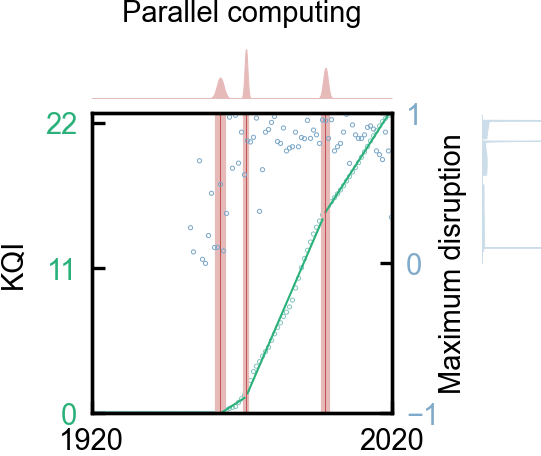}
    \includegraphics[height=4cm]{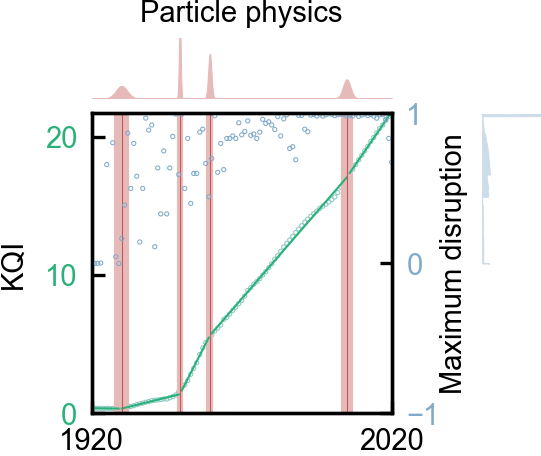}
    \includegraphics[height=4cm]{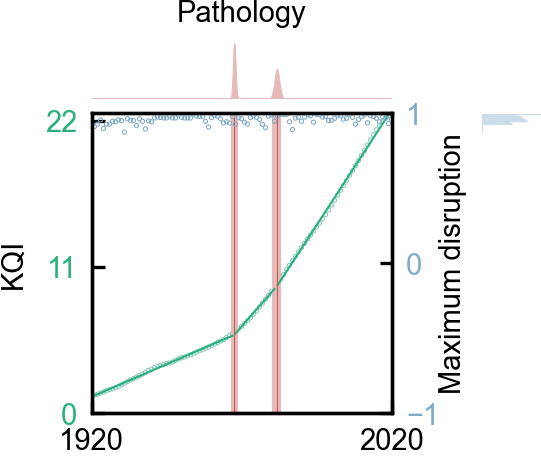}
    \includegraphics[height=4cm]{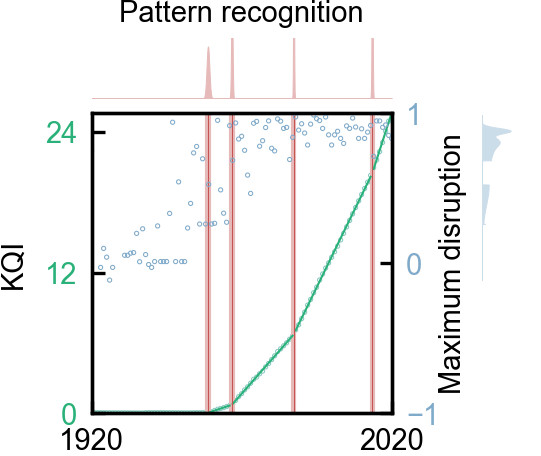}
    \includegraphics[height=4cm]{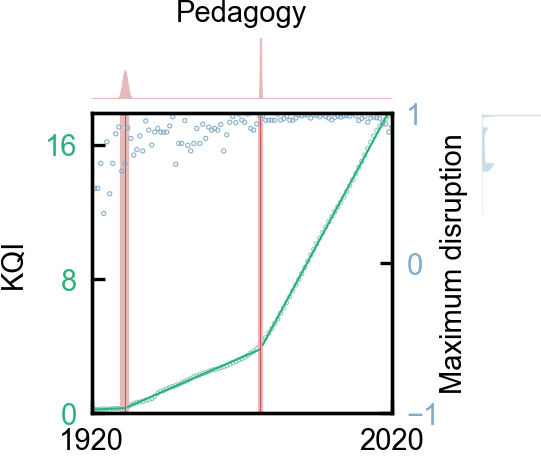}
    \includegraphics[height=4cm]{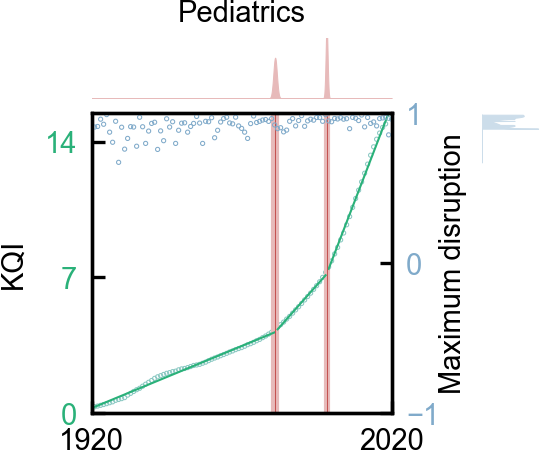}
\end{figure}
\begin{figure}[H]
    \centering
    \includegraphics[height=4cm]{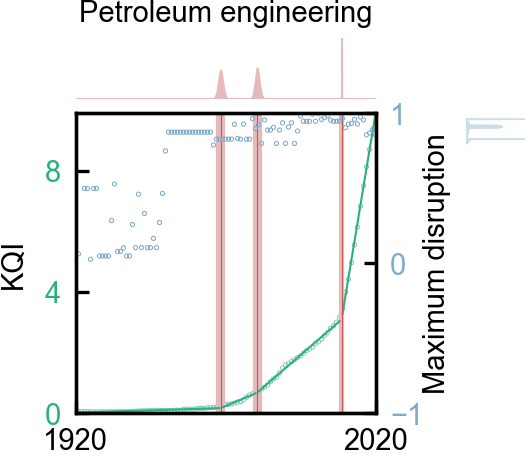}
    \includegraphics[height=4cm]{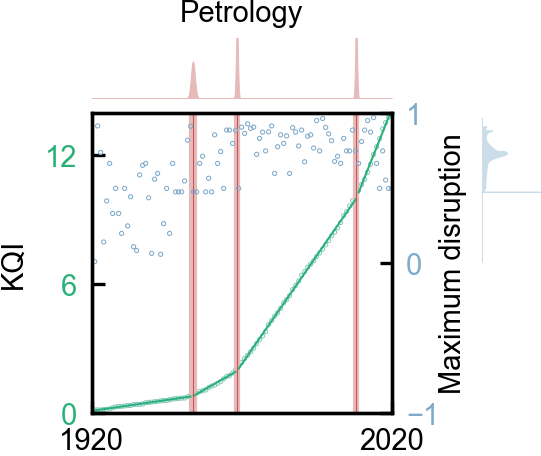}
    \includegraphics[height=4cm]{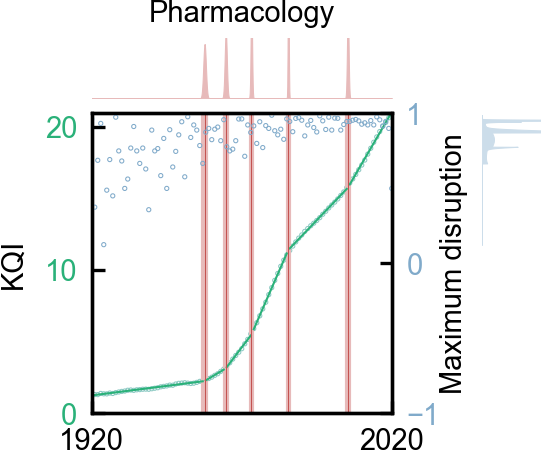}
    \includegraphics[height=4cm]{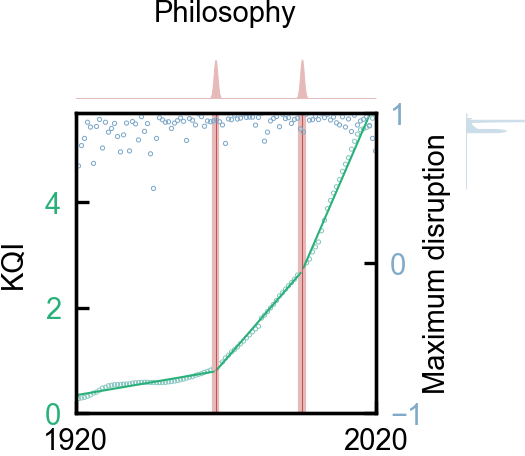}
    \includegraphics[height=4cm]{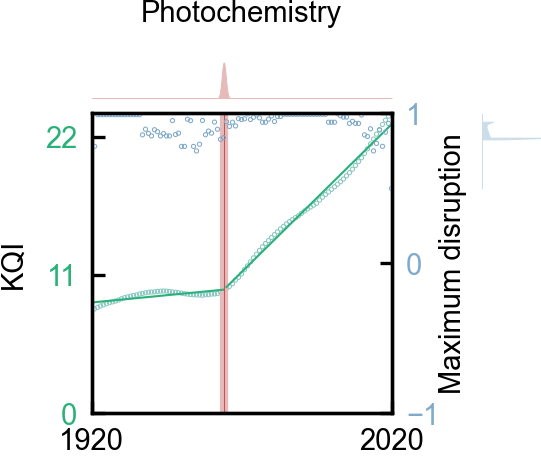}
    \includegraphics[height=4cm]{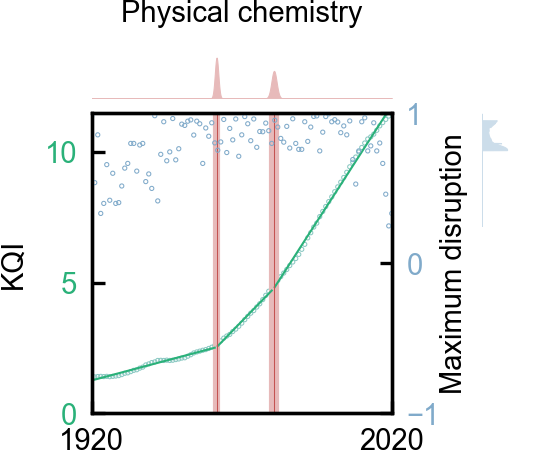}
    \includegraphics[height=4cm]{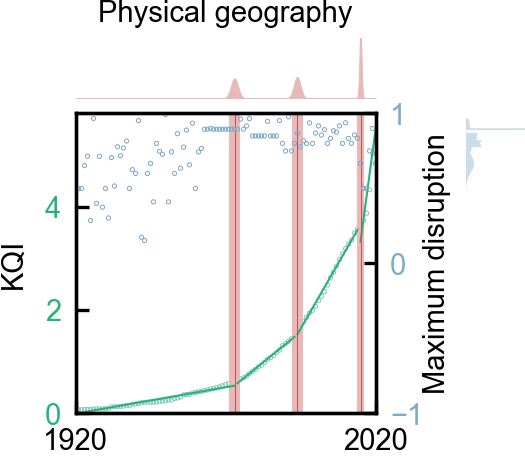}
    \includegraphics[height=4cm]{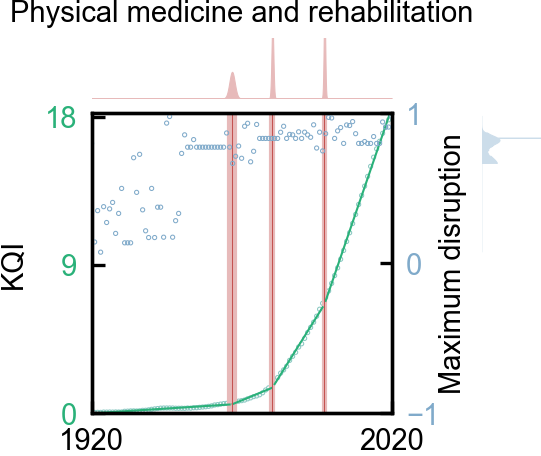}
    \includegraphics[height=4cm]{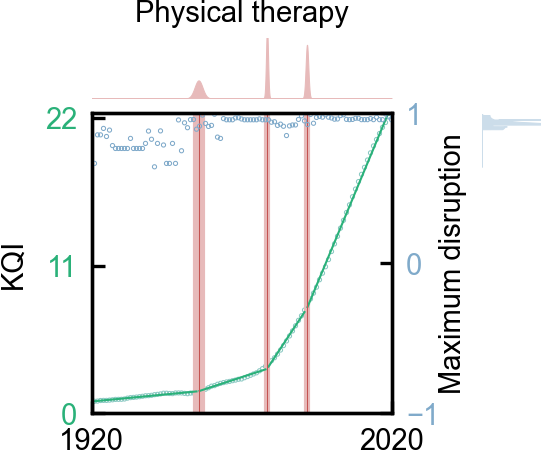}
    \includegraphics[height=4cm]{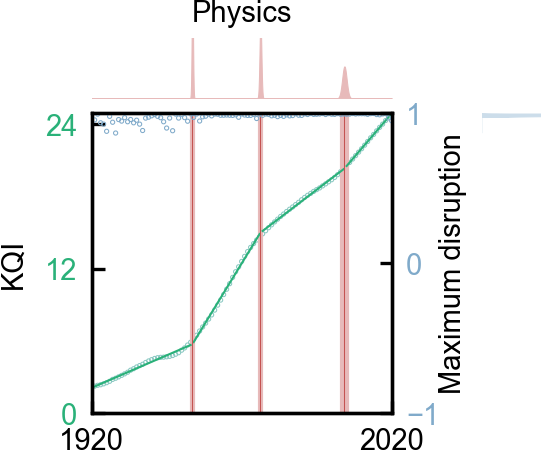}
    \includegraphics[height=4cm]{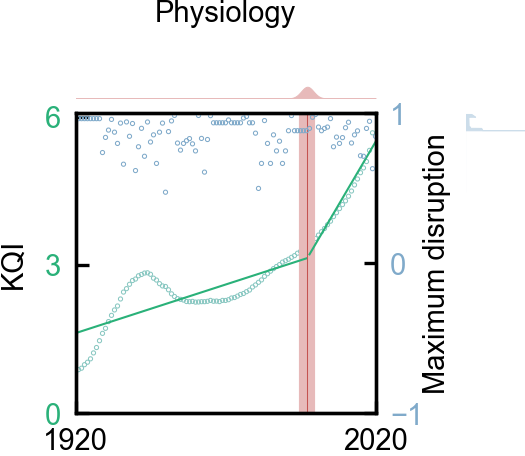}
    \includegraphics[height=4cm]{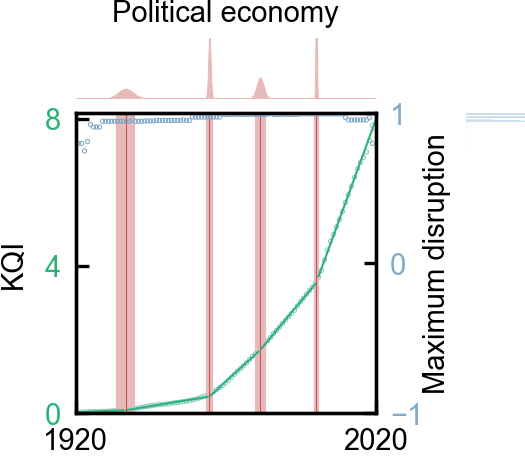}
    \includegraphics[height=4cm]{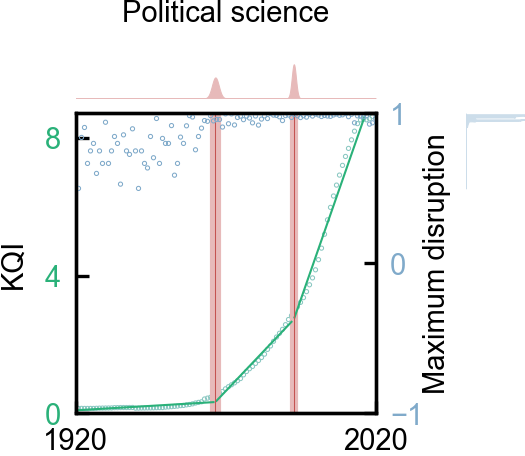}
    \includegraphics[height=4cm]{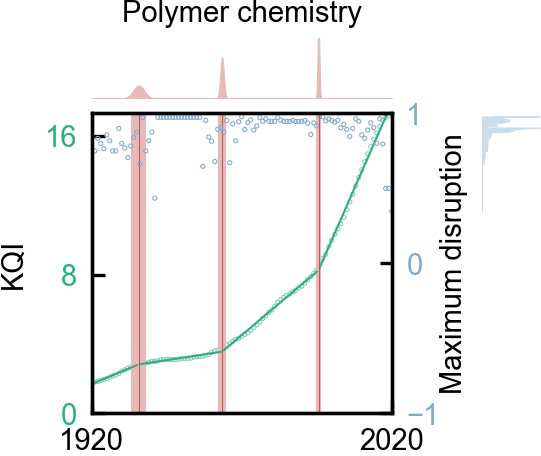}
    \includegraphics[height=4cm]{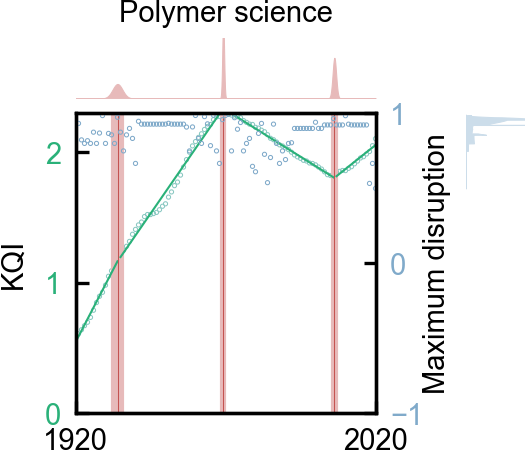}
\end{figure}
\begin{figure}[H]
    \centering
    \includegraphics[height=4cm]{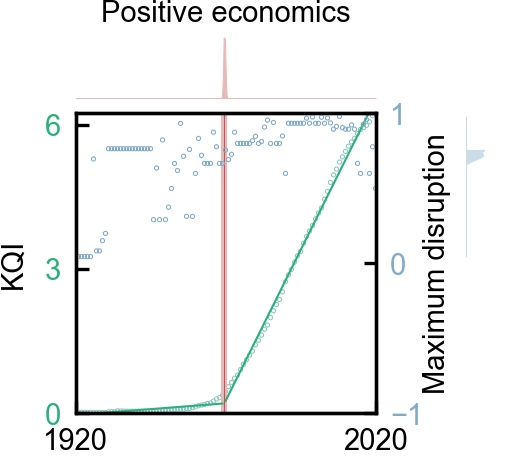}
    \includegraphics[height=4cm]{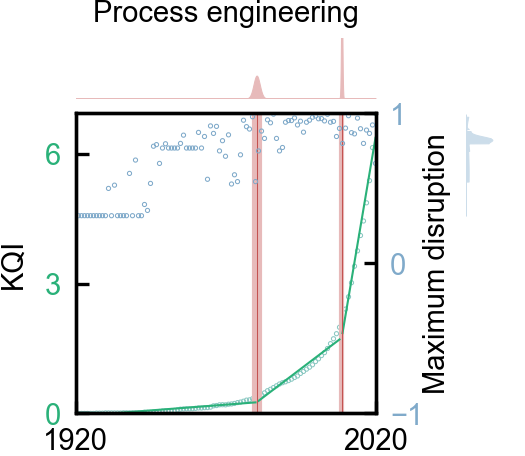}
    \includegraphics[height=4cm]{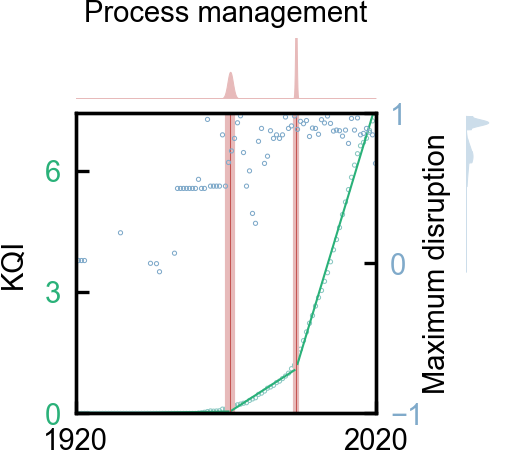}
    \includegraphics[height=4cm]{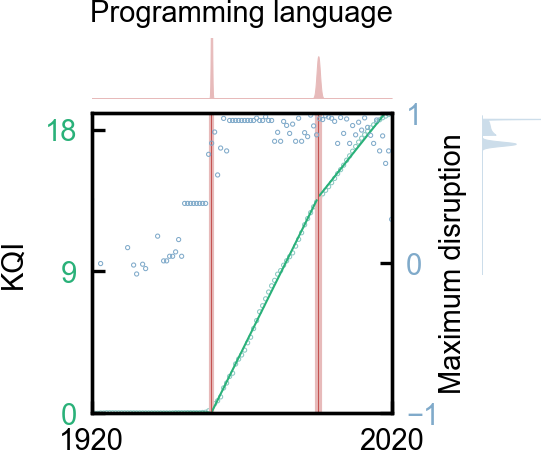}
    \includegraphics[height=4cm]{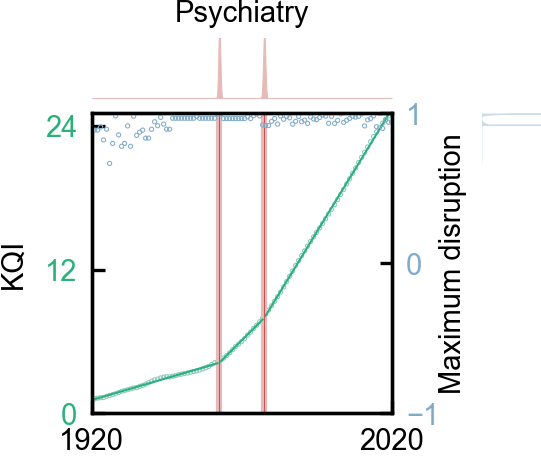}
    \includegraphics[height=4cm]{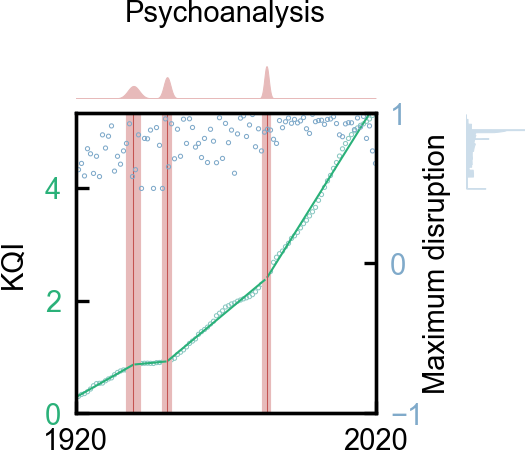}
    \includegraphics[height=4cm]{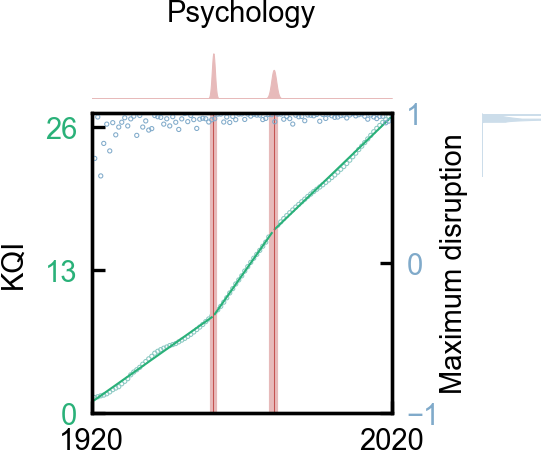}
    \includegraphics[height=4cm]{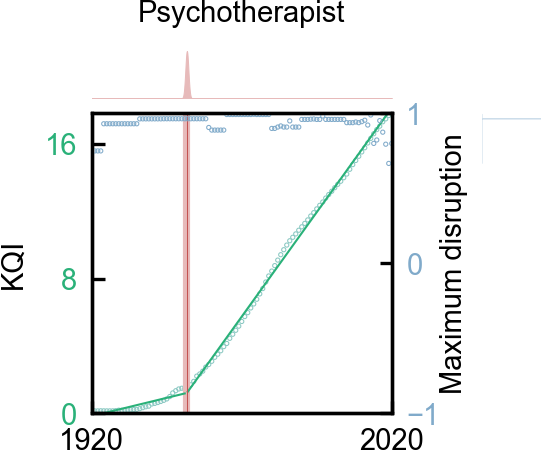}
    \includegraphics[height=4cm]{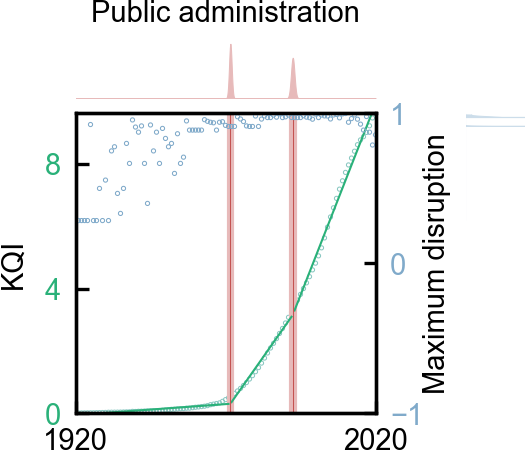}
    \includegraphics[height=4cm]{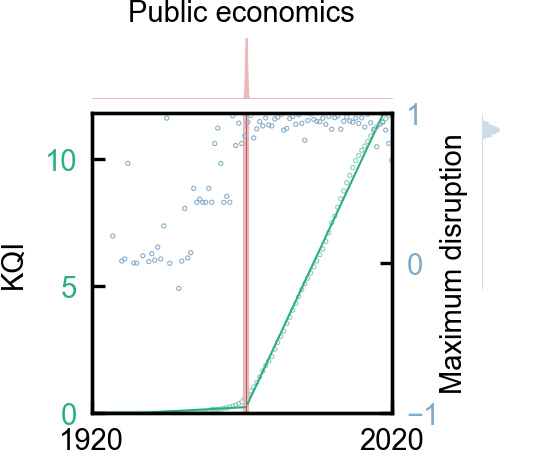}
    \includegraphics[height=4cm]{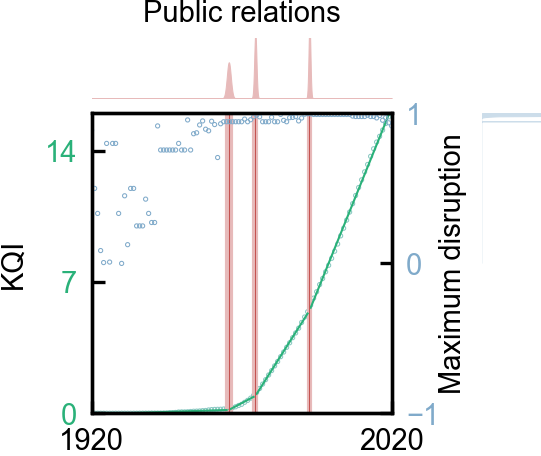}
    \includegraphics[height=4cm]{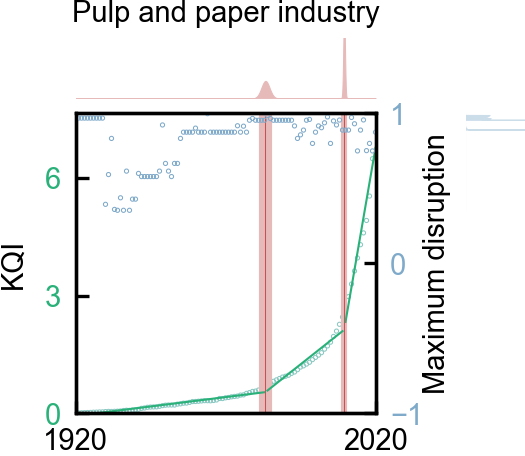}
    \includegraphics[height=4cm]{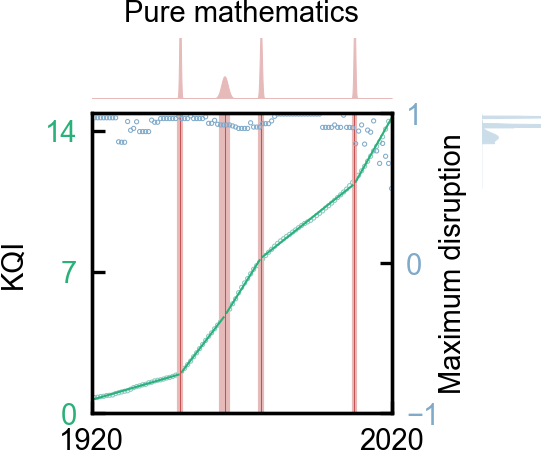}
    \includegraphics[height=4cm]{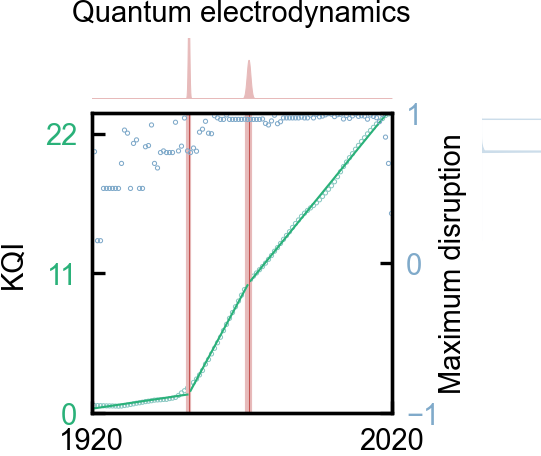}
    \includegraphics[height=4cm]{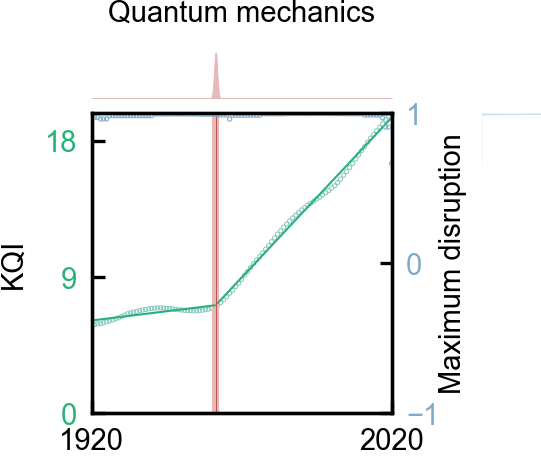}
\end{figure}
\begin{figure}[H]
    \centering
    \includegraphics[height=4cm]{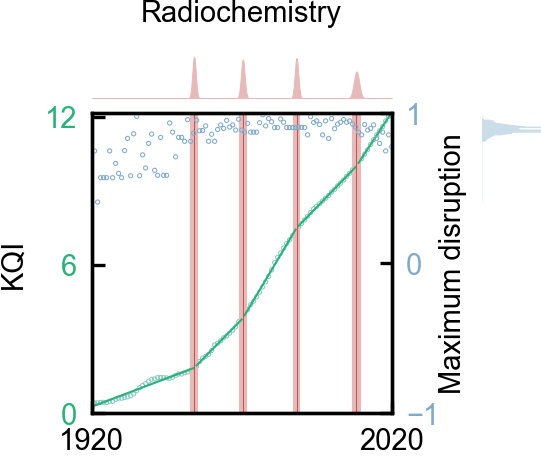}
    \includegraphics[height=4cm]{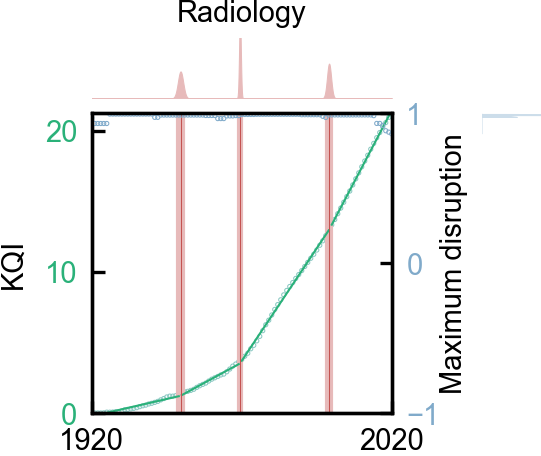}
    \includegraphics[height=4cm]{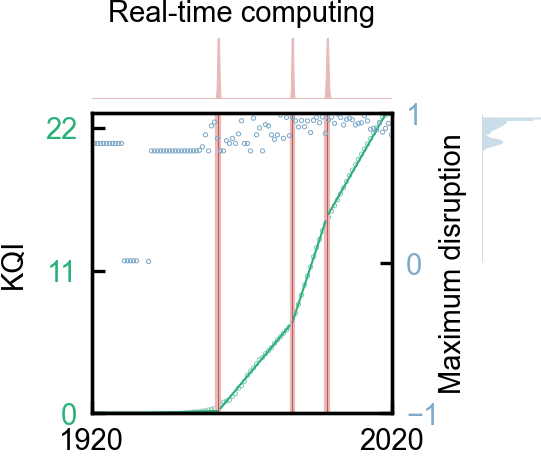}
    \includegraphics[height=4cm]{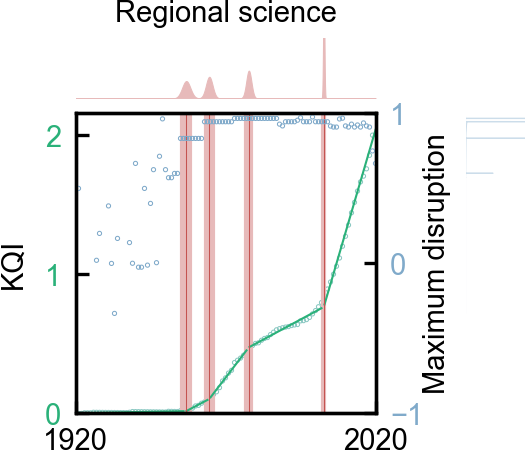}
    \includegraphics[height=4cm]{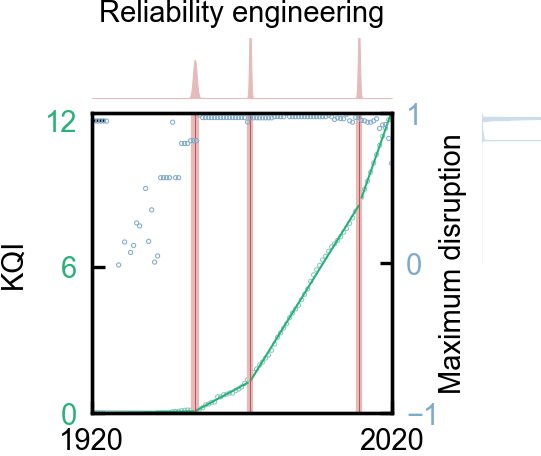}
    \includegraphics[height=4cm]{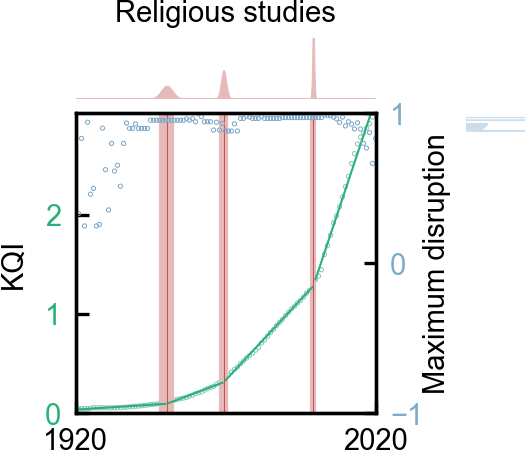}
    \includegraphics[height=4cm]{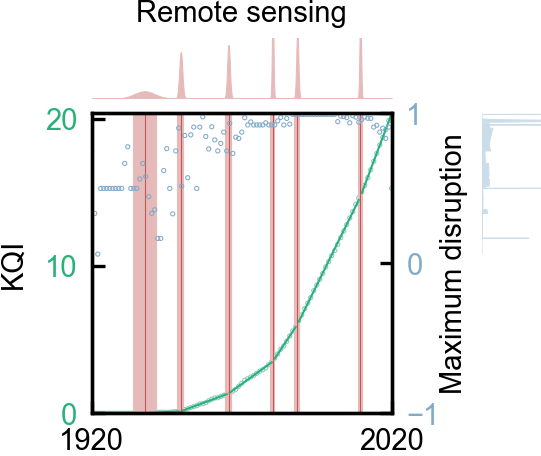}
    \includegraphics[height=4cm]{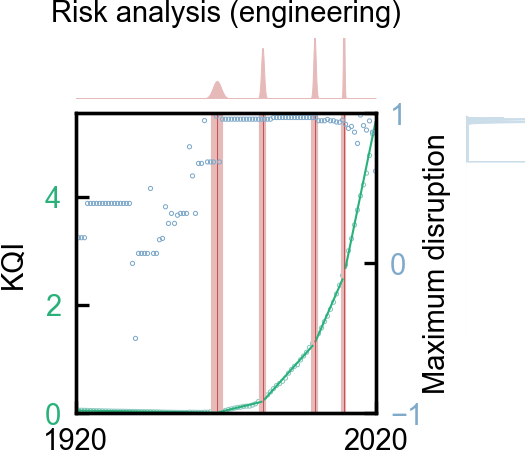}
    \includegraphics[height=4cm]{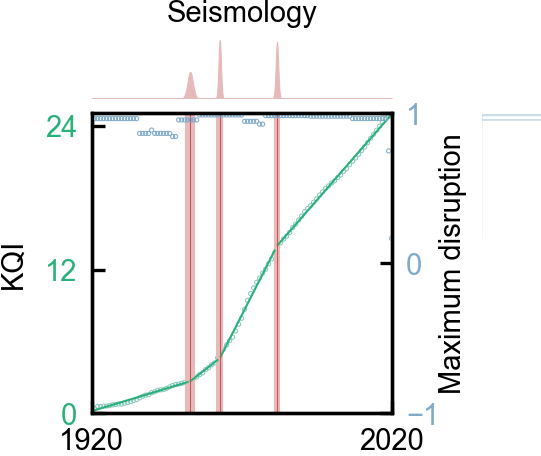}
    \includegraphics[height=4cm]{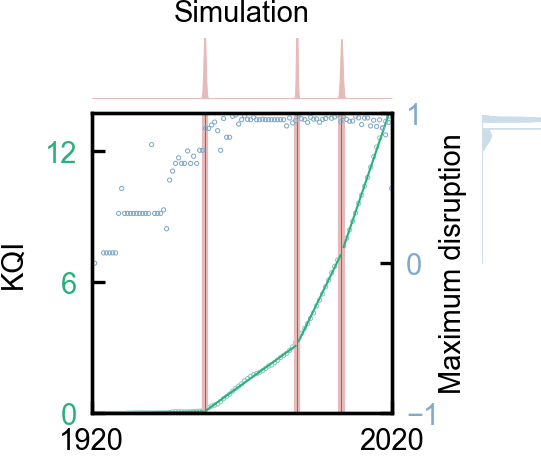}
    \includegraphics[height=4cm]{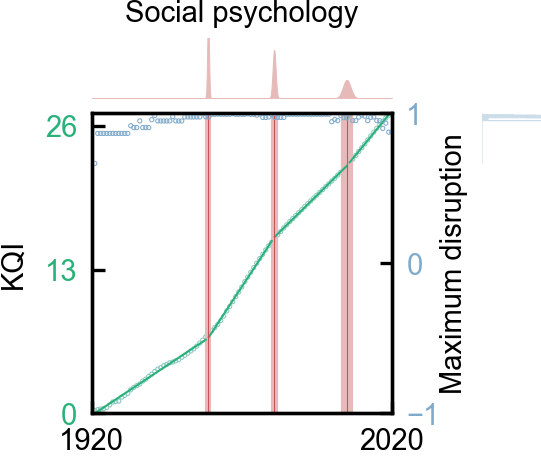}
    \includegraphics[height=4cm]{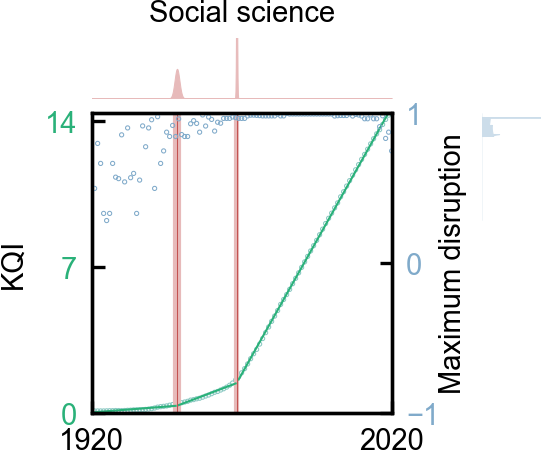}
    \includegraphics[height=4cm]{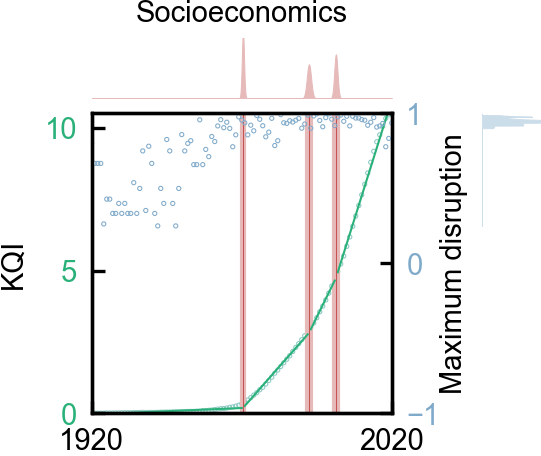}
    \includegraphics[height=4cm]{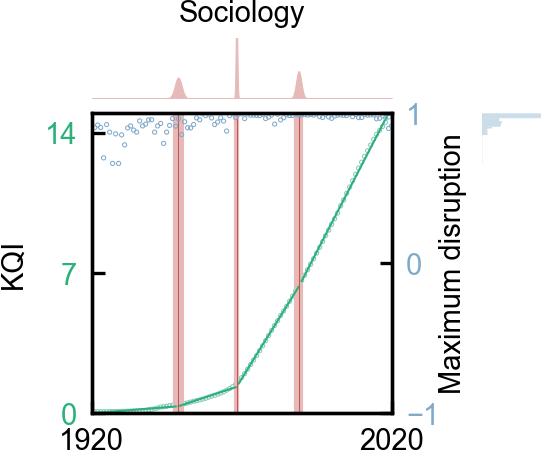}
    \includegraphics[height=4cm]{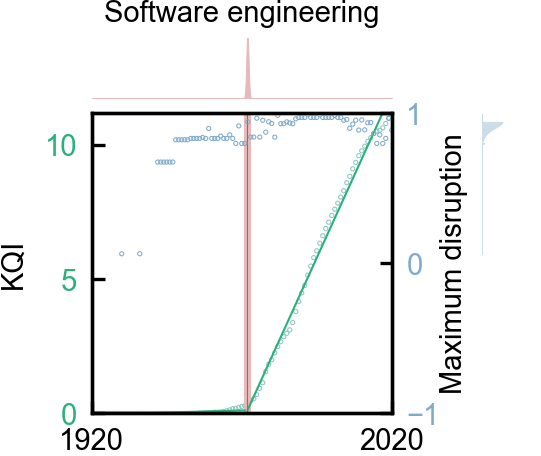}
\end{figure}
\begin{figure}[H]
    \centering
    \includegraphics[height=4cm]{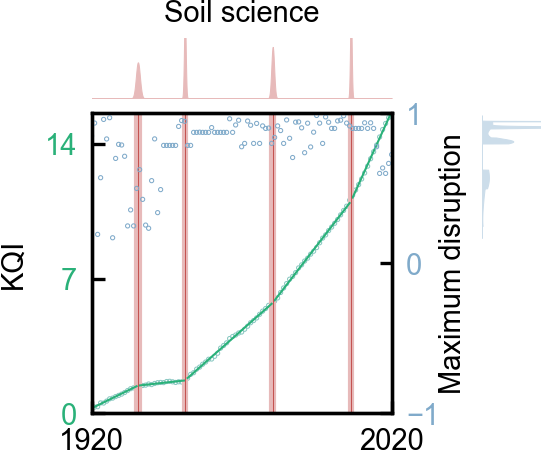}
    \includegraphics[height=4cm]{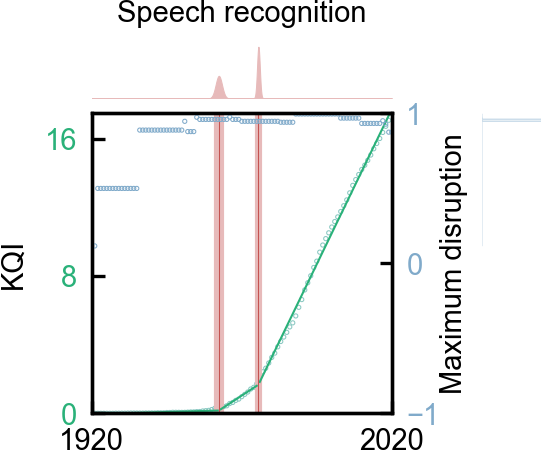}
    \includegraphics[height=4cm]{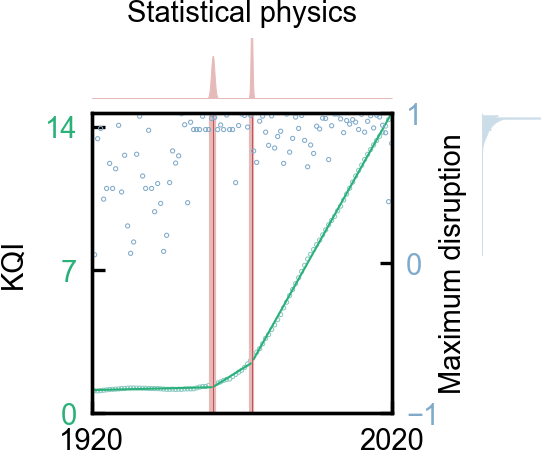}
    \includegraphics[height=4cm]{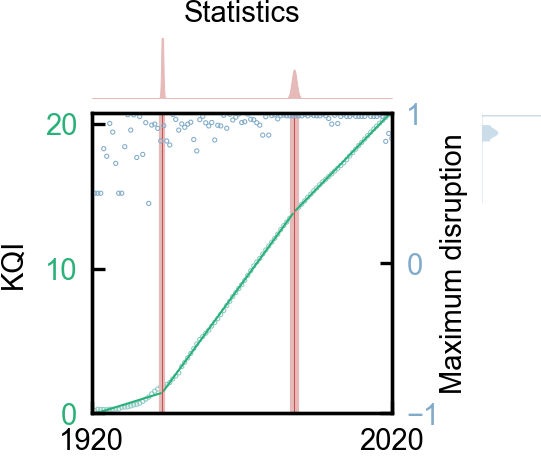}
    \includegraphics[height=4cm]{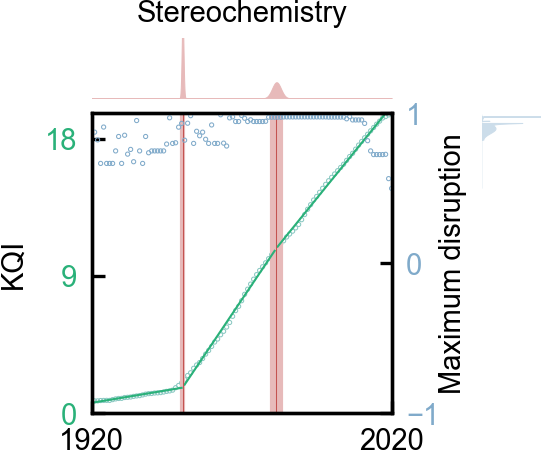}
    \includegraphics[height=4cm]{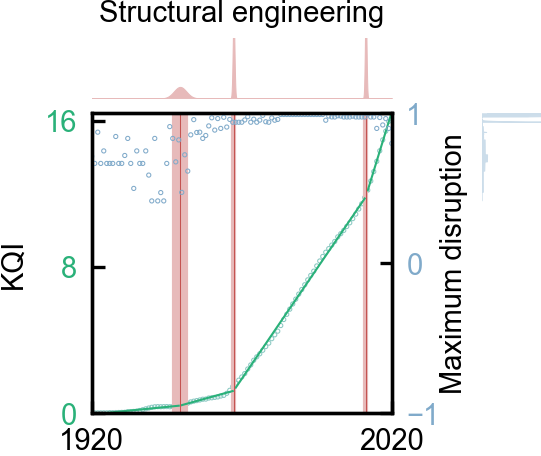}
    \includegraphics[height=4cm]{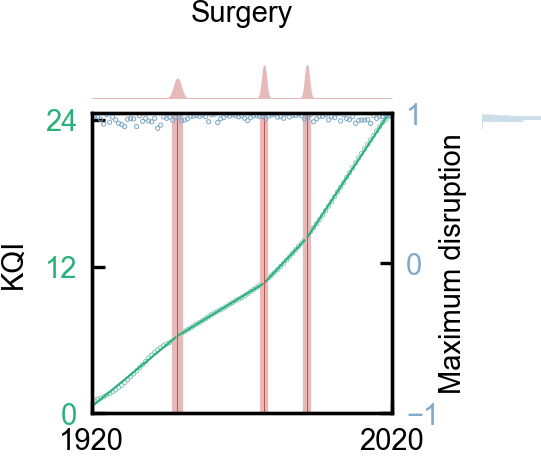}
    \includegraphics[height=4cm]{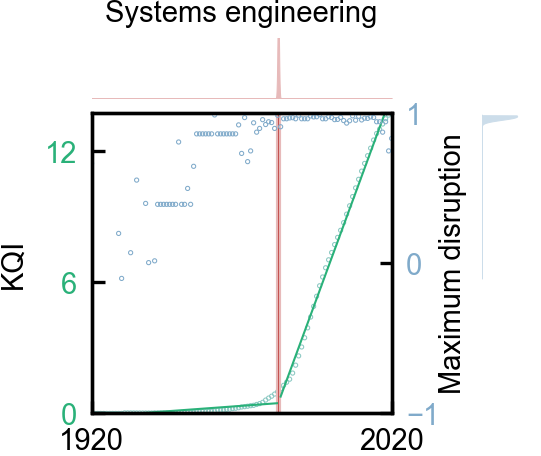}
    \includegraphics[height=4cm]{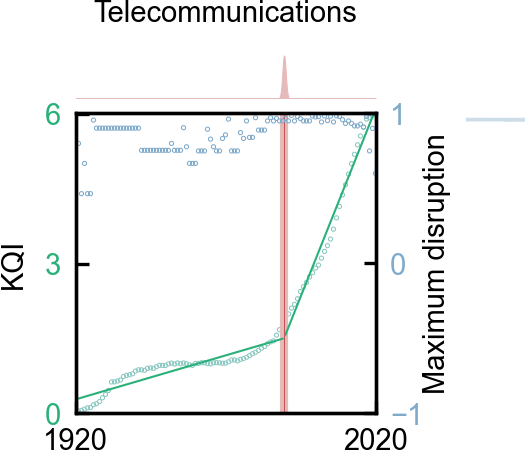}
    \includegraphics[height=4cm]{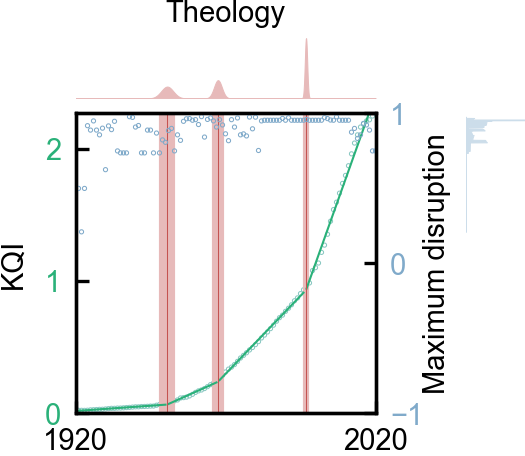}
    \includegraphics[height=4cm]{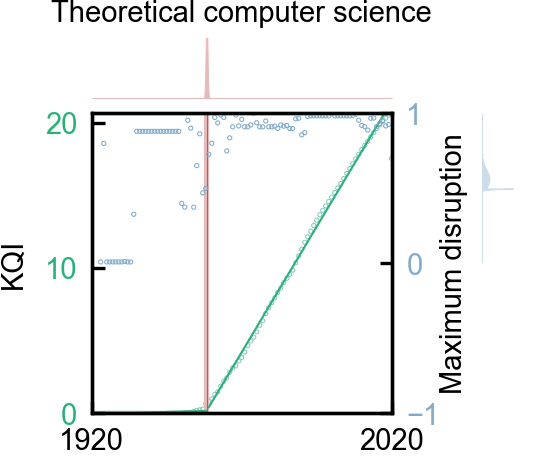}
    \includegraphics[height=4cm]{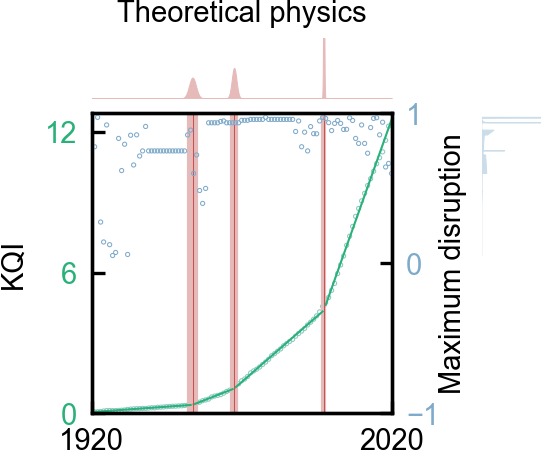}
    \includegraphics[height=4cm]{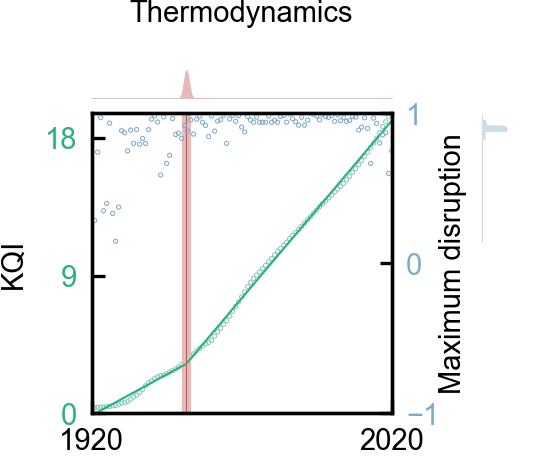}
    \includegraphics[height=4cm]{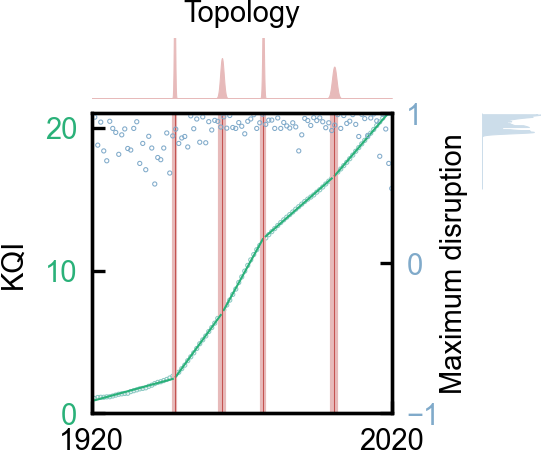}
    \includegraphics[height=4cm]{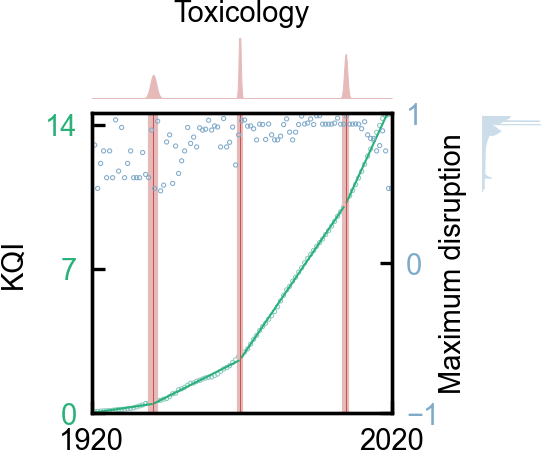}
\end{figure}
\begin{figure}[H]
    \centering
    \includegraphics[height=4cm]{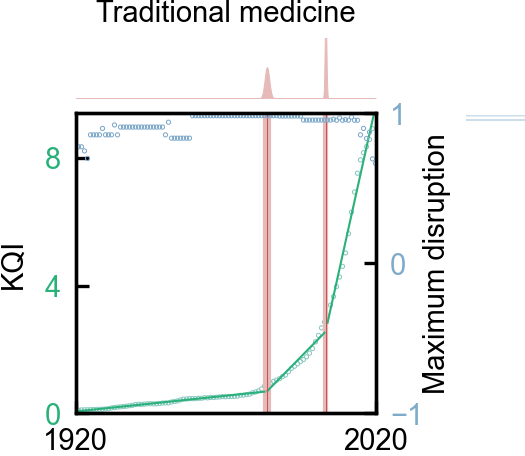}
    \includegraphics[height=4cm]{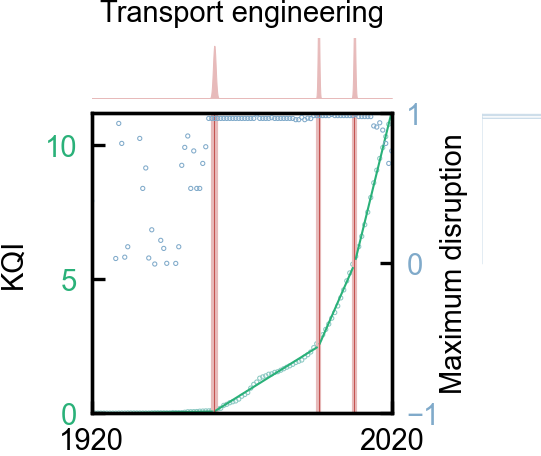}
    \includegraphics[height=4cm]{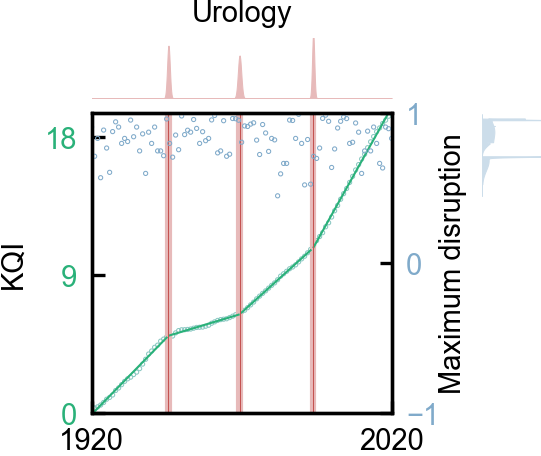}
    \includegraphics[height=4cm]{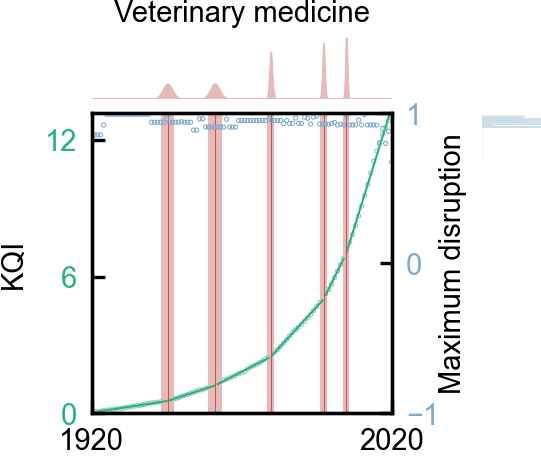}
    \includegraphics[height=4cm]{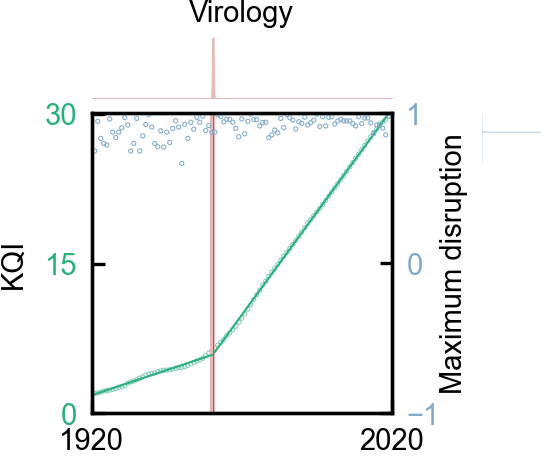}
    \includegraphics[height=4cm]{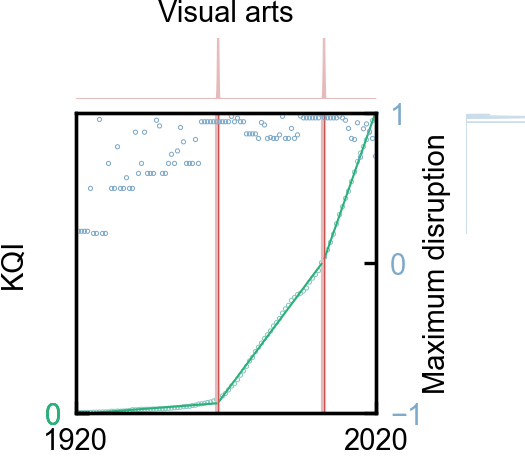}
    \includegraphics[height=4cm]{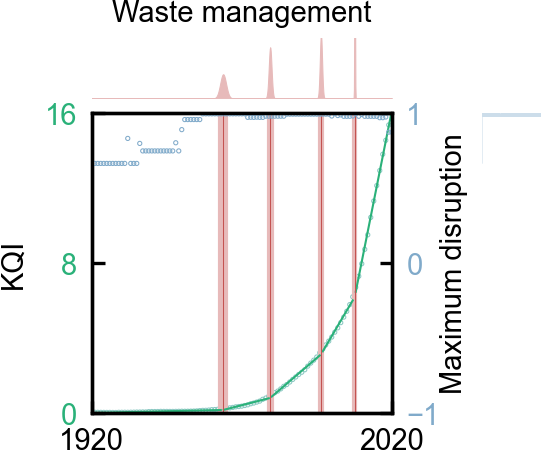}
    \includegraphics[height=4cm]{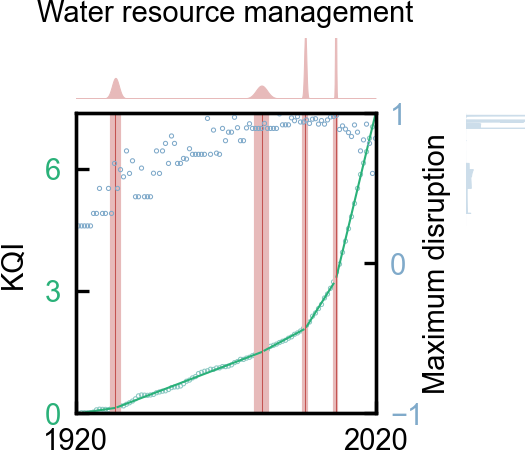}
    \includegraphics[height=4cm]{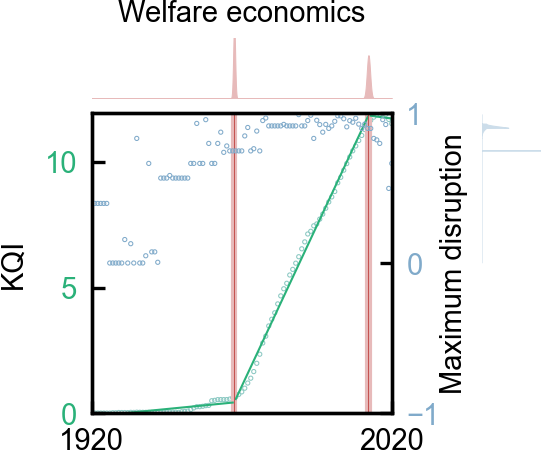}
    \includegraphics[height=4cm]{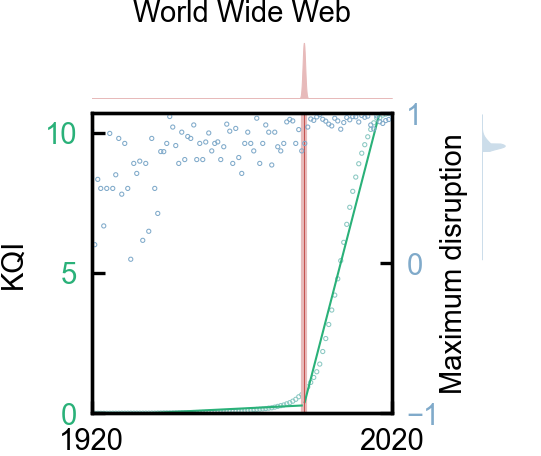}
    \includegraphics[height=4cm]{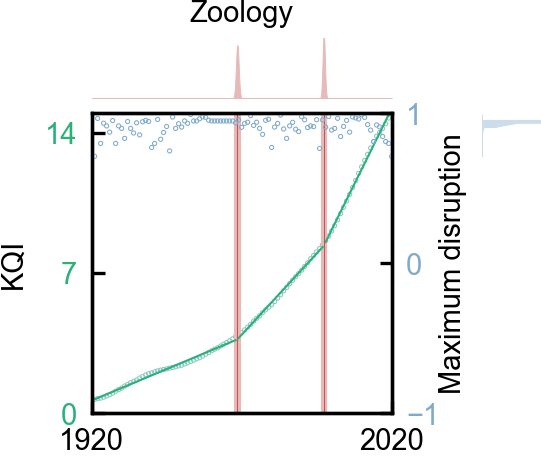}
\end{figure}

%%% Add this line AFTER all your figures and tables
% \FloatBarrier

% % \movie{Type legend for the movie here.}

% % \movie{Type legend for the other movie here. Adding longer text to show what happens, to decide on alignment and/or indentations.}

% % \movie{A third movie, just for kicks.}

\newpage

\subsection*{SI Dataset S1 (SourceDataFig.2.rar)}
\begin{itemize}
    \item[] Source data used in Figure 2.
\end{itemize}

\subsection*{SI Dataset S2 (SourceDataFig.3.rar)}
\begin{itemize}
    \item[] Source data used in Figure 3.
\end{itemize}

\subsection*{SI Dataset S3 (SourceDataFig.4.rar)}
\begin{itemize}
    \item[] Source data used in Figure 4.
\end{itemize}

\subsection*{SI Dataset S4 (SourceDataFig.5.rar)}
\begin{itemize}
    \item[] Source data used in Figure 5.
\end{itemize}

% \bibliography{pnas-sample}